\documentclass{aa}
\usepackage{txfonts}
\usepackage{graphicx}

\begin{document}

\title{A multi-wavelength de-blended {\it Herschel} view of the statistical properties of dusty star-forming galaxies across cosmic time}
\author{L. Wang\inst{1, 2}, W. J. Pearson\inst{1, 2}, W. Cowley\inst{2}, J. W. Trayford\inst{3}, M. B{\'e}thermin\inst{4}, C. Gruppioni\inst{5}, P. Hurley\inst{6}, M. J. Micha{\l}owski\inst{7}}

\institute{SRON Netherlands Institute for Space Research, Landleven 12, 9747 AD, Groningen, The Netherlands \email{l.wang@sron.nl} 
\and Kapteyn Astronomical Institute, University of Groningen, Postbus 800, 9700 AV Groningen, the Netherlands
\and Leiden Observatory, Leiden University, PO Box 9513, NL-2300 RA Leiden, the Netherlands
\and Aix Marseille Univ, CNRS, LAM, Laboratoire d'Astrophysique de Marseille, Marseille, France
\and Istituto Nazionale di Astrofisica (INAF) - Osservatorio Astronomico di Bologna, via Gobetti 93/3, I-40129 Bologna, Italy
\and Astronomy Centre, Department of Physics and Astronomy, University of Sussex, Falmer, Brighton BN1 9QH, UK
\and Astronomical Observatory Institute, Faculty of Physics, Adam Mickiewicz University, ul. S{\l}oneczna 36, 60-286 Poznan, Poland
}

\date{Received / Accepted}

\abstract
   {}
   {We aim to study the statistical properties of dusty star-forming galaxies across cosmic time, such as their number counts, luminosity functions (LF) and dust-obscured star-formation rate density (SFRD).}
   {We use state-of-the-art de-blended {\it Herschel} catalogue in the COSMOS field to measure the number counts and LFs at far-infrared (FIR) and sub-millimetre (sub-mm) wavelengths. The de-blended catalogue is generated by combining the probabilistic Bayesian source extraction tool XID+ and informative prior on the spectral energy distributions derived from the associated deep multi-wavelength photometric data. We compare our results with previous measurements and predictions from a range of empirical models and physically-motivated simulations.}
   {Thanks to our de-confusion technique and the wealth of deep multi-wavelength photometric information in COSMOS, we are able to achieve more accurate measurements of the number counts and LFs while at the same time probing roughly ten times below the {\it Herschel} confusion limit. Our number counts at 250 $\mu$m agree well with previous {\it Herschel} studies. However, our counts at 350 and 500 $\mu$m are considerably below previous {\it Herschel} results. This is due to previous {\it Herschel} studies suffering from source confusion and blending issues which is progressively worse towards longer wavelength. Our number counts at 450 and 870 $\mu$m show excellent agreement with previous determinations derived from single dish observations with SCUBA-2 on the JCMT and interferometric observations with ALMA and SMA. Our measurements of both the monochromatic LF at 250 $\mu$m and the total IR LF agree well with previous results in the overlapping redshift and luminosity range. The increased dynamic range of our measurements allows us to better measure the faint-end slope of the LF and measure the dust-obscured SFRD out to $z\sim6$. We find that the fraction of dust obscured star-formation activity is at its highest ($>80\%$) around $z\sim1$ which then decreases towards both low and high redshift. We do not find a shift of balance between $z\sim3$ and $z\sim4$ in the cosmic star-formation history from being dominated by unobscured star formation at higher redshift to obscured star formation at lower redshift. However, we do find the redshift range $3<z<4$ to be an interesting transition period as the fraction of the total SFRD that is obscured by dust is significantly lower at higher redshifts.}
   {}


\titlerunning{Number counts and luminosity functions at far-infrared and sub-millimetre wavelengths}

\titlerunning{A multi-wavelength de-blended {\it Herschel} view of the statistical properties of dusty star-forming galaxies}

\authorrunning{Wang et al.}

\maketitle

\section{Introduction}

About half of all the luminous power from stars and active galactic nuclei (AGN) which makes up the extra-galactic background was emitted in the far-infrared (FIR) and submillimetre (sub-mm), as a result of re-radiation of dust heated by ultraviolet (UV)/optical photons (Puget et al. 1996; Fixsen et al. 1998; Hauser et al. 1998; Lagache et al. 1999; Hauser \& Dwek 2001; Dole et al. 2006). Therefore, a complete understanding of the cosmic star-formation history (CSFH) depends critically on taking into account the dust-obscured star-formation activity from the local Universe out to the highest redshifts (e.g., Madau \& Dickinson 2014). For this purpose, it is of fundamental importance to accurately measure the statistical properties of FIR and sub-mm galaxies and their evolution with cosmic time. Number counts (also known as source counts), which is the number density of galaxies as a function of their intrinsic flux, and luminosity function (LF), which is the volume density of galaxies as a function of their intrinsic luminosity, are statistical descriptions of the galaxy populations at the most basic level and can provide strong constraints on models of galaxy formation and evolution (e.g., Granato et al. 2004; Baugh et al. 2005; Fontanot et al. 2007; Hayward et al. 2013; Cowley et al. 2015; Lacey et al. 2016).

Conducting observations at the IR and sub-mm wavelengths is challenging because of high background and the limited angular resolution of the single-dish instruments. Over the past three decades or so, tremendous progress has been made in our understanding of the properties of the IR and sub-mm galaxy population, thanks to a succession of breakthrough  space-based and ground-based telescopes, starting from the Infrared Astronomical Satellite (IRAS; Neugebauer et al. 1984), the Infrared Space Observatory (ISO; Kessler et al. 1996), the Submillimetre Common-User Bolometer Array (SCUBA; Holland et al. 1999) and the Submillimetre Common-User Bolometer Array-2 (SCUBA-2; Holland et al. 2013) camera on the James Clerk Maxwell Telescope (JCMT), the Spitzer Space Telescope  (Werner et al. 2004), the AKARI mission (Murakami et al. 2007), the Balloon-borne Large Aperture Submillimeter Telescope (BLAST; Pascale et al. 2008), the Large APEX Bolometer Camera (LABOCA; Siringo et al. 2009) on the Atacama Pathfinder Experiment telescope (APEX), and more recently the {\it Herschel} Space Observatory (Pilbratt et al. 2010). Together, these facilities have enabled the IR LF and its evolution to be successfully traced out to $z\sim4$. In particular, {\it Herschel} surveys allowed us for the first time to select statistically large samples of galaxies at or close to the rest-frame peak of the far-IR emission and gave us a direct measure of the bolometric dust emission across a wide redshift range (for a review, see Lutz 2014).

Despite the impressive progress, owing to great advances in the sensitivity of the instruments, because of the relatively large beam of single-dish instruments, the deepest FIR and sub-mm observations are severely limited by confusion which results from the blending of multiple sources within the same telescope beam (e.g., Dole et al. 2003). Confusion presents us with several significant challenges, such as contamination of flux density by neighbouring sources, lack of survey dynamic range, and ambiguity in multi-wavelength association (i.e. counterpart identification) and redshift determination. Indeed, follow-up high angular resolution observations of the bright SCUBA-2 850 $\mu$m sources and {\it Herschel}-selected sources with interferometric facilities such as the Atacama Large Millimeter/submillimeter Array (ALMA; Wootten \& Thompson 2009) and the Submillimeter Array (SMA; Ho, Moran \& Lo 2004) have already shown that a significant fraction\footnote{ The multiplicity rate varies from 15-20\% to $\sim70$\% in the literature, depending on factors such as sample selection and the exact definition of multiplicity.} are made up of multiple sources (e.g., Karim et al. 2013; Hodge et al. 2013; Simpson et al. 2015; Bussmann et al. 2015; Micha{\l}owski et al. 2017; Hill et al. 2018; Stach et al. 2018). As a result, the number counts measured with single-dish instruments and interferometers (e.g., Karim et al. 2013; Simpson et al. 2015) can differ strongly. However, interferometric follow-up observations from the ground are not possible or extremely difficult at the high frequencies of the {\it Herschel} surveys due to absorption by water vapour in the atmosphere. In order to probe galaxy populations below the confusion limit, various advanced statistical techniques were developed such as the stacking method (e.g., Dole et al. 2006; Marsden et al. 2009; B{\'e}thermin et al. 2010, 2012a; Viero et al. 2013; Wang et al. 2016) and the map statistics via the pixel intensity distribution, the so-called P(D) measurements (e.g., Condon 1974; Patanchon et al. 2009; Glenn et al. 2010). However, a common limitation of these statistical methods is that the properties of individual galaxies cannot be determined, which results in the loss of information about the detailed properties of the underlying galaxy population.

A different approach to overcome confusion noise without giving up measuring properties of individual sources is source extraction utilising the position prior information from galaxy catalogues extracted from imaging surveys with higher angular resolution conducted at other wavelengths (for example at Spitzer/MIPS 24 $\mu$m and  VLA 1.4 GHz) to disentangle the contribution to the total flux density from various sources within the telescope beam (e.g., Magnelli et al. 2009; B{\'e}thermin et al. 2010; Roseboom et al. 2010, 2012; Wang et al. 2013, 2014; Liu et al. 2018). However, most of these prior based techniques use a maximum-likelihood optimisation approach which have two major problems\footnote{Another potential issue with prior based source extraction methods (but not limited to methods which use a maximum-likelihood optimisation approach) is that there is a possibility one might miss sources which have significant flux densities at the {\it Herschel} wavelengths, due to their absence in the prior catalogues. One way to mitigate this potential problem is to use sufficiently deep prior catalogues which can account for most of the emission in the {\it Herschel} maps.}. The first problem is that variance and covariance of source fluxes cannot be properly estimated. The second problem is that of overfitting when many of the input sources are intrinsically faint. 

In this paper, we build on our success of developing a prior-based Bayesian probabilistic de-blending and source extraction tool called XID+ for confusion-dominated maps (Hurley et al. 2017) as part of the {\it Herschel} Extragalactic Legacy Project (HELP; Vaccari 2016, Oliver, in preparation) to study the statistical properties of galaxies over an unprecedented dynamic range in luminosity and redshift. Our method is based on using Bayesian inference techniques such as Markov Chain Monte Carlo (MCMC) methods to fully explore the posterior probability distribution and therefore to properly estimate the variance and covariance between sources (i.e. how the flux of sources affect each other). Because XID+ is built upon a Bayesian probabilistic framework, it also provides a natural way in which to introduce additional prior information. Subsequently, we introduced prior information on the flux densities themselves through extensive modelling of the spectral energy distributions (SED) and fitting to multi-wavelength imaging data of the galaxies under study and we were able to show that this SED prior enhanced XID+ significantly improves over the vanilla XID+, based on validation using high angular resolution data from interferometric observations (Pearson et al. 2017, 2018).

The structure of the paper is as follows. In Section 2, we first describe briefly how the SED prior enhanced XID+ de-confusion technique works and the salient features (such as source density, completeness limit, etc.) of our state-of-the-art de-blended catalogue in the COSMOS field. We then introduce the various theoretical models and simulations (including empirical models, semi-analytic simulations and hydrodynamic simulations) which will be used later on for comparison purposes. In Section 3, we present our measurements of the number counts, the monochromatic and total IR LFs, and the CSFH, using our de-blended source catalogue in COSMOS. We also show detailed comparisons between our results and previous measurements in the literature from single dish and interferometric observations as well as predictions from a range of theoretical models and simulations of galaxy formation and evolution. Finally, discussions and conclusions are presented in Section 4. Throughout the paper, we assume $\Omega_m=0.3$,  $\Omega_{\Lambda}=0.7$,  and $H_0=70$ km s$^{-1}$ Mpc$^{-1}$. Flux densities are corrected for Galactic extinction (Schlegel, Finkbeiner \& Davis 1998). Unless otherwise stated, we assume a Chabrier (2003) initial mass function (IMF) in this paper.

\section{Data}

\subsection{The SED prior enhanced XID+ de-confusion technique}

We have invested major effort in developing techniques which can use very deep optical/near-infrared(NIR)/mid-infrared(MIR) prior source catalogues to decompose {\it Herschel} images that suffer from source confusion, as the full power of {\it Herschel} can only be unleashed when combined with detailed knowledge of the physical properties of galaxies. A major breakthrough is our development of a prior-based de-blending and source extraction tool XID+ (Hurley et al. 2017), which is a Bayesian probabilistic framework in which to include prior information, and uses the Bayesian inference tool Stan (Stan Development Team 2015a, b) to obtain the full posterior probability distribution on flux estimates. Compared to the previous state-of-the-art de-confusion tool DESPHOT (Wang et al. 2014), we can probe much fainter sources (by a factor of $\sim10$) at the same flux accuracy of 10\%. 

In the original version, XID+ uses a flat prior in the flux density parameter space (between zero and the brightest value in a given segment in the map), along with the known source positions on the sky. More recently, we have introduced informative prior on the flux densities themselves in the vanilla XID+ through extensive modelling of the SED and fitting to multi-wavelength photometric data. Using ALMA continuum data as an independent validation, we have shown that, by including informative but still weak prior on SED, the performance of XID+ can be improved further (Pearson et al. 2017). The systematic bias in flux accuracy, characterised by the difference between our predicted 870 $\mu$m flux (based on the de-blended XID+ SPIRE fluxes) and the measured flux by ALMA, is reduced when using an informative flux prior, at an impressive depth of more than 10 times below the SPIRE $5\sigma$ confusion limit of $\sim30$ mJy (Nguyen et al. 2010).

Our SED prior enhanced XID+ is detailed in Hurley et al. (2017) and Pearson et al. (2017, 2018). Here we describe the main steps of our methodology for completeness:

\begin{itemize}

\item First, we use the SED modelling and fitting tool called Code Investigating GALaxy Emission (CIGALE; Burgarella et al. 2005; Noll et al. 2009; Serra et al. 2011; Boquien et al. 2019) to generate SEDs and to fit these SEDs to the multi-wavelength imaging data, from UV to IR, of the galaxies under study\footnote{CIGALE uses an energy balance approach between the attenuated UV/optical emission and the IR/sub-mm emission, allowing the estimation of the IR/sub-mm flux densities. The choices for the SED model components and parameters for the SPIRE band priors follow Pearson et al. (2018) and will be briefly repeated here. We use a delayed exponentially declining star-formation history (SFH) with an exponentially declining burst, Bruzual \& Charlot (2003) stellar emission, Chabrier (2003) IMF, Charlot \& Fall (2000) dust attenuation, the updated Draine et al. (2014) version of the Draine \& Li (2007) IR dust emission, and Fritz et al. (2006) AGN models.}. This step produces estimates for the flux densities and uncertainties in the {\it Herschel}-SPIRE wavebands at 250, 350, and 500 $\mu$m .

\item The second step is to incorporate the predicted SPIRE flux densities and uncertainties from CIGALE as informative but still weak flux density priors in the probabilistic de-blending and source extraction tool XID+ to estimate the flux densities in the SPIRE bands. Combined with positional information from the prior galaxy catalogue, XID+ then uses the Bayesian inference tool Stan (Stan Development Team 2015a,b) to obtain the full posterior probability distribution on flux estimates by modelling the confusion-limited SPIRE maps.

\item Once the de-blended XID+ SPIRE flux densities are extracted, these SPIRE data are added to the multi-wavelength photometric dataset and CIGALE is rerun to get estimates for physical properties such as the monochromatic luminosity, stellar mass ($M_*$) and star-formation rate (SFR) for each galaxy.  During this step, we also ask CIGALE to give the flux densities estimates and uncertainties at the desired infrared (IR) and sub-mm wavelengths (e.g. the ALMA 870 $\mu$m band) for each object. The same CIGALE SED models are used for the flux estimation in the first run and to obtain the physical parameters in the second run.

\end{itemize}

\subsection{The de-blended catalogue in the COSMOS field}

In the COSMOS field (Scoville et al. 2007), we used the COSMOS2015 catalogue (Laigle et al. 2016) containing photometric data in over 30 bands for around 1.2 million objects as our prior catalogue. We ran CIGALE to model the SEDs of all galaxies in the COSMOS2015 catalogue\footnote{Note that the COSMOS2015 catalogue does contain flux densities in the {\it Herschel}-SPIRE bands for $\sim18,000$ sources. These SPIRE flux densities were extracted via a maximum-likelihood optimisation approach using Spitzer/MIPS 24 $\mu$m sources as priors (Roseboom et al. 2010). In this paper, we do not use these SPIRE flux densities contained in the COSMOS2015 catalogue.} and generate flux density priors in the {\it Herschel}-SPIRE bands at 250, 350, and 500 $\mu$m. To account for as much dust emission as possible and at the same time keep the level of degeneracy as low as possible, we applied a cut of 0.7 mJy on the predicted flux density at 250 $\mu$m which left us with 205 958 objects over 2.15 square degrees (see Pearson et al. 2018 for more details). The 205 958 galaxies with a predicted 250 $\mu$m flux density above our flux cut were then used in XID+ to model the confusion-limited SPIRE maps and generate de-blended flux densities at 250, 350, and 500 $\mu$m. Finally, CIGALE was ran again combining the multi-wavelength photometric information and the de-blended SPIRE flux densities to generate estimates and uncertainties on quantities such as the flux density at 870 $\mu$m (observed-frame), rest-frame monochromatic luminosity at various FIR and sub-mm wavelengths, dust luminosity or the total IR luminosity\footnote{In CIGALE, dust luminosity is defined as the integrated infrared luminosity between 8 and 1000 $\mu$m, including contributions from both AGN (Active Galactic Nuclei) activity and star formation.}, stellar mass and SFR for each galaxy.

Due to our flux cut of 0.7 mJy at 250 $\mu$m, we also need to apply an equivalent flux cut at other wavelengths when comparing our number counts results with previous measurements in Section 3.1. To achieve this, we use the Simulated Infrared Dusty Extragalactic Sky (SIDES) empirical model which has the best match with existing FIR and sub-mm number counts measurements (see Section 2.3.1). We employ two methods to derive the equivalent flux cuts at other wavelengths. The first method uses the mean flux ratio from the SIDES model to convert the flux cut of 0.7 mJy at 250 $\mu$m to an equivalent flux cut of 0.8, 0.7, 0.6 and 0.3 mJy at 350, 450, 500 and 870 $\mu$m, respectively. However, due to the scatter present in the flux ratios, we also consider a second method which takes into account the broad correlation between the 250 $\mu$m flux density ($S_{250}$) and flux densities at other wavelengths. For example, to derive the flux cut at 350 $\mu$m, we compute the ratio of the number of objects with $S_{250}>0.7$ mJy and $S_{350}>x$ mJy to the total number of objects with $S_{350}>x$ mJy and then define the flux cut level at 350 $\mu$m as the value of $x$ when the ratio is equal to 95\%. Using this method, we can derive an equivalent flux cut  of 0.7, 0.7, 0.6 and 0.4 mJy at 350, 450, 500 and 870 $\mu$m, respectively. In this paper, we use the flux cut values derived from the second method.

\begin{figure}
\centering
\includegraphics[height=2.8in,width=3.6in]{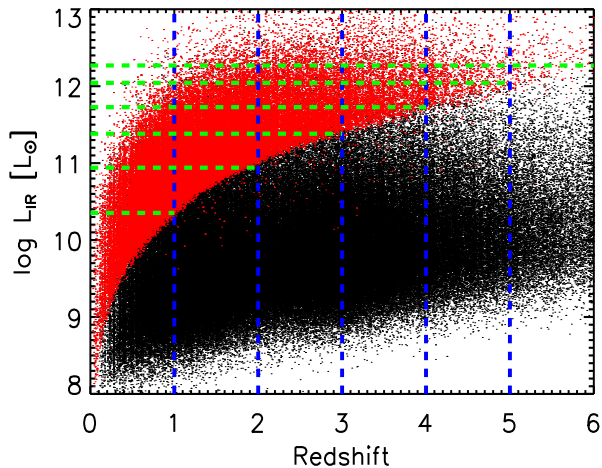}
\caption{The IR luminosity ($L_{IR}$) vs redshift from the SIDES simulation. The black dots are mock galaxies from the SIDES simulation. For clarity, only a random 20\% of the simulation is plotted. The red dots are mock galaxies with 250 $\mu$m flux density above 0.7 mJy. The vertical dashed blue lines indicate redshift bins. Note that these redshift bins are for illustration purpose only. We use a variety of redshift binning later on to facilitate the comparisons between our results and previous measurements or predictions from theoretical models. The horizontal dashed green lines indicate the IR luminosity limit above which the sample is 90\% complete.}
\label{LIRvsz}
\end{figure}

Also using the SIDES simulation, we can work out the corresponding limit on the total IR luminosity ($L_{IR}$) as a function of redshift due to the flux cut of 0.7 mJy at 250 $\mu$m. For a given redshift bin with lower limit ($z_{\rm lower~limit}$) and upper limit ($z_{\rm upper~limit}$), we compute the ratio of the number of galaxies with $L_{IR}>x~L_{\odot}$ and $S_{250}>0.7$ mJy to the total number of galaxies with $L_{IR}>x~L_{\odot}$ in that redshift bin and then define the IR luminosity limit as the value of $x$ at which the ratio is equal to 95\%. Fig.~\ref{LIRvsz} shows the IR luminosity $L_{IR}$ vs redshift for mock galaxies from the SIDES simulation. The black dots represent a random 20\% of all galaxies in the simulation. The red dots are mock galaxies with $S_{250}$ above 0.7 mJy. The vertical dashed blue lines indicate redshift bins which are for illustration purpose only as we use various redshift binning later on to compare with previous results. The horizontal dashed green lines indicate the IR luminosity limit above which the sample is 95\% complete.

It is worth pointing out that our prior catalogue, i.e., the COSMOS2015 catalogue, has limit on stellar mass due to a magnitude limit in the $K_s$ band. Section 3.3 in Pearson et al. (2018) details how the stellar mass completeness limit as a function of redshift is derived which basically follows the procedure in Pozzetti et al. (2010). Using the SIDES simulation, we checked that for our adopted limits on flux (or luminosity), most of our selected galaxies have stellar masses above the stellar mass limit inherent in the COSMOS2015 catalogue. As a result, we will ignore the impact of the stellar mass limit in our results in Section 3.

\subsection{Empirical models and simulations of galaxy formation physics}

Broadly speaking, there exist two different kinds of models,  empirical models which are designed to reproduce observations but contain minimal information regarding galaxy formation physics and physically-motivated models which can be tuned by observations to varying degrees. The latter includes the Durham semi-analytic model (SAM) which uses simplified flow equations for bulk components and the EAGLE numerical hydrodynamic simulation which solves the equations of gravity, hydrodynamics, and thermodynamics at the same time (see Somerville \& Dav{\'e} 2015 for a review). 

\subsubsection{Empirical models}

We use two different empirical models. The publicly available Simulated Infrared Dusty Extragalactic Sky (SIDES) simulation\footnote{\url{http://cesam.lam.fr/sides/}} is a simulation of the extragalactic sky from the FIR to the sub-mm, including clustering based on empirical prescriptions. The method used to build this simulated catalogue is described in detail in B{\'e}thermin et al. (2017). Briefly, a lightcone covering 2 deg$^2$ was produced from the Bolshoi-Planck simulation (Klypin et al. 2016; Rodr{\'{\i}}guez-Puebla et al. 2016). To populate dark-matter halos with galaxies, an abundance-matching technique was used (e.g., Vale \& Ostriker 2004). The luminous properties of the galaxies are generated by using an updated version of the 2SFM (2 star-formation modes) model (Sargent et al. 2012; B{\'e}thermin et al. 2012b, 2013), which is based on the observed evolution of the galaxy star-formation main sequence\footnote{The galaxy star-formation main sequence refers to the observed tight correlation (with an intrinsic scatter of $\sim0.2$ to 0.3 dex) between SFR and stellar mass of star-forming galaxies which exists both in the local Universe and at high redshifts (e.g., Noeske et al. 2007; Elbaz et al. 2007; Daddi et al. 2007; Whitaker et al. 2012; Wang et al. 2013; Speagle et al. 2014; Lee et al. 2015; Schreiber et al. 2015; Tomczak et al. 2016; Pearson et al. 2018).} and the observed evolution of the SEDs with redshift. The SIDES simulation reproduces a large set of observables, such as number counts and their evolution with redshift and cosmic IR background power spectra. The SIDES simulated lightcone contains information such as redshift, halo mass, stellar mass, SFR, SED shape, and flux densities at various wavelengths between 24 and 2000 $\mu$m (observed-frame).

The Empirical Galaxy Generator (EGG; Schreiber et al. 2017) is a tool for generating mock galaxy catalogues with realistic fluxes and simple morphological types, developed by the ASTRODEEP collaboration. By construction, EGG is designed to match current observations from the UV to the sub-mm at $0<z<7$. EGG generates mock galaxies which are composed of two broad populations of star-forming (SFGs) and quiescent galaxies (QGs), based on the observed stellar mass functions of each population. SFRs are assigned (with random scatter) to mock galaxies based on the galaxy star-formation main sequence. Other properties such as optical colours, morphologies and SEDs are assigned using empirical relations derived from {\it Hubble} and {\it Herschel} observations from the Cosmic Assembly Near-infrared Deep Extragalactic Legacy Survey (CANDELS) fields (Grogin et al. 2011; Koekemoer et al. 2011).

\subsubsection{The Durham semi-analytic models}

In the Durham SAM of galaxy formation, GALFORM, galaxies populate dark matter halo merger trees according to simplified prescriptions for the baryonic physics involved (gas cooling, star formation, feedback etc.), which result in a set of coupled differential equations that track the exchange of mass and metals between the different baryonic components (stars, cold gas etc.) of a galaxy. Here we use the version of GALFORM presented in Lacey et al. (2016) with a minor recalibration due to the model being implemented in an updated \emph{Planck} cosmology (Baugh et al. 2018, see also Cowley et al. 2018). This model is calibrated to reproduce a large set of observational data at ($z\lesssim6$), including sub-mm galaxy number counts such as those presented in this work. For predicting sub-mm fluxes the star formation histories and galaxy properties predicted by GALFORM are coupled with the spectrophotometric code GRASIL (Silva et al. 1998) which computes the absorption and re-emission of stellar radiation by interstellar dust resulting in self-consistent UV-to-mm SEDs for each simulated galaxy.

One of the main features of the model relevant for the predictions shown here is that it assumes a top-heavy IMF for bursts of star formation triggered by a dynamical process, either disc instabilities, major mergers or some gas-rich minor mergers, though sub-mm bright galaxies in this model are majoritively triggered by disc instabilities. This feature was first introduced into GALFORM by Baugh et al. (2005) so that the model could simultaneously reproduce observational constraints such as the 850 $\mu$m galaxy number counts and optical/near-IR luminosity functions at $z=0$. A top-heavy IMF is extremely efficient at boosting sub-mm flux due to (i) the increased UV radiation through having more massive stars per unit star formation and (ii) an increased dust mass available to absorb and re-emit this UV radiation due to faster recycling of material back into the interstellar medium as these massive stars go supernovae. It is a combination of these two effects that allows the model to reproduce the galaxy number counts shown here whilst simultaneously reproducing many other observational datasets (see e.g. Lacey et al. 2016).

\subsubsection{The EAGLE hydrodynamic simulations}

The EAGLE simulation suite (described fully in Schaye et al.2015; Crain et al. 2015) comprises cosmological hydrodynamical simulations of periodic cubic volumes  with a range of sizes and numerical resolutions , using a modified version of the Gadget-3 TreeSPH code (an update to Gadget-2, Springel et al. 2005) and a $\Lambda$CDM cosmology, with the cosmological parameters derived in the initial Planck release (Planck Collaboration et al. 2014). We use the Ref-100 run, which simulates the evolution of a volume with mean cosmic and a side length of 100 Mpc using the fiducial EAGLE model at a standard resolution. Models for unresolved physical process are implemented to treat star formation and stellar evolution, photoheating and radiative cooling of gas, energetic feedback by supernovae and AGN, and the chemical enrichment of the interstellar medium by stars.

Virtual observables are generated for EAGLE galaxies in post-processing  using the Monte Carlo dust radiative transfer code SKIRT (Baes et al. 2011; Camps \& Baes 2015). The SKIRT modelling approach is  briefly summarised here, with a full description in Camps et al. (2016) and Trayford et al. (2017).  Emission from stellar populations older than 10 Myr is represented by the  Bruzual \& Charlot (2003) spectral libraries. As dust is not included explicitly in the EAGLE simulations, the diffuse dust distribution is taken to trace that of the sufficiently cool, enriched ISM gas that emerges in EAGLE galaxies, assuming 30\% the metal mass is in dust grains and a Zubko et al. (2004) model for grain properties. To include dust associated with the unresolved birth clouds of stars, emission from populations younger than 10 Myr are represented using the HII region spectral libraries of Groves et al. (2008). FIR emission is then produced by the iterative absorption and re-emission of UV-FIR radiation by dust as well as directly from the HII region SEDs, and measured in multiple broad bands in both rest- and observer-frames. The fraction of metals in dust grains, photo-dissociation region covering fraction in HII regions, and temperature threshold for  is set  in order to reproduce FIR properties of the local galaxies in the Herschel Reference Survey (Boselli et al. 2010, Cortese et al. 2012) for a K-band matched sample (Camps et al. 2016). The virtual observations measured at a number of discrete redshifts for galaxies with more than 250  are all publicly accessible via the EAGLE database (McAlpine et al. 2016, Camps et al. 2017), and $L_{\rm IR}$ values are estimated by integrating the available broad bands over the 8-1000 $\mu$m range.

In contrast to GALFORM, EAGLE assumes a universal Chabrier (2003) IMF. Also, while the simulated volume of EAGLE is large for a  hydrodynamic simulation of its resolution, the volume is small relative to that achievable for SAMs\footnote{The Lacey et al. (2016) GALFORM model has a $\sim$125 times larger volume, i.e. $500 h^{-1}$ Mpc on a side.} and empirical models. As a result, more massive systems (i.e. high-mass groups and clusters) are not captured by the simulation, and thus the brightest sources in the FIR are potentially absent in projected counts.  It is unclear whether the EAGLE model would produce the correct density of extreme starbursts in a larger volume simulations, or whether it would be necessary to appeal to something like IMF variations to reproduce observed counts.

\section{Results}

In this section, we present our results of the number counts at various FIR and sub-mm wavelengths (250, 350, 450, 500 and 870 $\mu$m)\footnote{We also provide our number counts and LFs in a tabular format in the Appendix.}, the monochromatic LFs at 250 $\mu$m (rest-frame) and the total IR LFs as a function of redshift, and the CSFH out to $z\sim6$. We also compare our results with previous measurements in the literature as well as predictions from empirical models and physically-motivated simulations.

\subsection{Number counts}

\subsubsection{The 250, 350 and 500 $\mu$m number counts}
 
 \begin{figure}
\centering
\includegraphics[height=2.8in,width=3.6in]{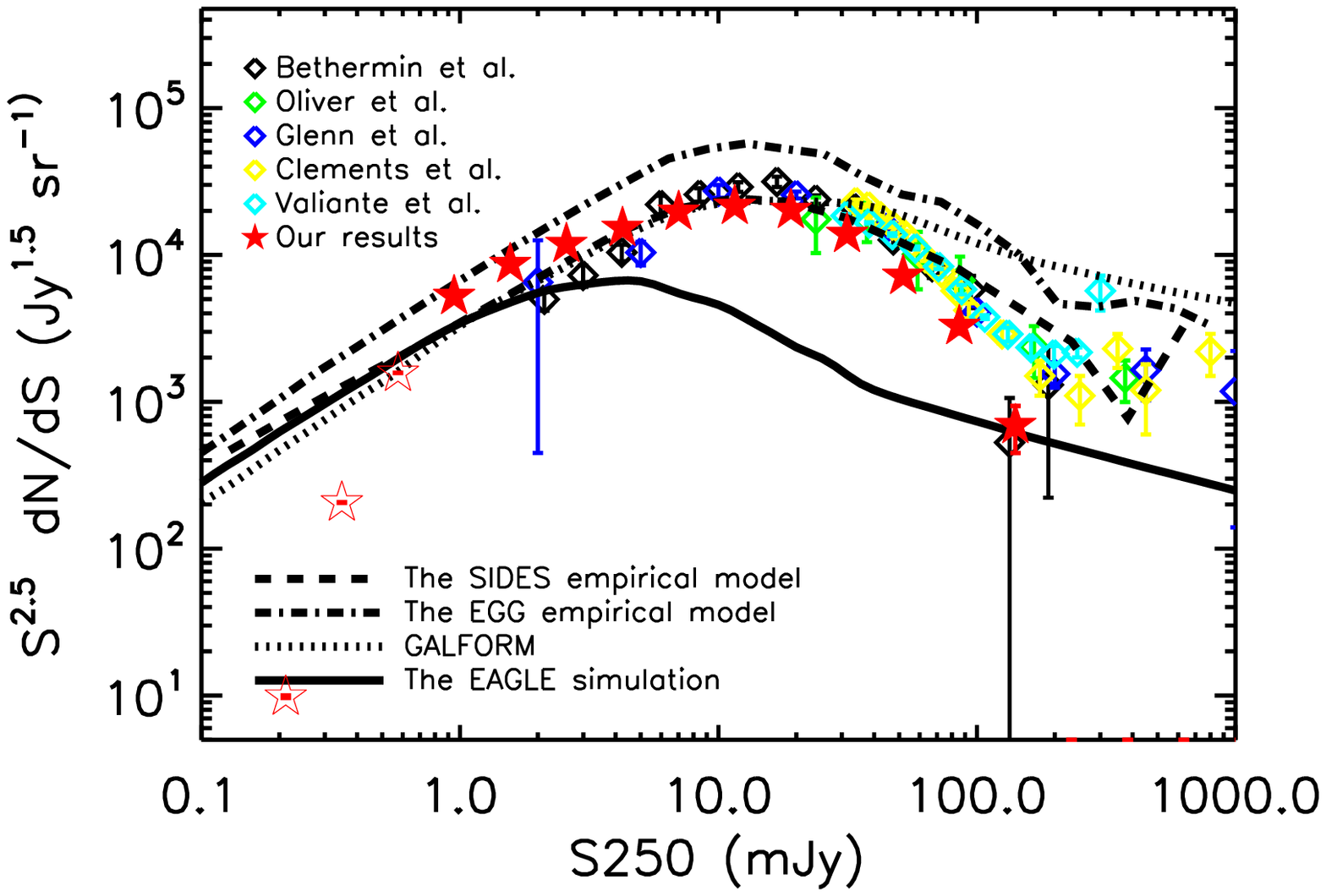}
\includegraphics[height=2.8in,width=3.6in]{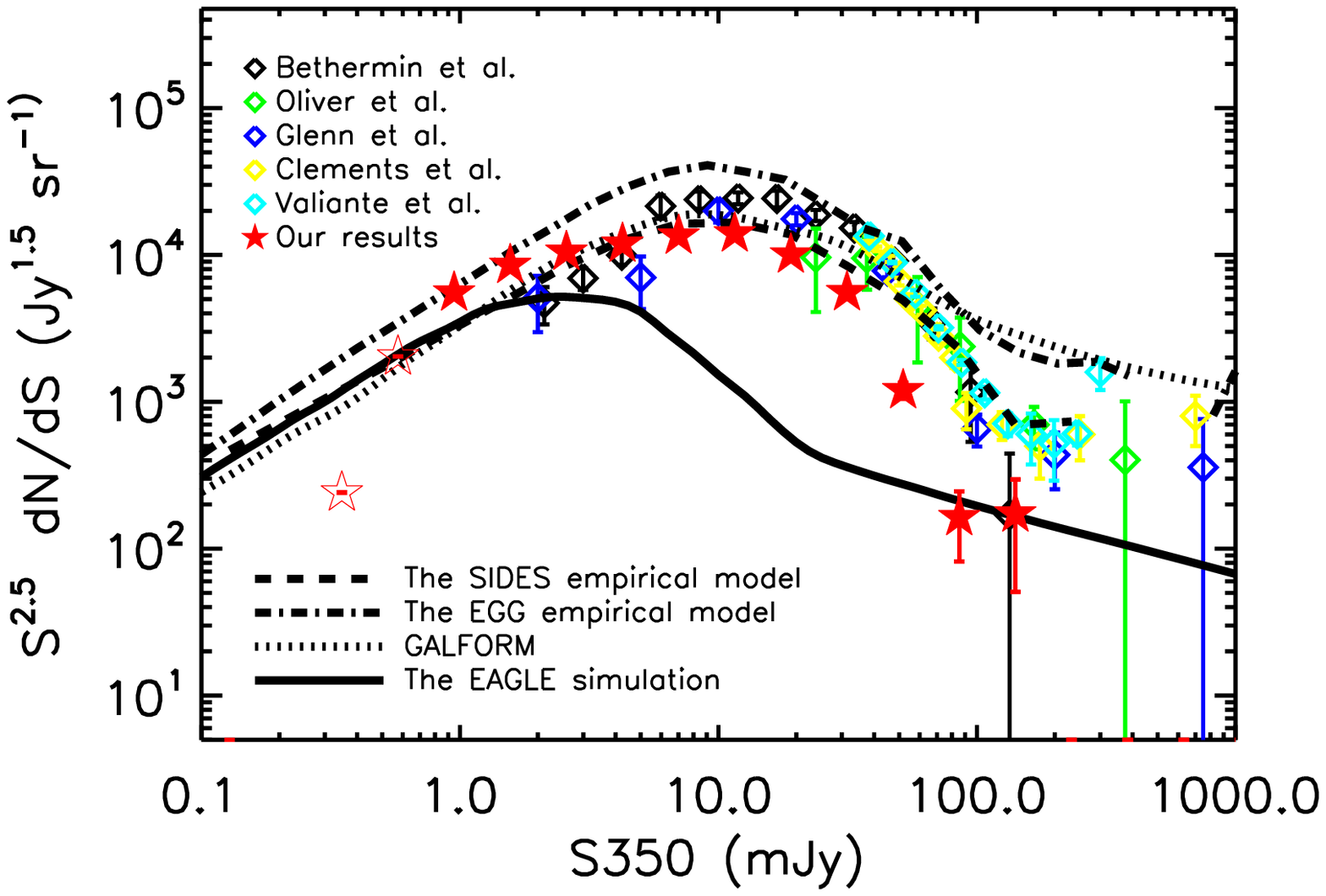}
\includegraphics[height=2.8in,width=3.6in]{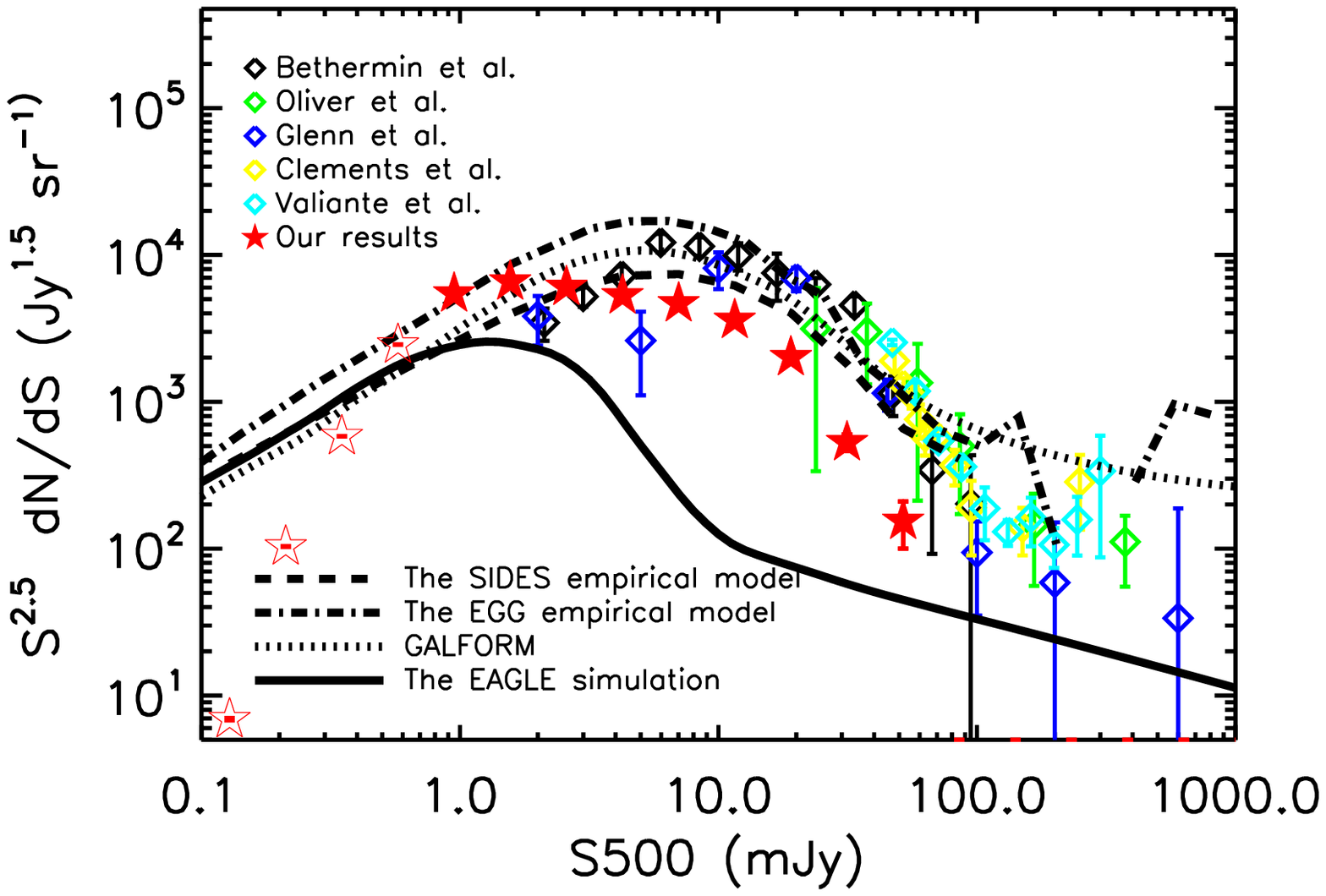}
\caption{The differential number counts at 250, 350 and 500 $\mu$m. Our number counts derived from the SED prior enhanced XID+ de-blended catalogue in COSMOS are shown as red stars (filled red stars: our number counts above the flux cut; empty red stars: our number counts below the flux cut). Error bars on the red stars only represent Poisson errors. The lines show the predicted number counts from various models and simulations (including SIDES, EGG, GALFORM and the EAGLE hydrodynamic simulation). The empty diamonds are previous measurements based on {\it Herschel} observations.}
\label{NCSPIRE}
\end{figure}

In this subsection, we present our measurements of the {\it Herschel}-SPIRE 250, 350 and 500 $\mu$m differential number counts using the SED prior enhanced XID+ de-blended catalogue in the COSMOS field. 

Fig.~\ref{NCSPIRE} compares our counts with previous measurements based on {\it Herschel} observations\footnote{These observations include the {\it Herschel} Multi-tiered Extragalactic Survey (HerMES; Oliver et al. 2012) and the {\it Herschel} Astrophysical Terahertz Large Area Survey (H-ATLAS; Eales et al. 2010).} and predictions from various models and simulations (including the empirical models SIDES and EGG, GALFORM and the EAGLE hydrodynamic simulation). The number counts are multiplied by a factor of $S^{2.5}$ to reduce the dynamic range of the plot and to highlight the plateau at high flux densities where the Euclidian approximation is valid. In Fig.~\ref{NCSPIRE}, previous measurements based on {\it Herschel} observations are shown as empty diamonds. Our number counts derived from the de-blended catalogue in COSMOS are shown as red stars. The filled red stars correspond to our counts above the flux cut adopted in Section 2.2 and the empty red stars correspond to our counts below the flux cut. Note that as discussed in Section 2.2, our flux cut is based on the predicted 250 $\mu$m flux and therefore should be treated as a guide as it is not likely to be exact. However, we can see that our number counts below the flux cut (i.e. the empty red stars) begin to drop rapidly which indicate that our flux cut is a reasonable representation of the completeness limit. Our error bars only include Poisson errors (and so likely to be underestimated) while other studies have included uncertainties due to field-to-field variations. 

The top panel in Fig.~\ref{NCSPIRE} shows that at 250 $\mu$m our measurements agree very well with previous observational results derived using various techniques such as blind source extraction, stacking and a P(D) analysis using pixel flux distribution. There is a lack of very bright sources with $S_{250}>\sim100$ mJy in our counts which is due to the limited size of the COSMOS field. Our counts are more than 10 times deeper compared to counts derived from blind source extraction which probes the counts above the confusion limit (Oliver et al. 2010; Clements et al. 2010; Valiante et al. 2016) and roughly two times deeper compared to counts derived from P(D) or stacking (Glenn et al. 2010; Bethermin et al. 2012a) which can also probe the counts below the confusion limit. 

At the longer wavelengths 350 and 500 $\mu$m, the agreement between our number counts and previous measurements gets worse, which is understandable as the effects of confusion and source blending become progressively more important. In general, our counts are lower than previous {\it Herschel} measurements at $S_{350}>\sim8$ and $S_{500}>\sim5$ mJy. At 500 $\mu$m, our counts are lower by as much as $\sim0.5$ dex. In addition, our counts indicate the turnover in the number counts occurs at fainter flux levels and the shift is more pronounced at 500 $\mu$m than at 350 $\mu$m. This could suggest that either previous  {\it Herschel} counts measurements still suffer confusion and source blending problems which is progressively worse at longer wavelength due to the larger beam size or our results could be over de-blended. As will be discussed in Section 3.1.2 and Section 3.1.3, the excellent agreement between our number counts at 450 $\mu$m and the SCUBA-2 450 $\mu$m counts, and between our number counts at 870 $\mu$m and the SCUBA-2 850 $\mu$m, SMA  860 $\mu$m and ALMA 870 $\mu$m number counts (derived from single dish observations with much higher angular resolution and interferometric observations with arcsec or even sub-arcsec resolution) strongly supports the former interpretation.

Apart from the EAGLE hydrodynamic simulation, all other models give roughly similar predictions of the number counts at the SPIRE wavelengths and agree well with measurements from previous {\it Herschel} studies. The SIDES empirical model is the best at reproducing the number counts from previous {\it Herschel}-based studies. The EGG empirical model tends to over-predict (by a factor of $\sim2$) the number density of galaxies at all fluxes compared to the SIDES empirical model and GALFORM. GALFORM, although very closely reproducing the predictions from SIDES, does over-predict the number of bright galaxies and this effect is increasingly pronounced towards shorter wavelength. Again, our measurements are lower (increasingly so towards longer wavelength) than predictions from these three models (SIDES, EGG and GALFORM). This is not surprising because these models are more or less designed to reproduce the previous {\it Herschel} measurements. SIDES and EGG are purely empirical models and therefore by construction they will have a good agreement with the input observational constraints. However, the downside of these empirical models is that they provide little physical insight about the galaxy formation physics (e.g., physical processes driving the formation and evolution of dusty star-forming galaxies).

In comparison, the predicted number counts from the EAGLE hydrodynamic simulation are much lower (increasingly so towards longer wavelength) compared to all observational measurements and predictions from the empirical models and GALFORM, at all wavelengths. This is caused by the fact that, unlike GALFORM, the statistical properties of the dusty star-formation galaxies are not used to tune the EAGLE simulation. In general, physically-motivated models of galaxy formation struggle to reconcile the statistical properties (such as the number counts and LFs) of sub-mm galaxies without appealing to something like a top-heavy IMF in starburst galaxies (e.g., Baugh et al. 2005; Lacey et al. 2016). However, alternative solutions to solve the problem of matching the observed sub-mm number counts, such as changes to prescriptions for star formation and feedback, have also been suggested in the literature (e.g., Hayward et al. 2013; Safarzadeh et al. 2017). In addition, as described in Section 2.3.3, the limited volume of EAGLE (which is 125 times smaller than GALFORM) means that it misses rarer objects such as luminous starbursts which make a significant contribution to the number counts at the bright end. In order to assess how much the issue of volume can affect the number counts, we checked a smaller simulation box with side length of 50 Mpc. We conclude that the relatively small volume of EAGLE does affect the counts at the bright end, but it is unlikely to account for the large discrepancy with the observations.

\subsubsection{The 450 $\mu$m number counts}

\begin{figure}
\centering
\includegraphics[height=3.in,width=3.8in]{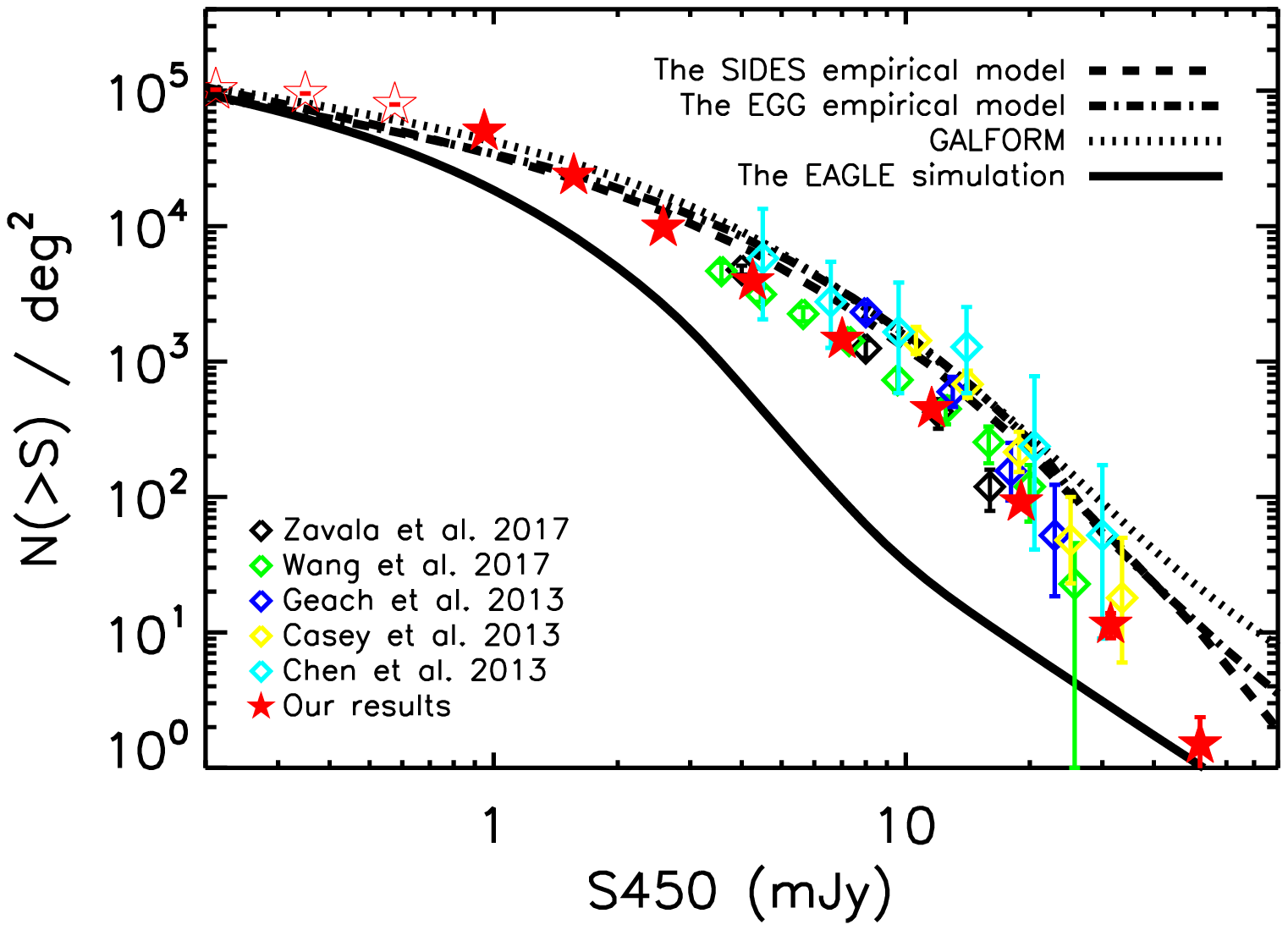}
\includegraphics[height=3.in,width=3.8in]{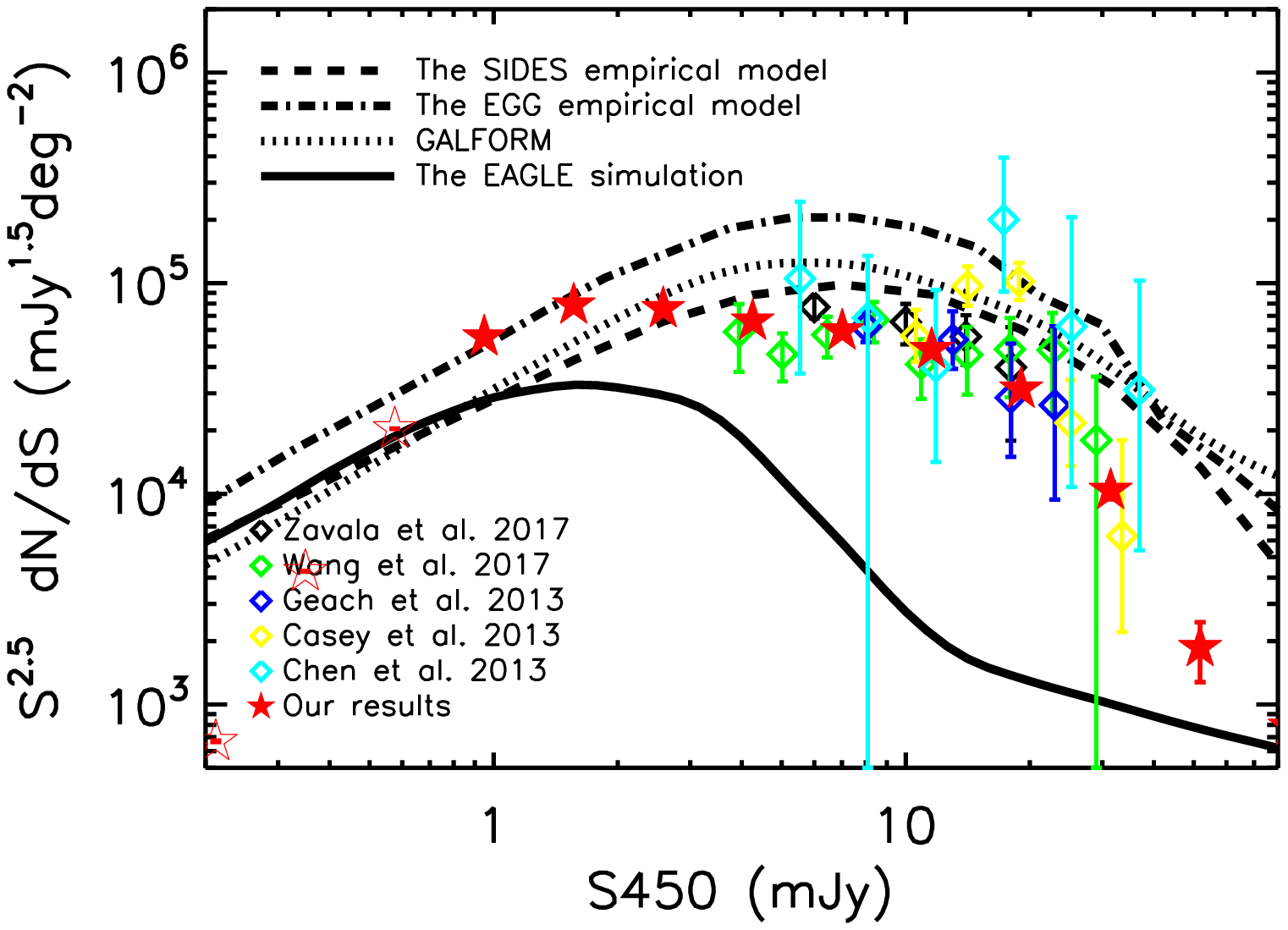}
\caption{Upper panel: The cumulative (or the integral) number counts at 450 $\mu$m. Our results derived from the de-blended catalogue in the COSMOS field are plotted as red stars (filled red stars: our number counts above the flux cut; empty red stars: our number counts below the flux cut). Error bars on the red stars only represent Poisson errors. The lines show the predicted number counts from various models and simulations (including SIDES, EGG, GALFORM and the EAGLE hydrodynamic simulation). The empty diamonds are previous results based on JCMT SCUBA-2 observations. Lower panel: The differential number counts at 450 $\mu$m.}
\label{NC450}
\end{figure}

In this subsection, we present our measurements of the 450 $\mu$m number counts using the SED prior enhanced XID+ de-blended catalogue in the COSMOS field. Our 450 $\mu$m number counts are generated using the de-blended 500 $\mu$m flux densities after applying a scaling factor of the flux ratio $S_{450}/S_{500}=0.86$, derived from the SIDES simulation. The standard deviation of the flux ratio $S_{450}/S_{500}$ is very small ($\sim0.079$) and therefore is ignored here.

Fig.~\ref{NC450} compares our counts with previous measurements using observations carried out with the SCUBA-2 camera on the 15 metre JCMT. In comparison, the primary mirror of the {\it Herschel} satellite is 3.5 metres in diameter. Chen et al. (2013) presented SCUBA-2 450 $\mu$m observations in the field of the massive lensing cluster A370 (total survey area $>100$ arcmin$^2$) and 20 detected sources with $S/N>4$. The intrinsic number counts derived in Chen et al. (2013) which probes flux levels below the {\it Herschel} confusion limit are plotted as light blue diamonds in Fig.~\ref{NC450}. The first deep blank-field cosmological 450 $\mu$m imaging covering an area of 140 arcmin$^2$ of the COSMOS field was conducted as part of the SCUBA-2 Cosmology Legacy Survey (S2CLS) and presented in Geach et al. (2013). Consequently, Geach et al. (2013) made the first number counts at 450 $\mu$m from an unbiased blank-field survey at a flux density limit $S_{450}>5$ mJy which are plotted as dark blue diamonds in Fig.~\ref{NC450}. Later on, Casey et al. (2013) studied the number counts at 450 $\mu$m using 78 sources detected from a wider and shallower 394 arcmin$^2$ area in COSMOS observed with SCUBA-2 with more uniform coverage. Their counts are plotted as yellow diamonds. More recently, Zavala et al. (2017), using deep observations in the Extended Groth Strip (EGS) field taken  as part of the S2CLS, detected 57 sources at 450 $\mu$m. They presented one of the deepest number counts available so far derived using directly extracted sources from only blank-field observations, which are plotted as black diamonds. Wang et al. (2017) , using a new program on the JCMT, i.e., the SCUBA-2 Ultra Deep Imaging EAO (East Asian Observatory) Survey (STUDIES), detected $\sim100$ 450 $\mu$m sources in the COSMOS field and presented determination of the number counts down to a flux limit of 3.5 mJy (green diamonds in Fig.~\ref{NC450}).

As can be seen Fig.~\ref{NC450}, previous observational measurements of the 450 $\mu$m counts from blank-field and lensing-cluster surveys carried out with SCUBA-2 on the JCMT are more or less consistent with each other within errors, especially if the counts derived from the lensing cluster observations by Chen et al. (2013) with very large uncertainties are excluded. These SCUBA-2 number counts results are all significantly lower compared to the {\it Herschel} counts at 350 and 500 $\mu$m. As pointed out in previous studies such as Wang et al. (2017), the much higher {\it Herschel} counts are mostly due to confusion and source blending which is more severe when sources under the beam are strongly clustered. At 450 $\mu$m, the angular resolution achievable with JCMT is roughly a factor of 5 better than {\it Herschel} at 500 $\mu$m. Because of its much higher angular resolution and much fainter confusion limit ($\sim7$ times fainter), SCUBA-2 counts at 450 $\mu$m do not suffer as much from confusion and source blending issues.

Our measurements of the 450 $\mu$m number counts which are lower than previous {\it Herschel} results (as shown in Section 3.1.1) are now in excellent agreement with the number counts derived from the higher resolution SCUBA-2 observations. We also extend the number counts measurements by a factor of $\sim4$ compared to the deepest SCUBA-2 study by Wang et al. (2017). Previous SCUBA-2 measurements did not find any evidence of a turnover in the number counts at the faint end, while {\it Herschel} studies suggest a turn over at around 5 mJy at 500 $\mu$m (or around 4 mJy at 450 $\mu$m, using the colour ratio of $S_{450}/S_{500}=0.86$). Thanks to the increased dynamic range, our number counts derived using the SED prior enhanced XID+ de-blended catalogue in COSMOS suggest a turnover in the counts at around 2 mJy.

The empirical models SIDES and EGG and GALFORM generally over-predict the number counts compared to the SCUBA-2 measurements and our measurement. This is not surprising as the models are more or less tuned to reproduce the {\it Herschel}/SPIRE number counts results, as discussed in Section 3.1.1. Again, the predicted number counts from the EAGLE hydrodynamic simulation (which have not been tuned to match the statistical properties of dusty star-forming galaxies) are much lower compared to the observations and predictions from the empirical models and GALFORM.

\subsubsection{The 870 $\mu$m number counts}

\begin{figure}
\centering
\includegraphics[height=3.in,width=3.8in]{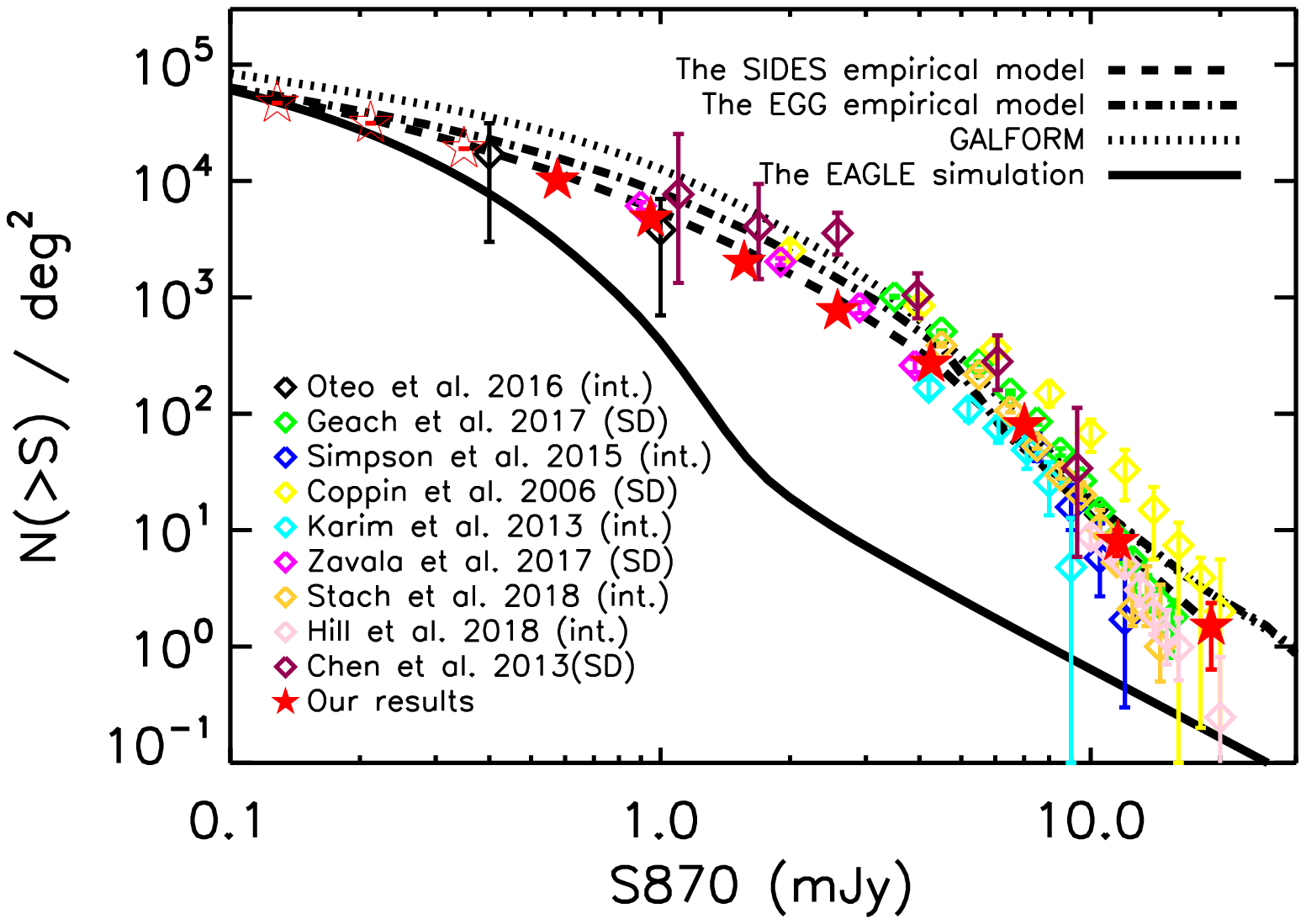}
\includegraphics[height=3.in,width=3.8in]{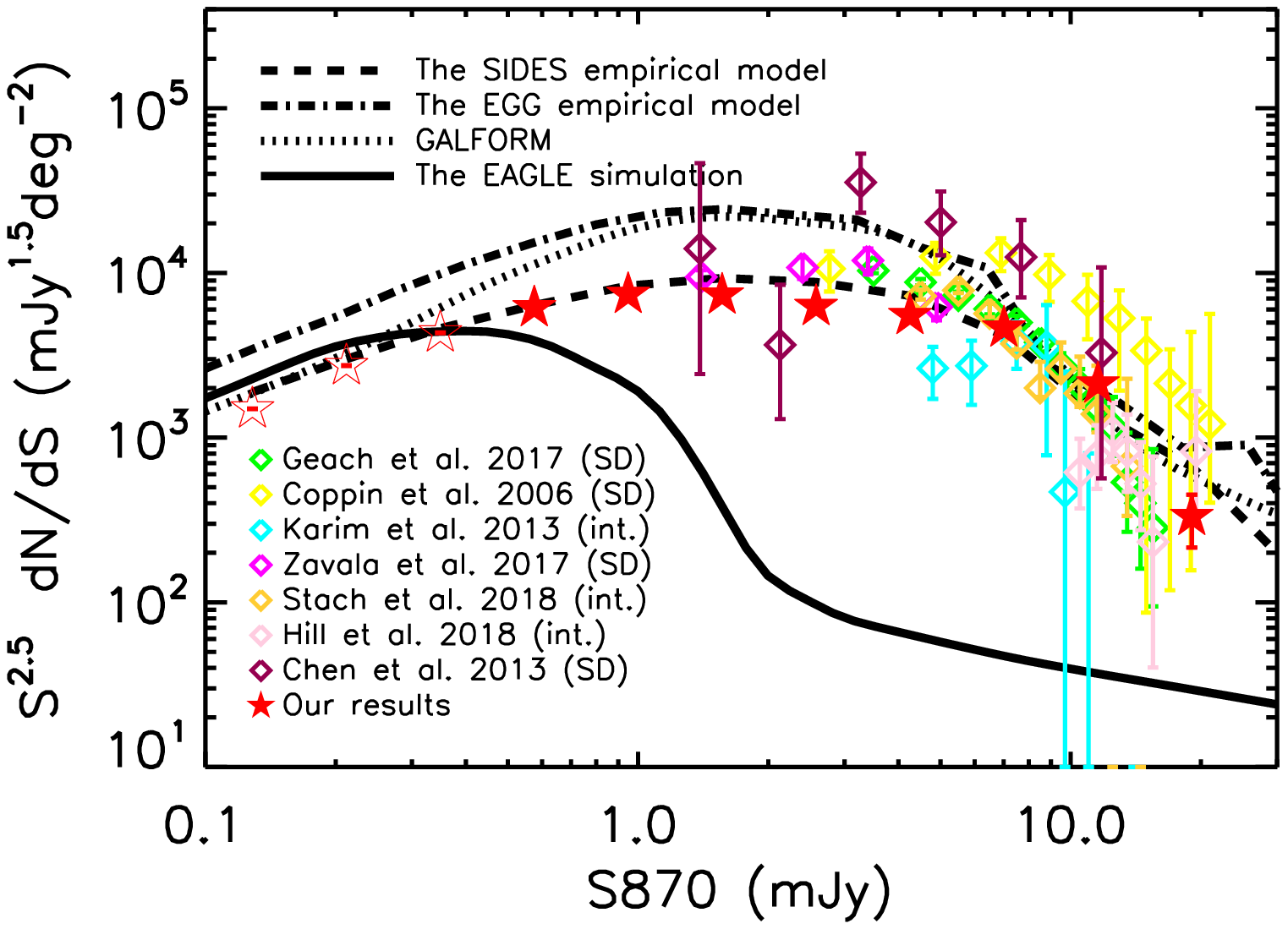}
\caption{Upper panel: The cumulative number counts at 870 $\mu$m. Our results derived from the de-blended catalogue in COSMOS are plotted as red stars (filled red stars: our number counts above the flux cut; empty red stars: our number counts below the flux cut). Error bars on the red stars only represent Poisson errors. The lines show the predicted number counts from various models and simulations (including SIDES, EGG, GALFORM and the EAGLE hydrodynamic simulation). The empty diamonds are previous results using either single dish (SD) observations or interferometric (int.) observations. Note that the SCUBA-2 number counts are as a function of 850 $\mu$m flux density and the SMA number counts are as a function of 860 $\mu$m flux density, but we will ignore the slightly different effective wavelengths of the different instruments. Lower panel: The differential number counts at 870 $\mu$m.}
\label{NC850}
\end{figure}

In this subsection, we present our measurements of the 870 $\mu$m number counts using the SED prior enhanced XID+ de-blended catalogue in COSMOS. As described in Section 2.1 and Section 2.2, in the second run of CIGALE which combined the multi-wavelength photometric information with the de-blended XID+ SPIRE flux densities, we also generated the predicted flux densities and uncertainties at 870 $\mu$m. Therefore, we can compare our predicted 870 $\mu$m number counts with previous measurements using either single dish (e.g. SCUBA-2 on the JCMT) and interferometric observations (e.g., ALMA and SMA). In general, single dish observations cover much larger area compared to interferometric observations, but with limited angular resolution and depth. In this paper, we will ignore the small differences due to the slightly different effective wavelengths of different instruments. 

Fig.~\ref{NC850} compares our results with previous measurements using single dish observations from SCUBA-2, and interferometric observations from SMA and ALMA. The first estimates of the 850 $\mu$m number counts were presented in Coppin et al. (2006) using $>100$ detected sources from the SCUBA HAlf Degree Extragalactic Survey (SHADES; Mortier et al. 2005; van Kampen et al. 2005) over an area of 720 arcmin$^2$. Using SCUBA-2 observations of a field ($>100$ arcmin$^2$) lensed by the massive cluster A370, Chen et al. (2013) detected 26 sources at 850 $\mu$m with a signal-to-noise ratio $>4$. Thanks to the effect of gravitational lensing, Chen et al. (2013) were able to probe fainter galaxies compared to sources detected in Coppin et al. (2006). More recent single dish observations come from Geach et al. (2017) and Zavala et al. (2017). Geach et al. (2017) detected $\sim3,000$ sub-mm sources at S/N $>3.5$ at 850 $\mu$m over $\sim$5 deg$^2$ surveyed as part of the S2CLS. This is the largest survey of its kind at this wavelength which increases the sample size selected at 850 $\mu$m by an order of magnitude. As a result, Geach et al. (2017) were able to measure the number counts at 850 $\mu$m with unprecedented accuracy. In particular, the large area of the survey enabled better determination of the counts at the bright end. Zavala et al. (2017), using deep observations with SCUBA-2 in the EGS as part of the S2CLS detected 90 sources at 850 $\mu$m with S/N $>3.5$ over 70 arcmin$^2$ and derived the deepest number counts from blank field single-dish observations at $S_{850}>0.9$ mJy. 

Karim et al. (2013) reported the first determination of the number counts at 870 $\mu$m based on arcsecond resolution observations with ALMA for a sample of 122 sub-mm sources selected from the LABOCA Extended {\it Chandra} Deep Field South Submillimetre Survey (LESS; Wei{\ss} et al. 2009). They found that the ALMA derived number counts are in broad agreement with previous determinations from single dish observations. Following Karim et al. (2013), Simpson et al. (2015) presented high-resolution 870 $\mu$m ALMA observations of a representative sample of the brightest 30 sub-mm sources in the UKIDSS UDS field which are selected from the S2CLS. 52 sub-mm galaxies were at S/N $>4$. They found that the level of multiplicity present in their observations boost the number counts from single dish observations by 20\% at $S_{870}>7.5$ mJy and by 60\% at $S_{870}>12$ mJy. Oteo et al. (2016), by exploiting sub-arcsec resolution ALMA calibration observations in a variety of frequency bands and array configurations, were able to reach lower noise levels (25$\mu$Jy beam$^{-1}$) to detect faint dusty star-forming galaxies. Oteo et al. (2016) presented cumulative number counts at 870 $\mu$m based on 11 sub-mm sources detected in ALMA band 7 at S/N $>5$. Following Simpson et al. (2015), Stach et al. (2018) reported the first results of the recently completed ALMA 870 $\mu$m continuum survey of a complete sample of over 700 sources from the UKIDSS/UDS field (50 arcmin$^2$). They were able to derive the number counts at $S_{870}>4$ mJy and confirm that the number counts derived from single dish SCUBA-2 observations are about 28\% too high in comparison. Hill et al. (2018) observed the brightest sources (down to $S_{850}\sim8$ mJy) in the S2CLS with SMA at 860 $\mu$m at an average syntheiszed beam of 2.4 arcsec. Their number counts are consistent with previous single dish results but the cumulative counts are systematically lower by $\sim14$\%.

Apart from the earliest measurement from Coppin et al. (2006) and the measurement from the lensing cluster field Chen et al. (2013), all other estimates more or less agree well with each other within errors. There is also excellent agreement between our predicted 870 $\mu$m number counts and previous measurements based on SCUBA-2 850 $\mu$m, SMA 860 $\mu$m and ALMA 870 $\mu$m observations. Thanks to our de-confusion technique and the wealth of deep multi-wavelength photometric information in COSMOS, we are able to extend the 870 $\mu$m counts measurements down to fainter flux levels (by a factor of $\sim2$) compared to the deepest observations carried out by SCUBA-2. 

The SIDES empirical model  has the best agreement with the observational measurements. In addition, there is an excellent agreement between our 870 $\mu$m number counts and the predicted counts from SIDES across the entire dynamic range where our measurements are available. Both the EGG empirical model and GALFORM over-predicted the number counts, especially in the flux range between $\sim0.3$ and $\sim6$ mJy. Again, the predicted number counts from the EAGLE hydrodynamic simulation are much lower compared to the observations and predictions from the empirical models and GALFORM. It is also interesting to see that the under-prediction of EAGLE number counts compared to the observed counts worsens towards longer wavelength.

\subsection{Luminosity functions and their evolution}

\begin{figure*}
\centering
\includegraphics[height=2.6in,width=3.3in]{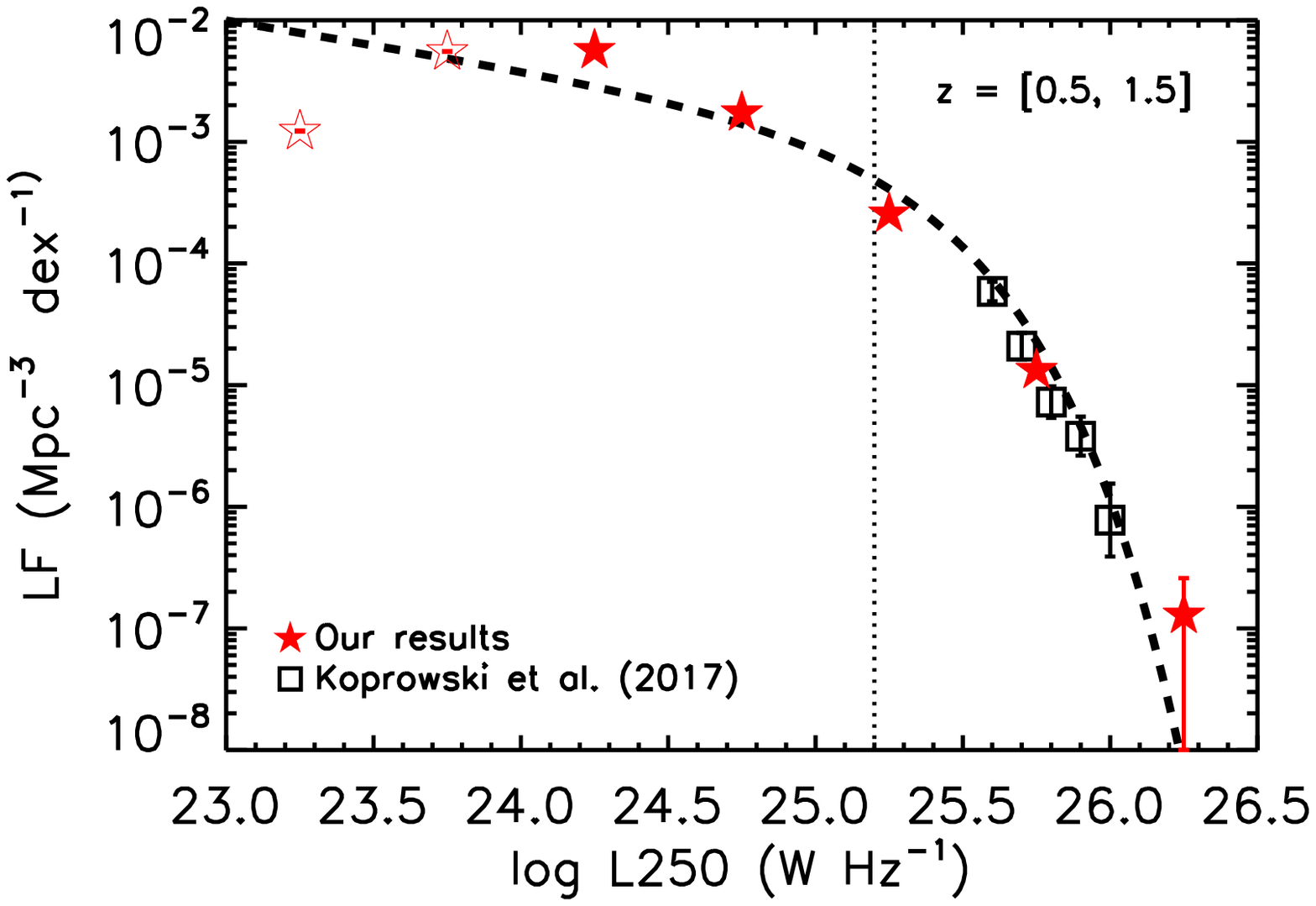}
\includegraphics[height=2.6in,width=3.3in]{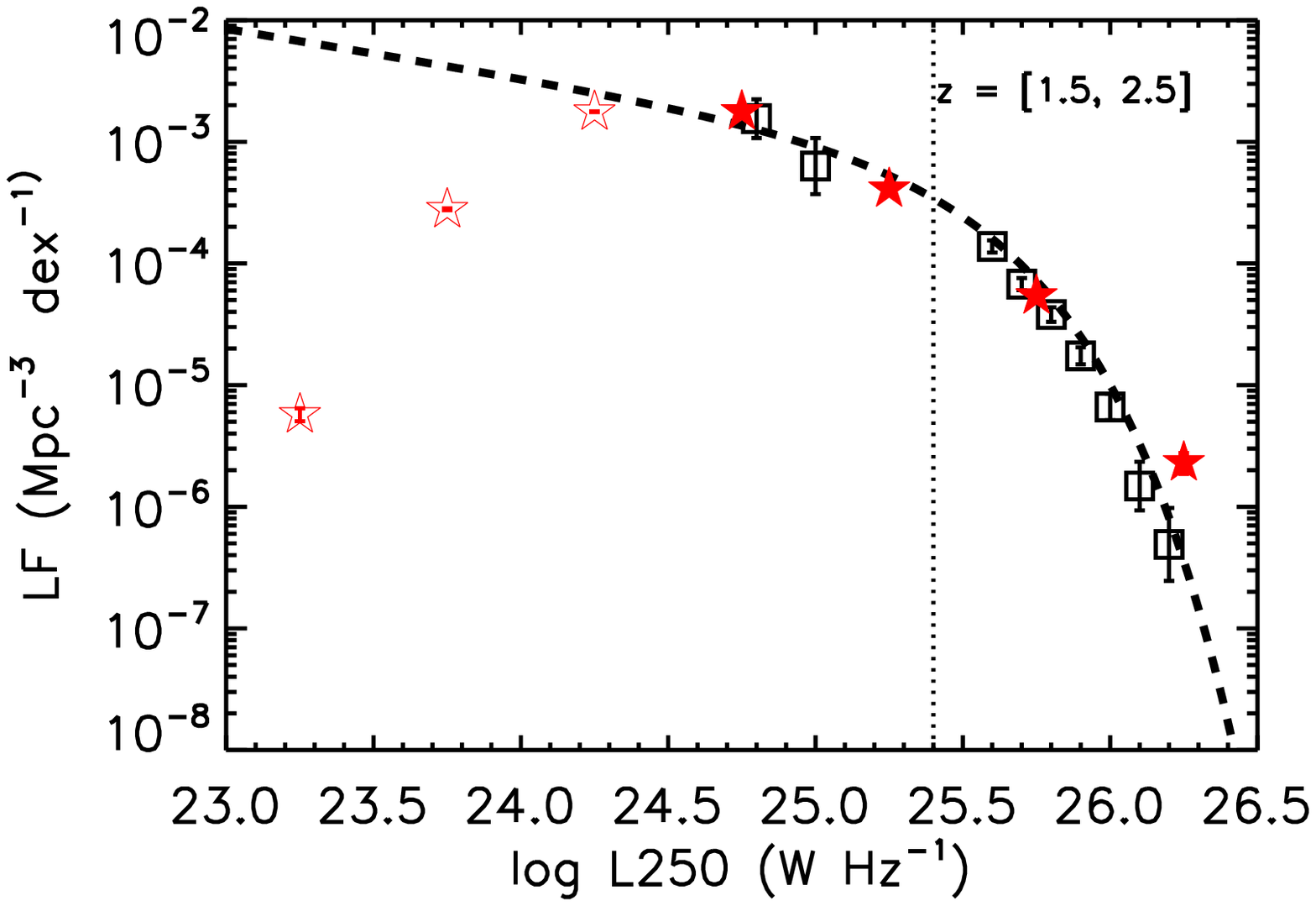}
\includegraphics[height=2.6in,width=3.3in]{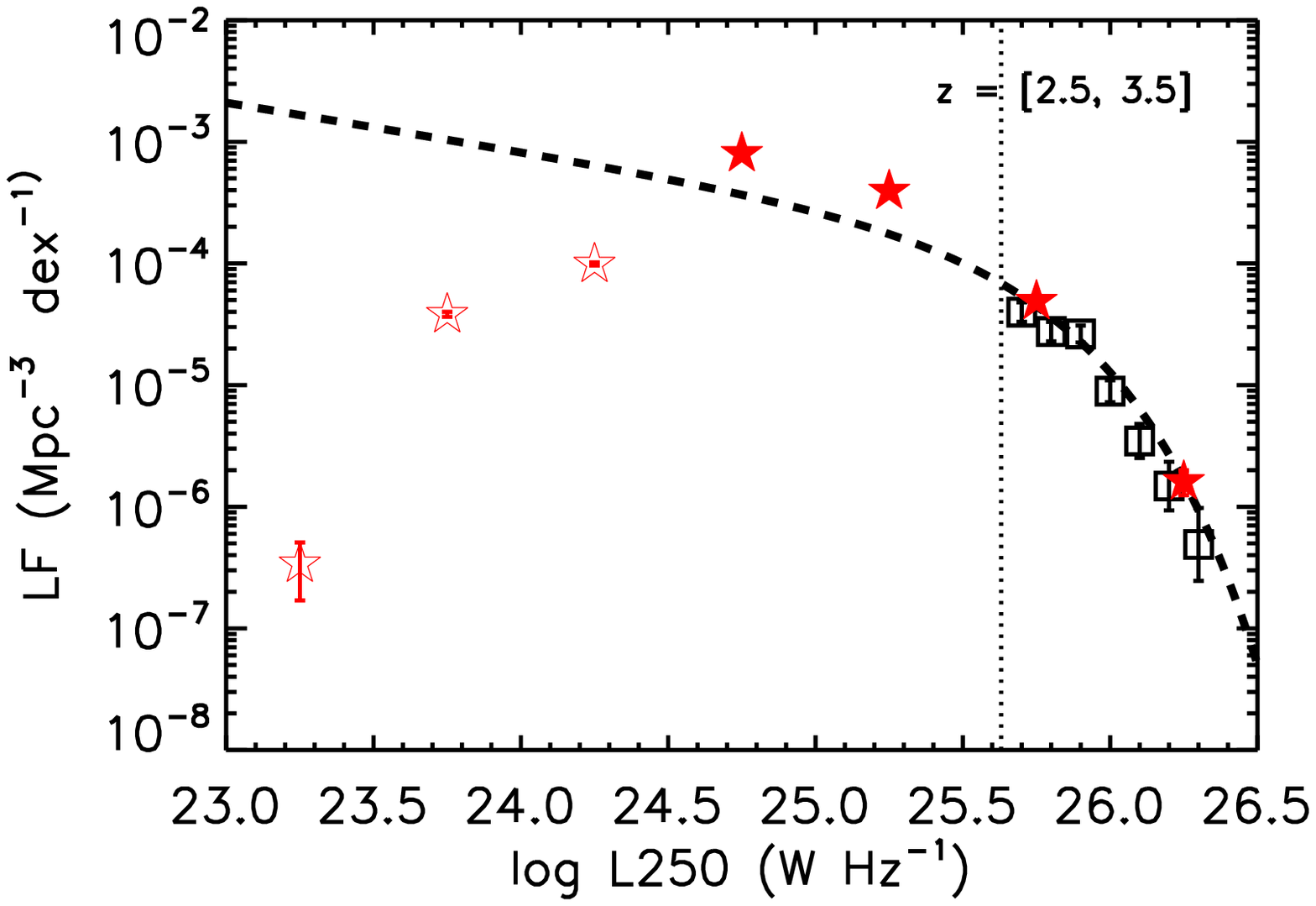}
\includegraphics[height=2.6in,width=3.3in]{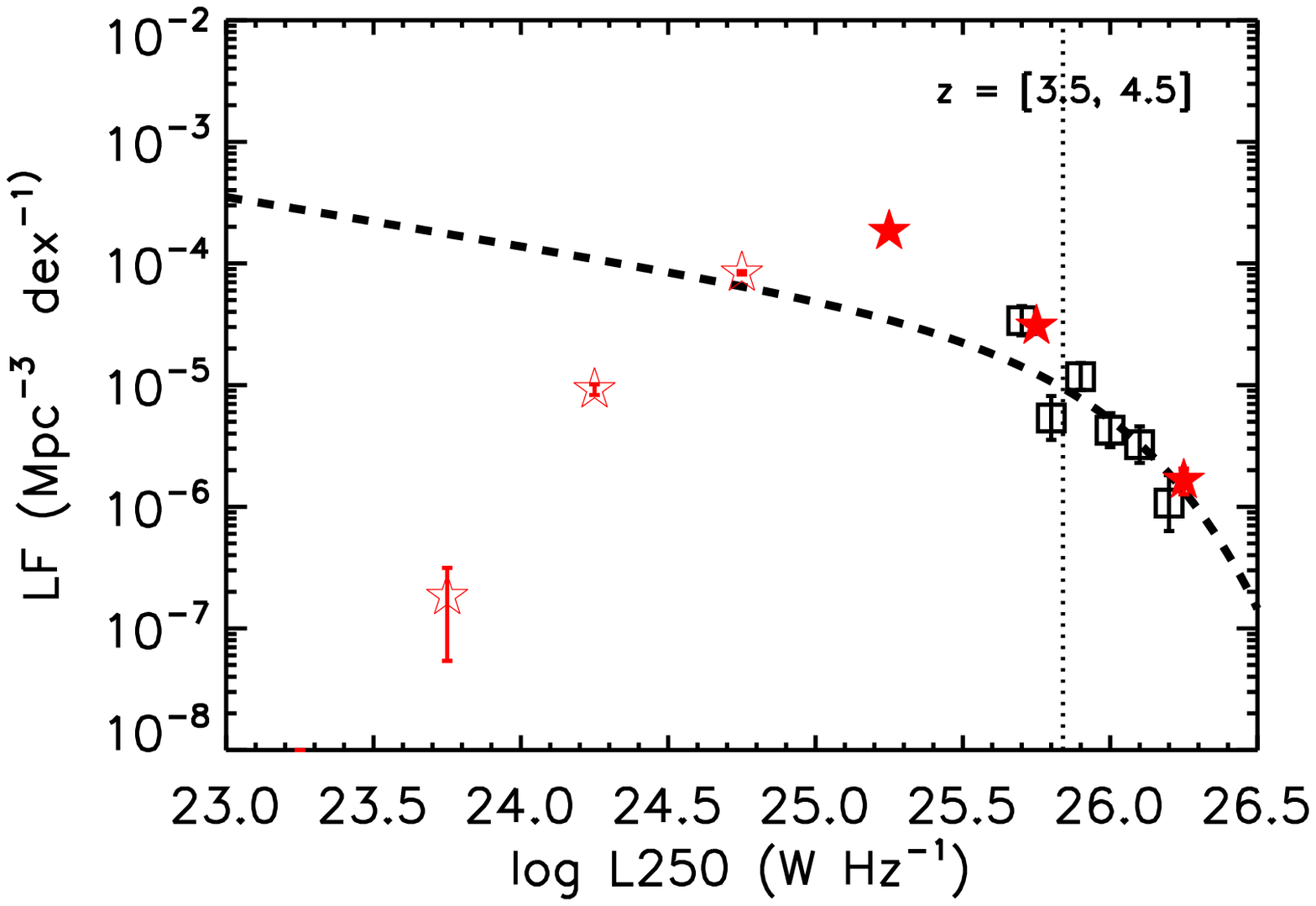}
\caption{The rest-frame 250 $\mu$m luminosity function. The red stars are derived from our de-blended catalogue in COSMOS (filled red stars: our LF above the completeness limit; empty red stars: our LF below the completeness limit). Error bars on the red stars only represent Poisson errors. The black empty squares are taken from Koprowski et al. (2017), based on SCUBA-2 850 $\mu$m observations of the COSMOS and UKIDSS-UDS fields as part of the S2CLS. The two faintest points in the Koprowski et al. (2017) measurements in the $1.5<z<2.5$ redshift bin (which has the largest dynamic range) were derived using the ALMA 1.3 mm data. The dashed line is the best-fit Schechter function adopted in Koprowski et al. (2017). Note that the faint-end slope of the Schechter function was found to be $\alpha=-0.4$ in the $1.5<z<2.5$ redshift bin in Koprowski et al. (2017) and was kept fixed in the remaining three redshift bins. The vertical dotted line indicates the location of the characteristic luminosity, i.e. $L_*$ in Eq (1), as derived in Koprowski et al. (2017).}
\label{LF250}
\end{figure*}

In this subsection, we first present our results on the monochromatic rest-frame 250 $\mu$m LF and then the total IR LF (integrated from 8 to 1000  $\mu$m)  in various redshift bins. We will also compare our results with previous measurements as well as predictions from the Durham SAM and the EAGLE hydrodynamic simulations.

\subsubsection{The monochromatic rest-frame 250 $\mu$m luminosity function}

In Fig.~\ref{LF250}, we compare our monochromatic rest-frame 250 $\mu$m LF in four redshift bins from $z\sim0.5$ to $z\sim4.5$ with those in Koprowski et al. (2017). Koprowski et al. (2017), using SCUBA-2 850 $\mu$m observations in the COSMOS and UKIDSS-UDS fields from the S2CLS together with ALMA 1.3 mm imaging data of the HUDF (Dunlop et al. 2017), determined the rest-frame 250 $\mu$m LFs out to redshift $z\sim5$. As the mean redshift of the population of their 850 $\mu$m detected sources is around 2.5 (probing rest-frame around 250 $\mu$m at the mean redshift), the average sub-mm galaxy template from Micha{\l}owski et al. (2010) was adopted to convert the observed-frame 850 $\mu$m to the rest-frame 250 $\mu$m flux density.  Koprowski et al. (2017) also presented the best-fitting Schechter functions which are parametrised as
\begin{equation}
\phi(L, z) = \phi_* \left(\frac{L}{L_*}\right)^{\alpha} \exp{\left(\frac{-L}{L_*}\right)}
\end{equation}
where $\phi_*$ is the normalisation parameter, $\alpha$ is the faint-end slope and $L_{*}$ is the characteristic luminosity. As can be seen from Fig.~\ref{LF250}, our rest-frame 250 $\mu$m LF in the four redshift bins agree well with the measurements from Koprowski et al. (2017) in the overlapping luminosity range, but our measurements also extend to much fainter luminosities (roughly 10 times fainter). In the redshift bin $1.5<z<2.5$, the dynamic range in luminosity probed by our study is the same as by Koprowski et al. (2017). This is because the two faintest points in Koprowski et al. (2017) in the $1.5<z<2.5$ bin were derived using the ALMA 1.3 mm data.  Note that in Koprowski et al. (2017), the faint-end slope $\alpha$ is derived to be $\alpha=-0.4$ in the redshift bin $1.5<z<2.5$ and is kept fixed at this value for the remaining three redshift bins. While our rest-frame 250 $\mu$m LF measurements agree well with the best-fitting Schechter functions presented in Koprowski et al. (2017) in the two lowest redshift bins, our measurements indicate higher volume densities towards the faint end at higher redshifts.

It is worth noting at this point that Koprowski et al. (2017) found their total IR LF measurements based on SCUBA-2 observations have a much smaller number of bright sources at all redshifts compared to the {\it Herschel}-based studies of Magnelli et al. (2013) and Gruppioni et al. (2013). However, the Koprowski et al. (2017) study used a single SED (i.e. the average sub-mm galaxy SED template from Micha{\l}owski et al. 2010) in order to convert the observed 850 $\mu$m flux density into a total IR flux (integrated between 8 and 1000 $\mu$m). Given that we agree reasonably well with the monochromatic rest-frame 250 $\mu$m LF from Koprowski et al. (2017) and also with the total IR LF from Magnelli et al. (2013)  and Gruppioni et al. (2013), as will be seen in Section 3.2.2, we conclude that the likely cause for the disagreement between the SCUBA-2 based study and the {\it Herschel}-based studies is the use of a single SED shape instead of the full wide range of SEDs present in the dusty star-forming galaxy population. Gruppioni \& Pozzi (2019) conducted a thorough investigation into the large discrepancies seen in the total IR LF from Koprowski et al. (2017)  and the {\it Herschel}-based measurements of Magnelli et al. (2013) and Gruppioni et al. (2013). They concluded that the discrepancy is mainly caused by the use of a single template in  Koprowski et al. (2017) and sample incompleteness as SCUBA-2 surveys are biased against galaxies with "warm" SED shapes.

\subsubsection{The total infrared luminosity function}

\begin{figure*}
\centering
\includegraphics[height=1.7in,width=2.3in]{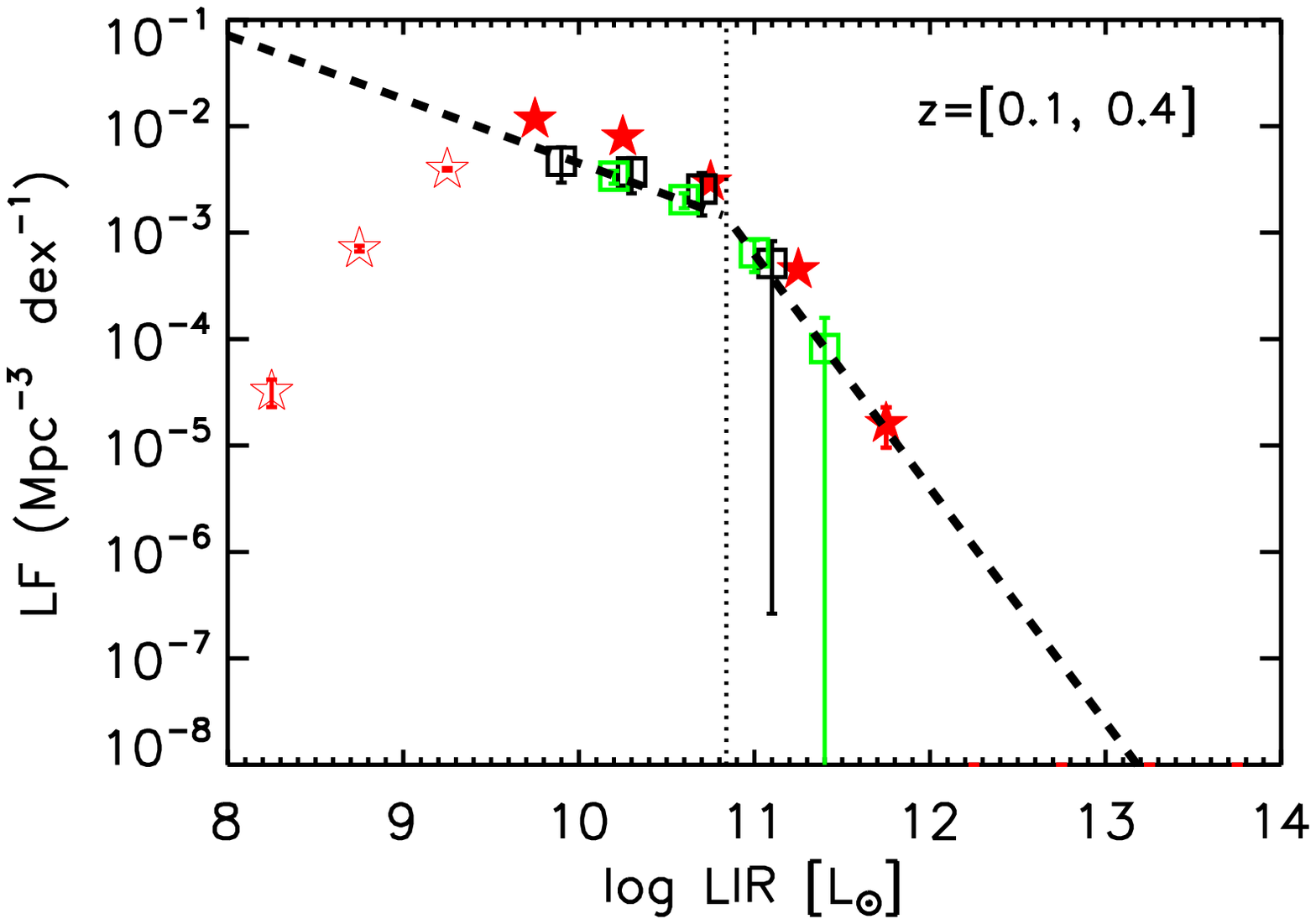}
\includegraphics[height=1.7in,width=2.3in]{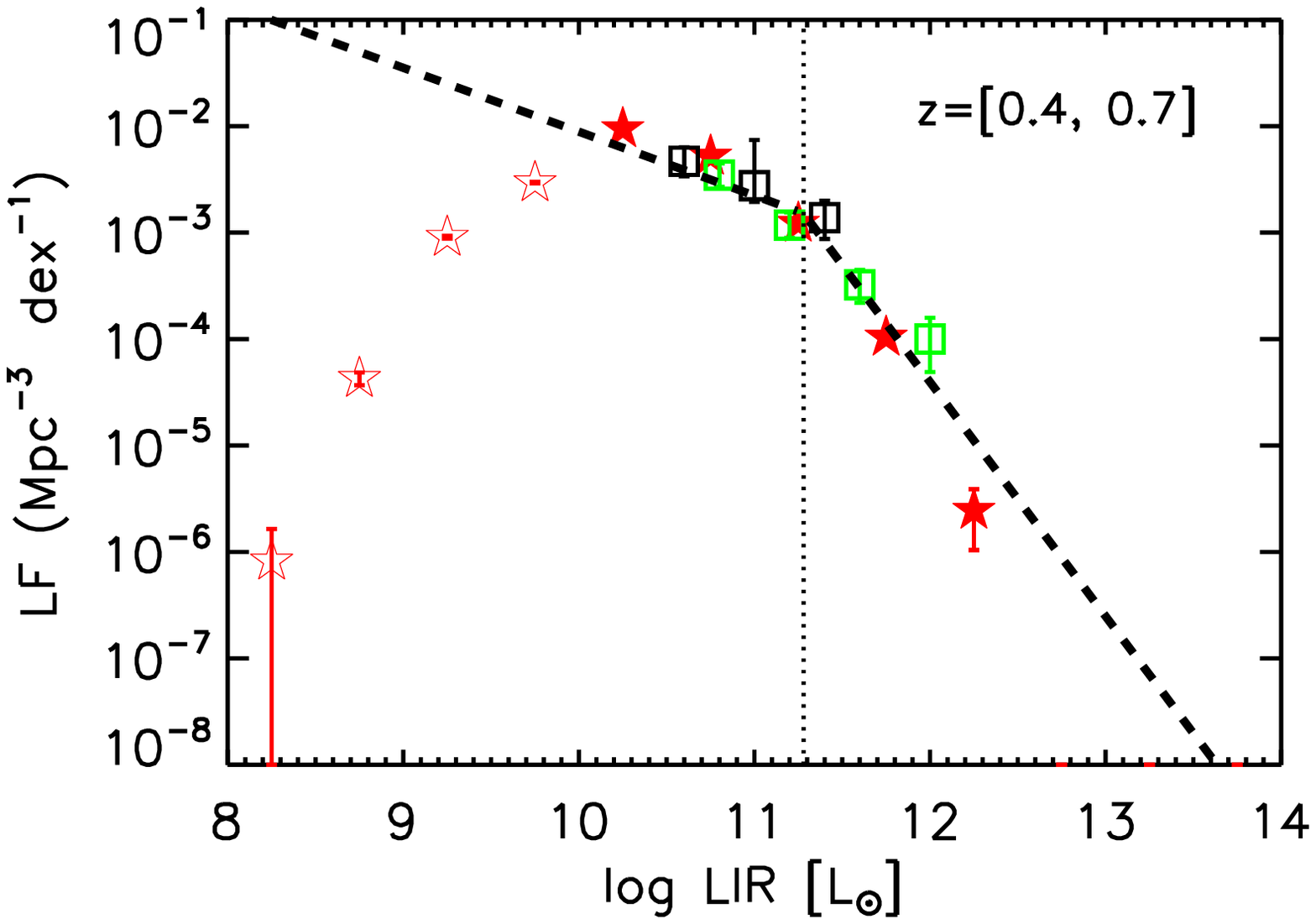}
\includegraphics[height=1.7in,width=2.3in]{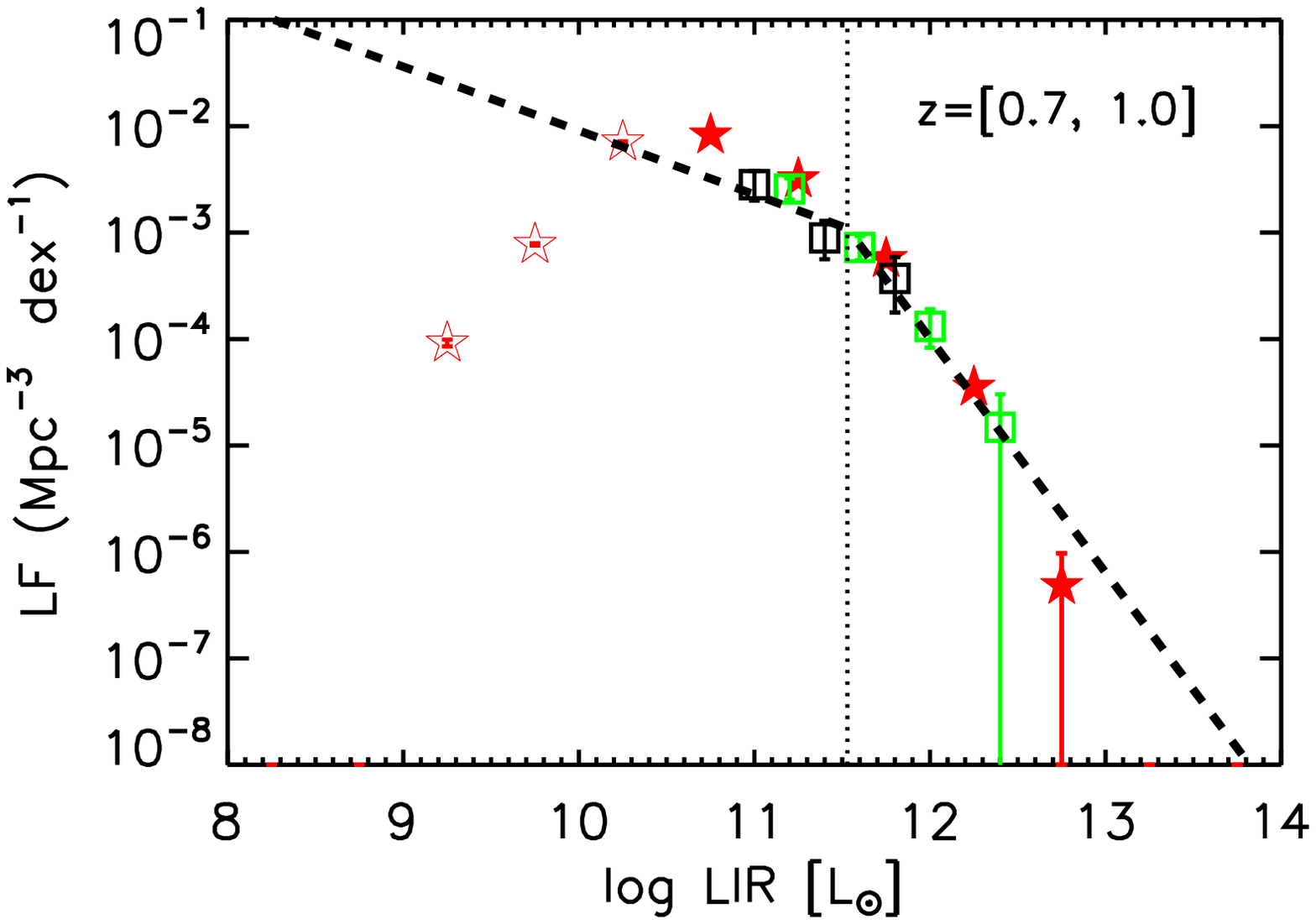}
\includegraphics[height=1.7in,width=2.3in]{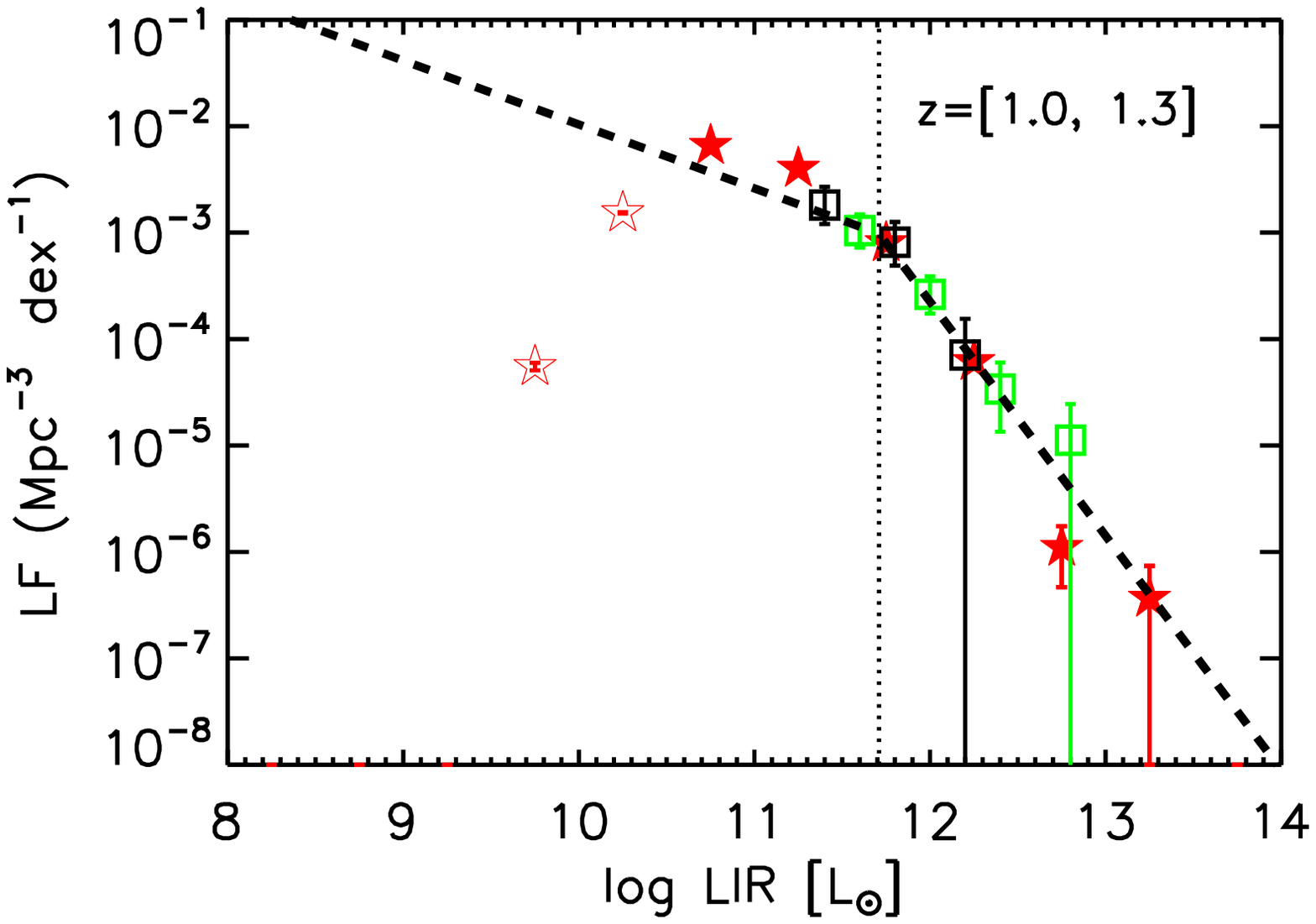}
\includegraphics[height=1.7in,width=2.3in]{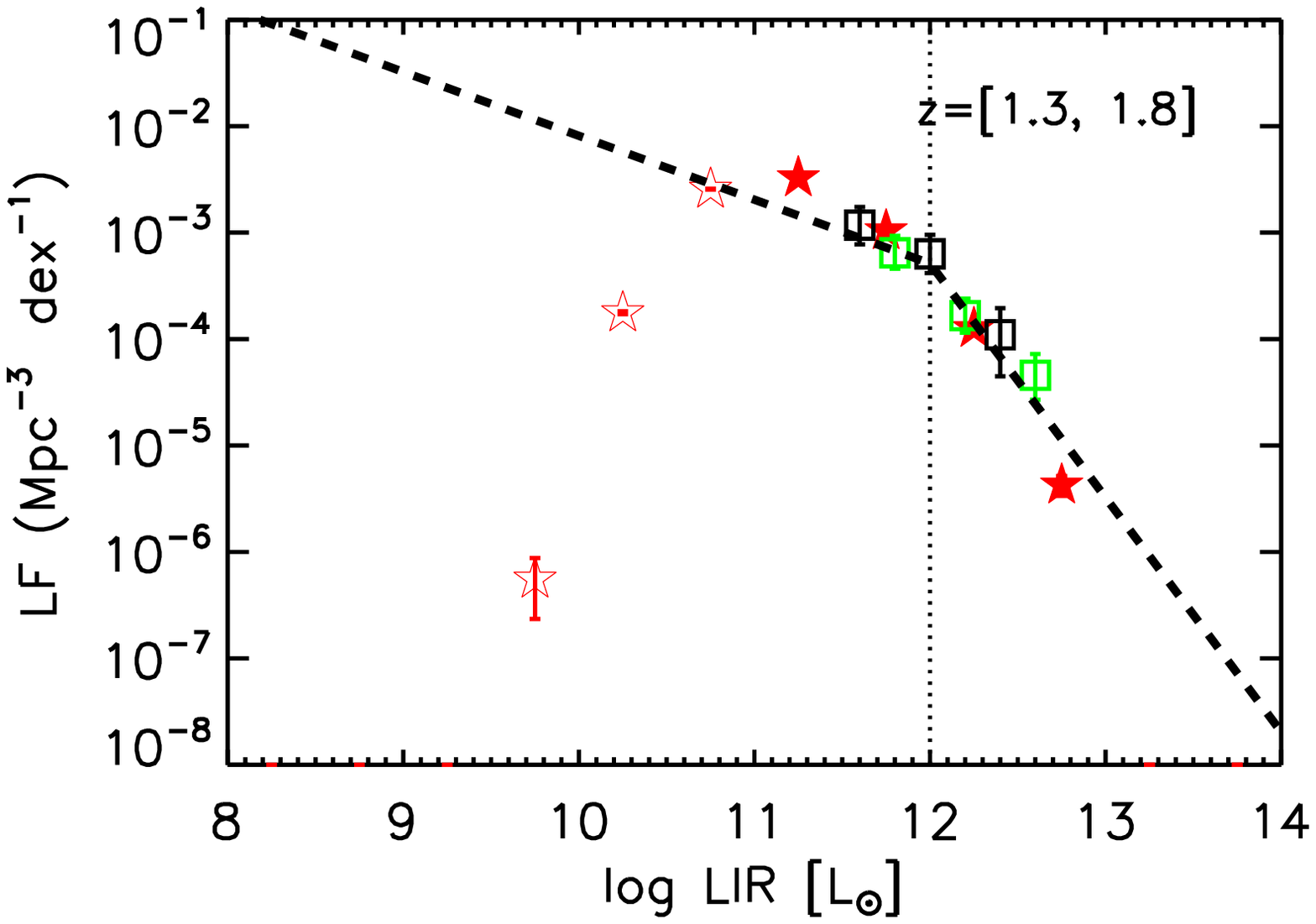}
\includegraphics[height=1.7in,width=2.3in]{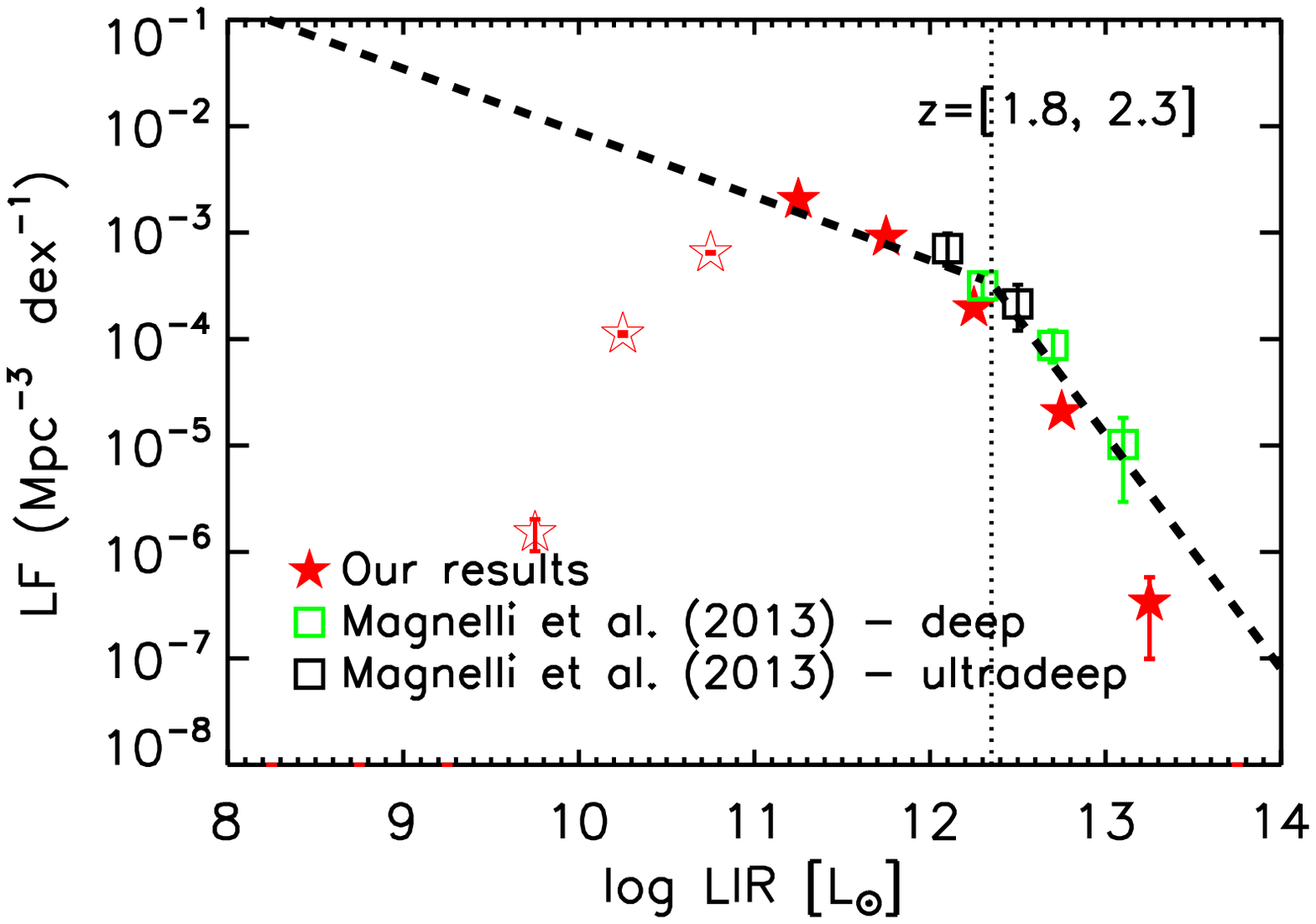}
\caption{The total IR luminosity function out to $z\sim2.3$. The red stars are derived from our de-blended catalogue in COSMOS (filled red stars: our LF above the completeness limit; empty red stars: our LF below the completeness limit). Error bars on the red stars only represent Poisson errors. The black squares are from Magnelli et al. (2013) based on observations of the GOODS-S ultradeep field. The green squares are also from Magnelli et al. (2013) but based on observations of the GOODS-N/S deep fields. The dashed line in each panel is the best-fit double power law from Magnelli et al. (2013).  The vertical dotted line indicates the location of the transition luminosity, i.e. $L_{knee}$ in Eq. (2) and (3), as derived in Magnelli et al. (2013).}
\label{LF_mag}
\end{figure*}

\begin{figure*}
\centering
\includegraphics[height=1.8in,width=2.3in]{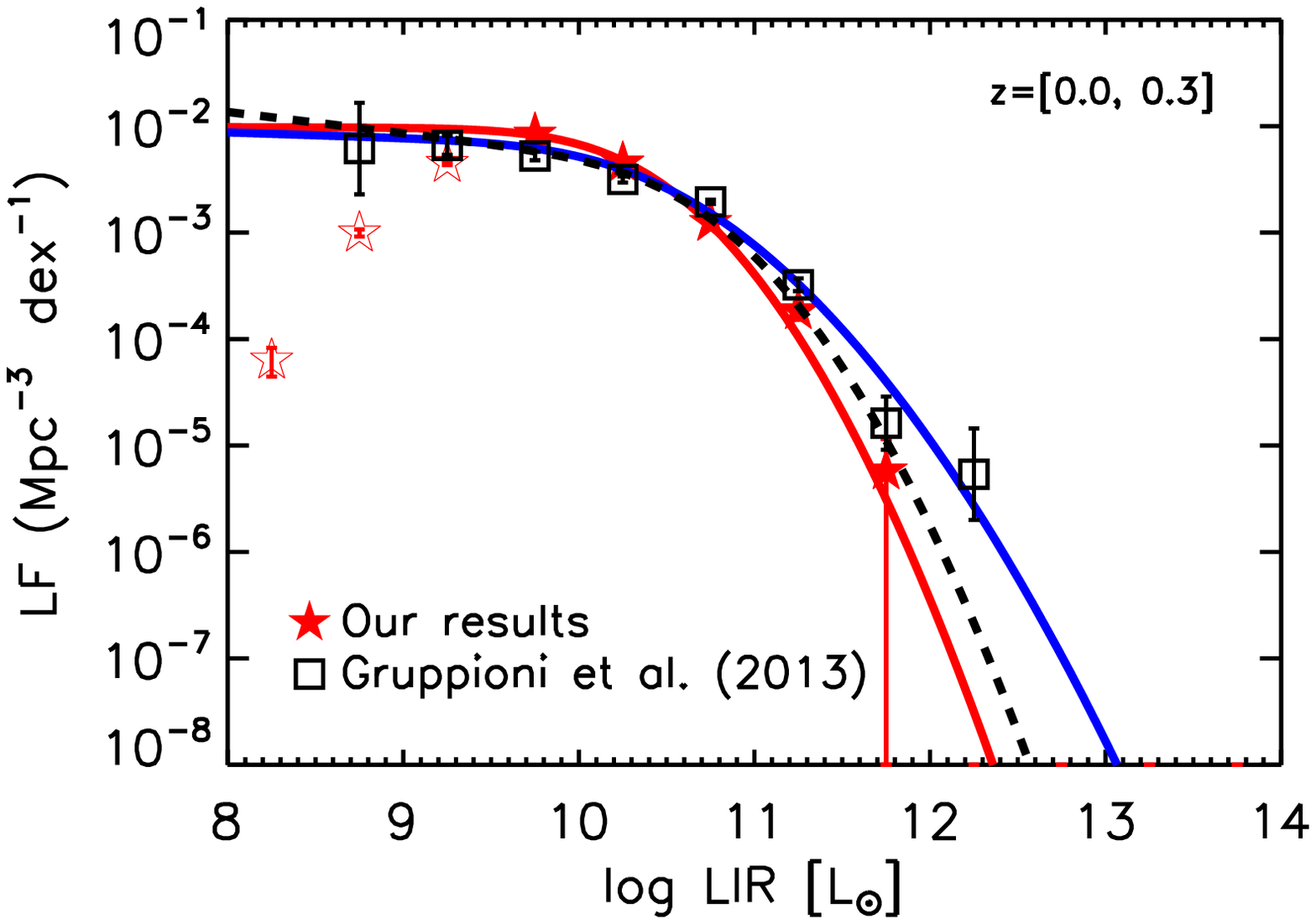}
\includegraphics[height=1.8in,width=2.3in]{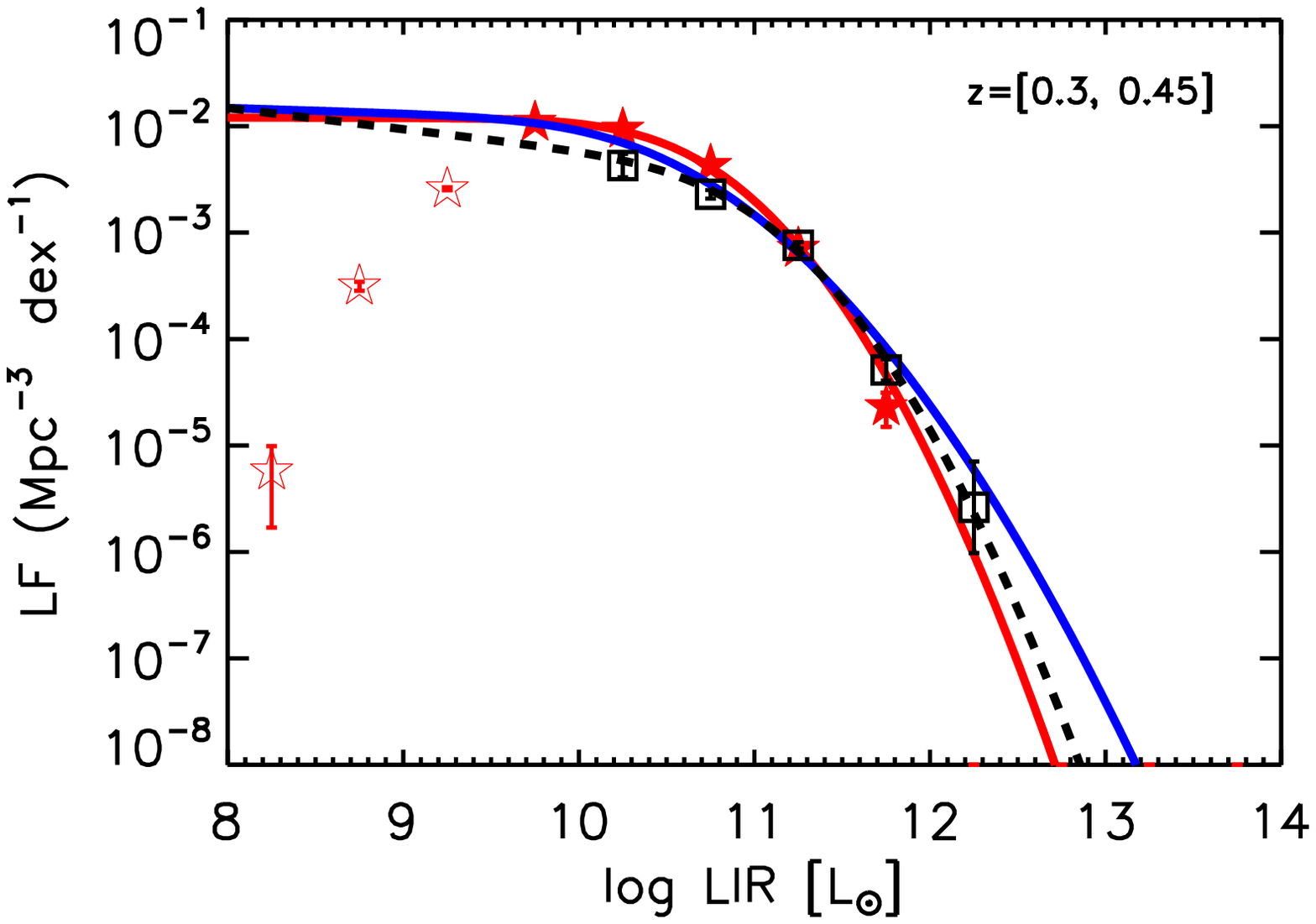}
\includegraphics[height=1.8in,width=2.3in]{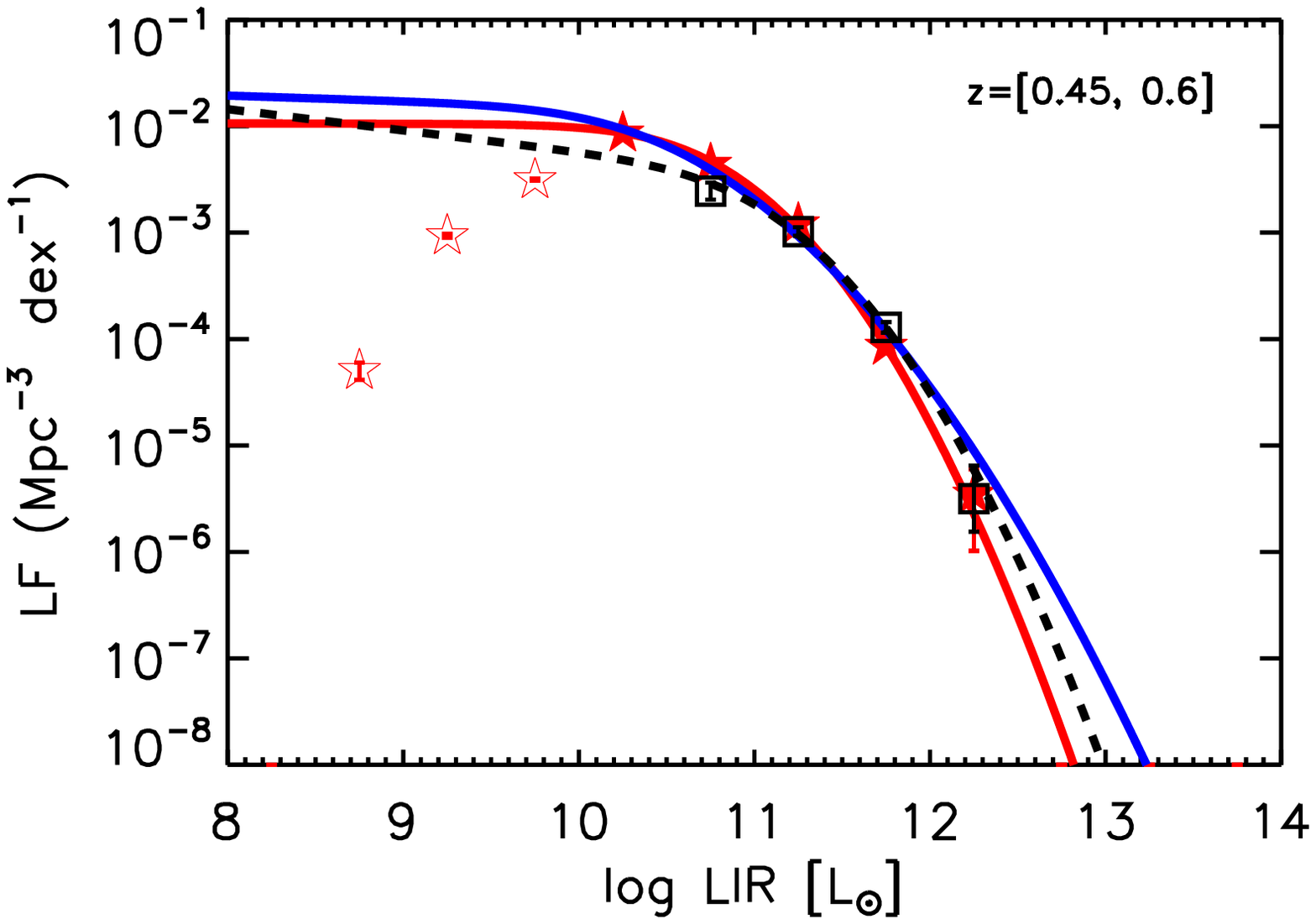}
\includegraphics[height=1.8in,width=2.3in]{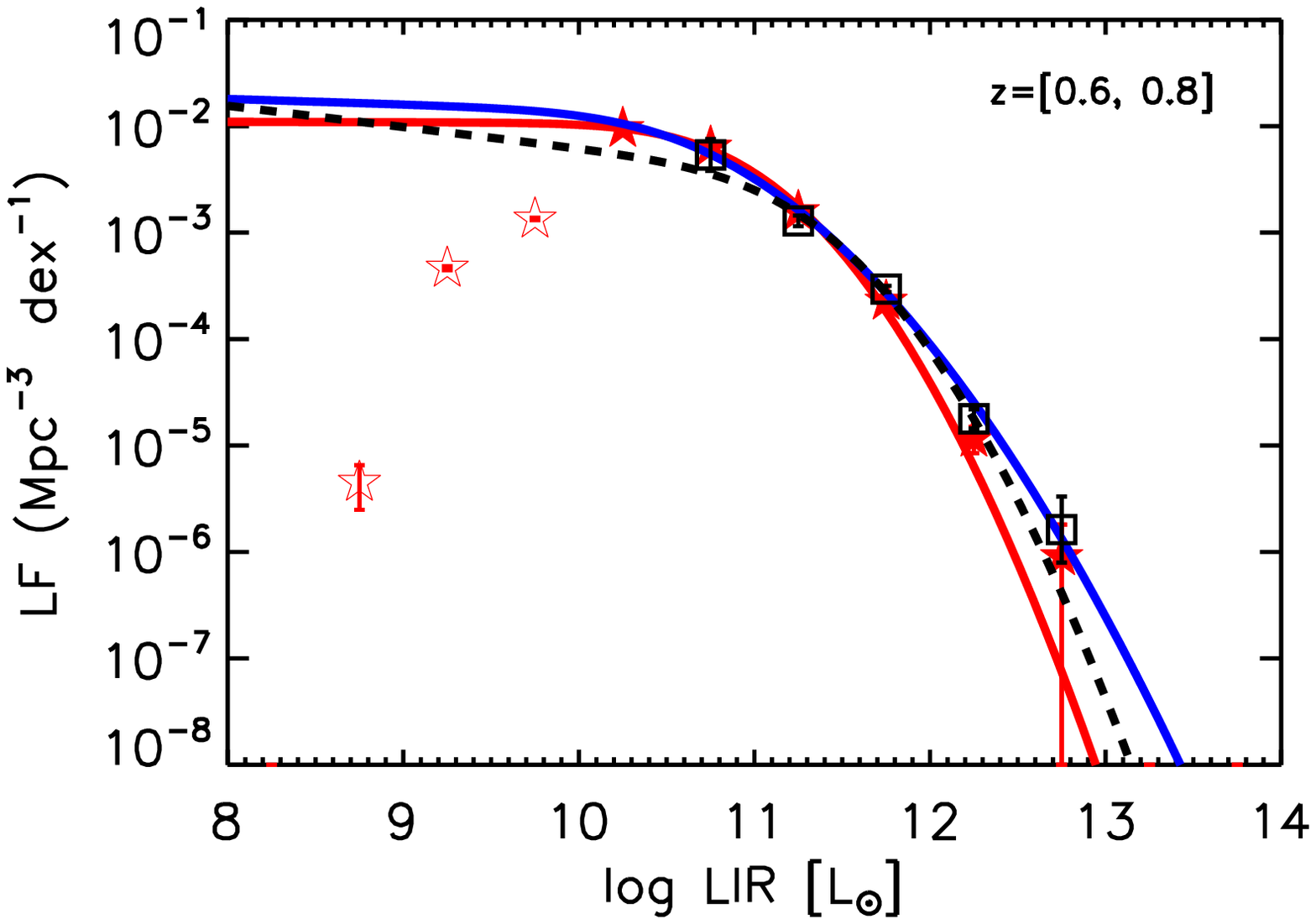}
\includegraphics[height=1.8in,width=2.3in]{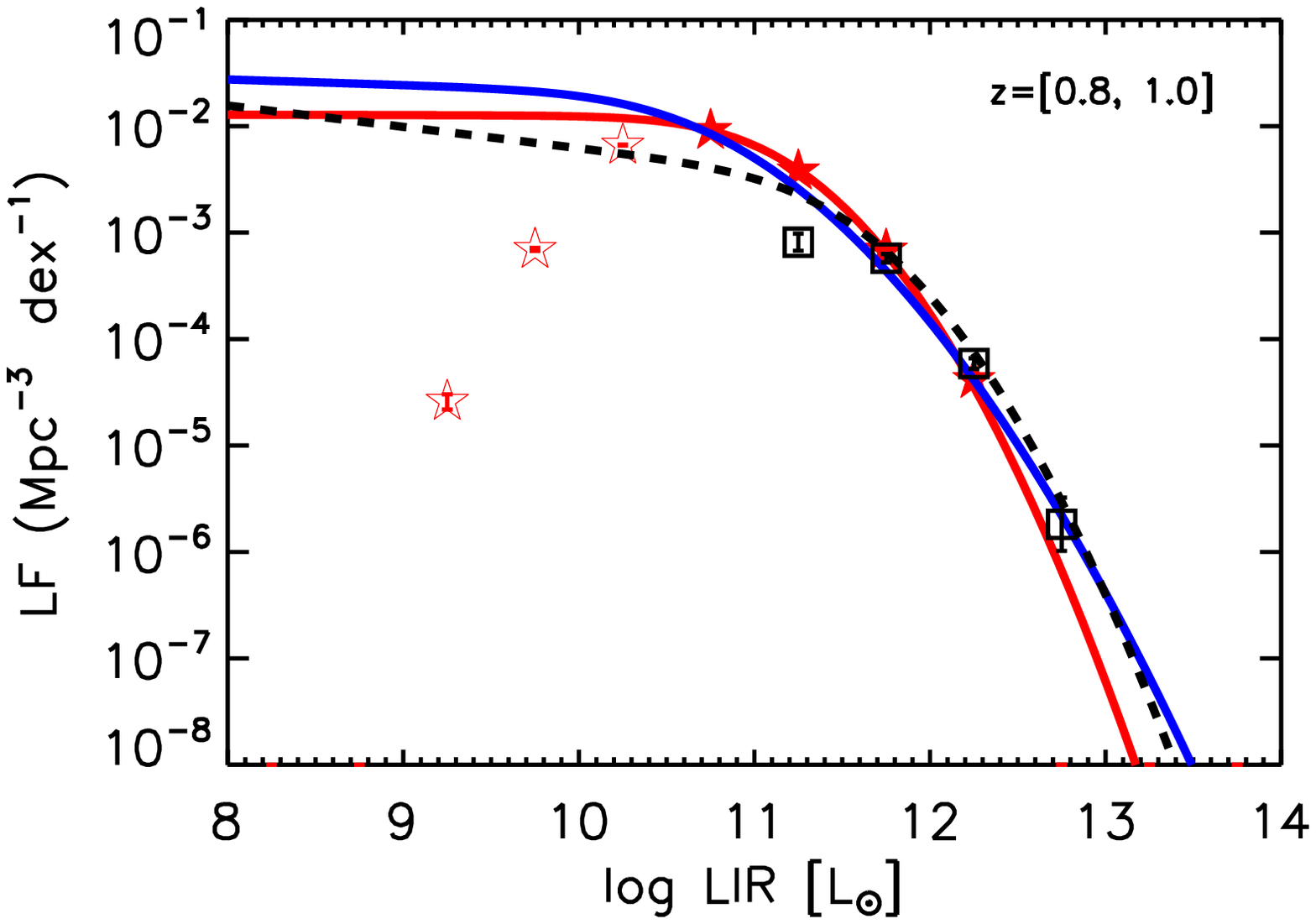}
\includegraphics[height=1.8in,width=2.3in]{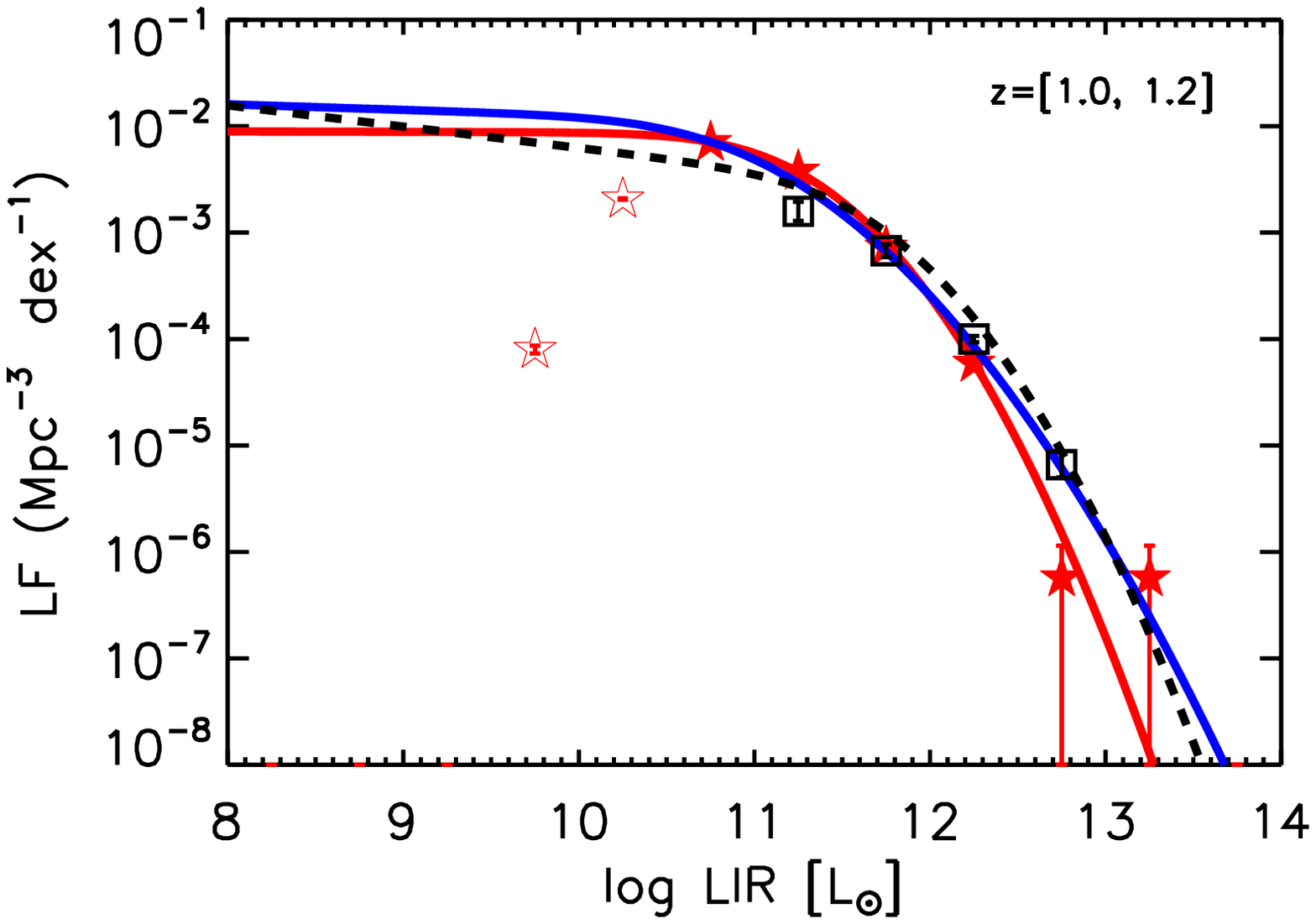}
\includegraphics[height=1.8in,width=2.3in]{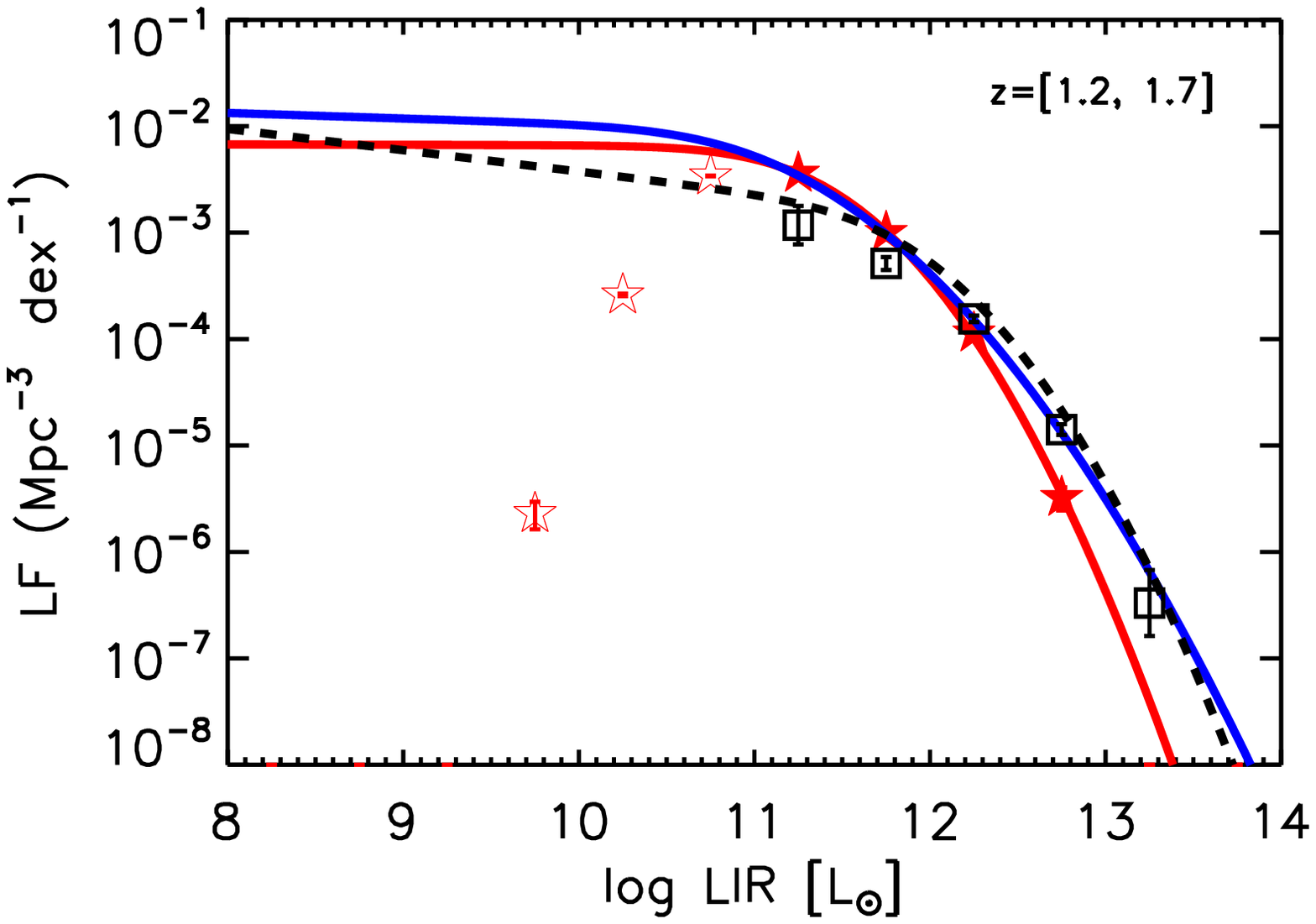}
\includegraphics[height=1.8in,width=2.3in]{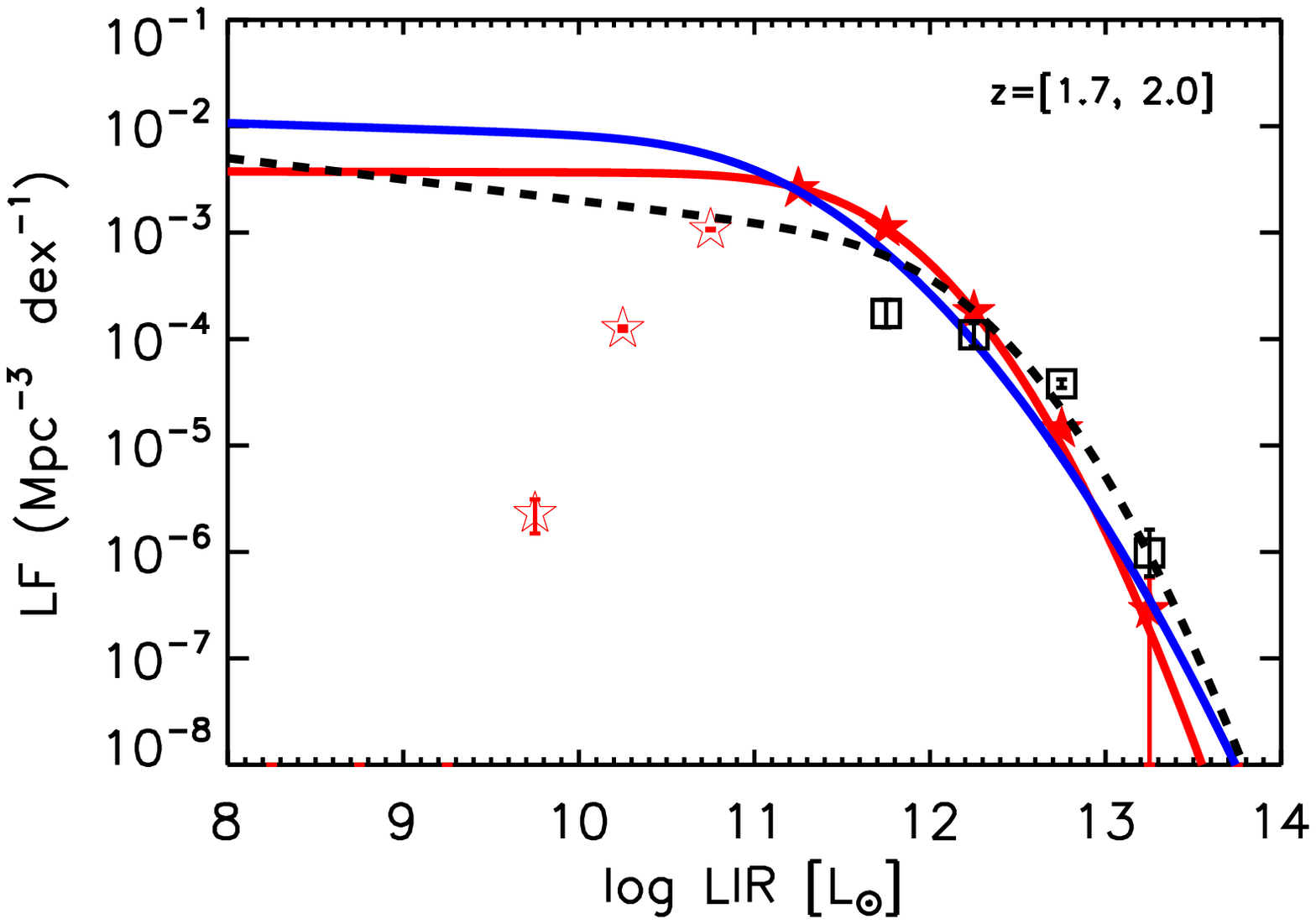}
\includegraphics[height=1.8in,width=2.3in]{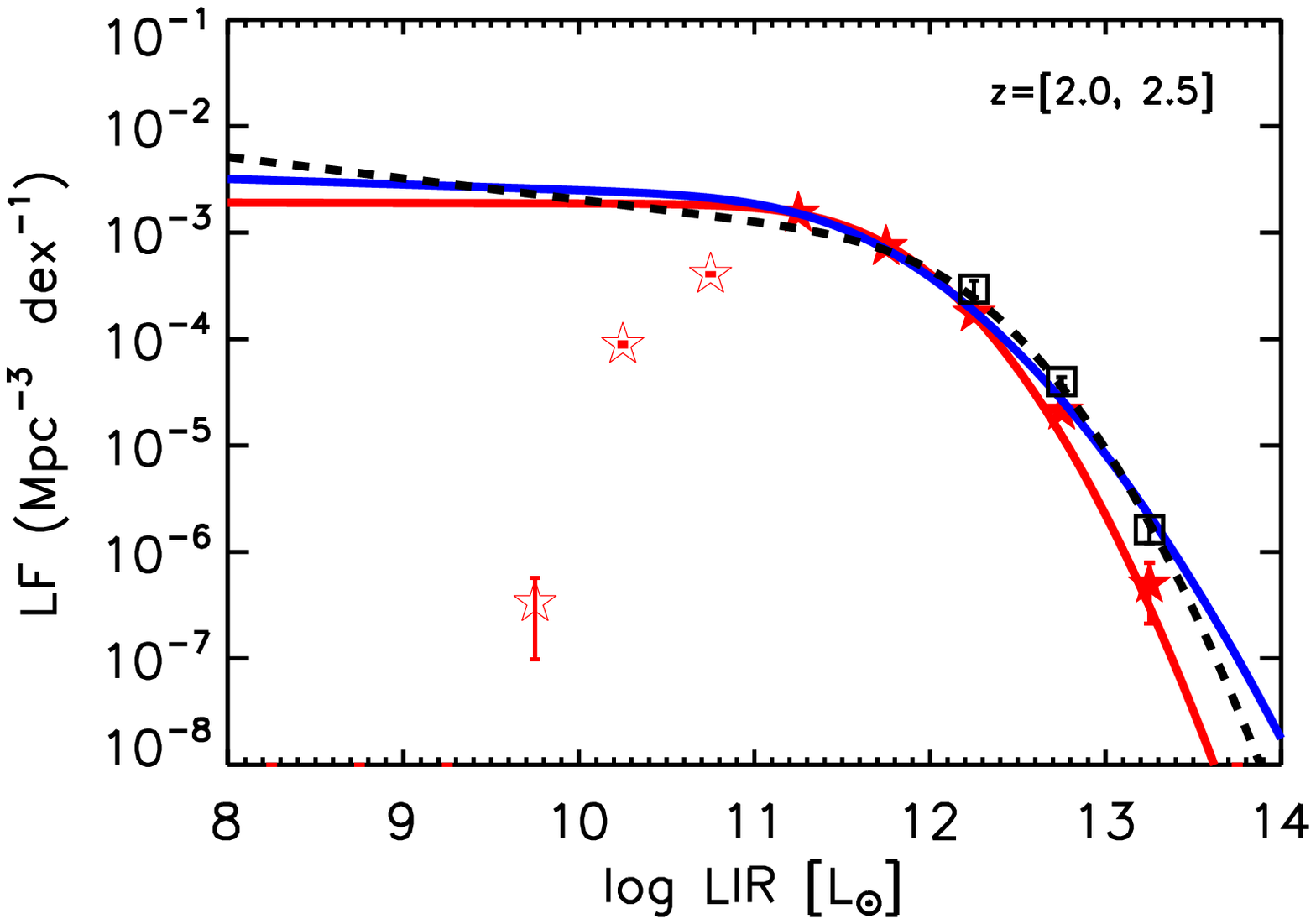}
\includegraphics[height=1.8in,width=2.3in]{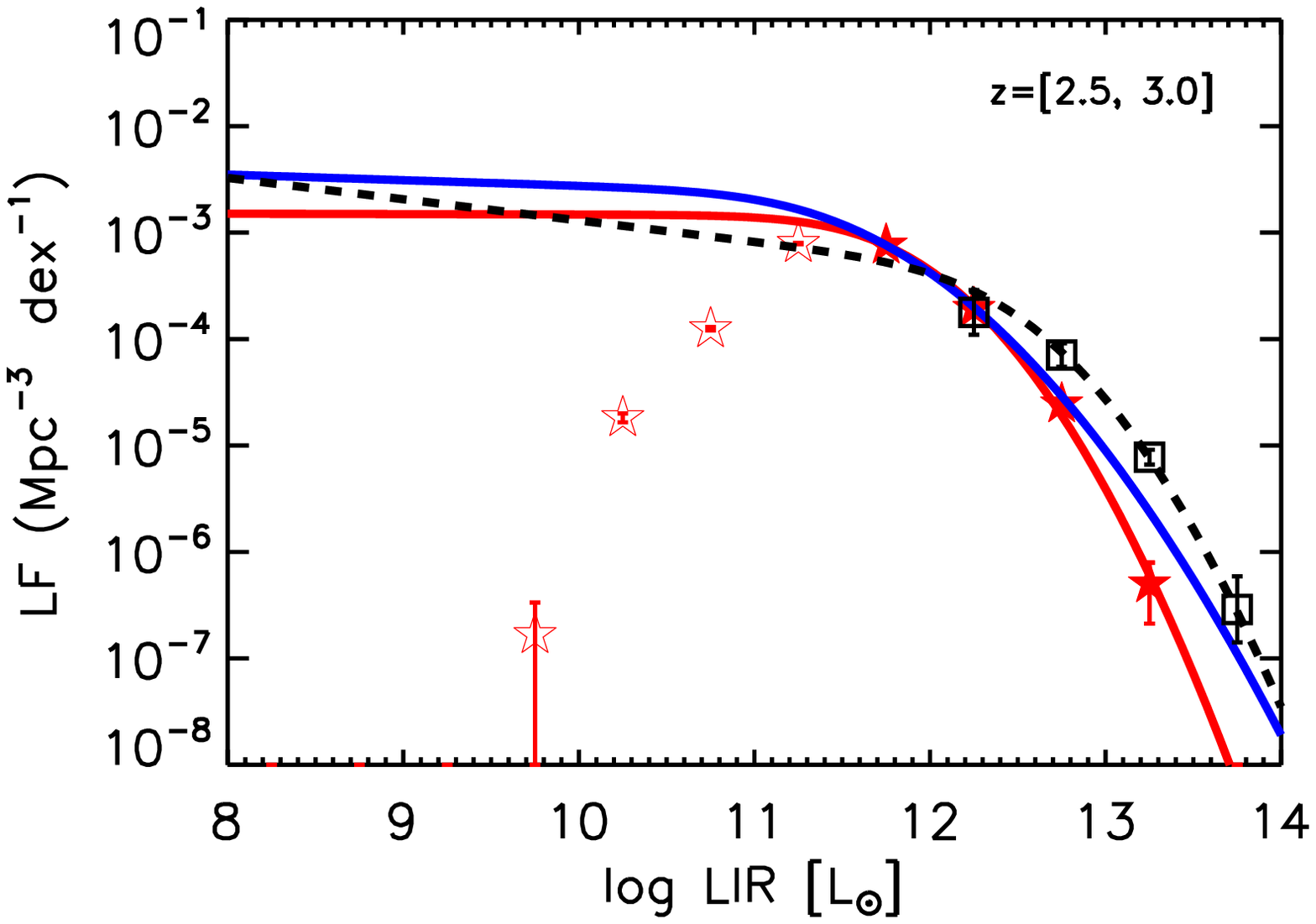}
\includegraphics[height=1.8in,width=2.3in]{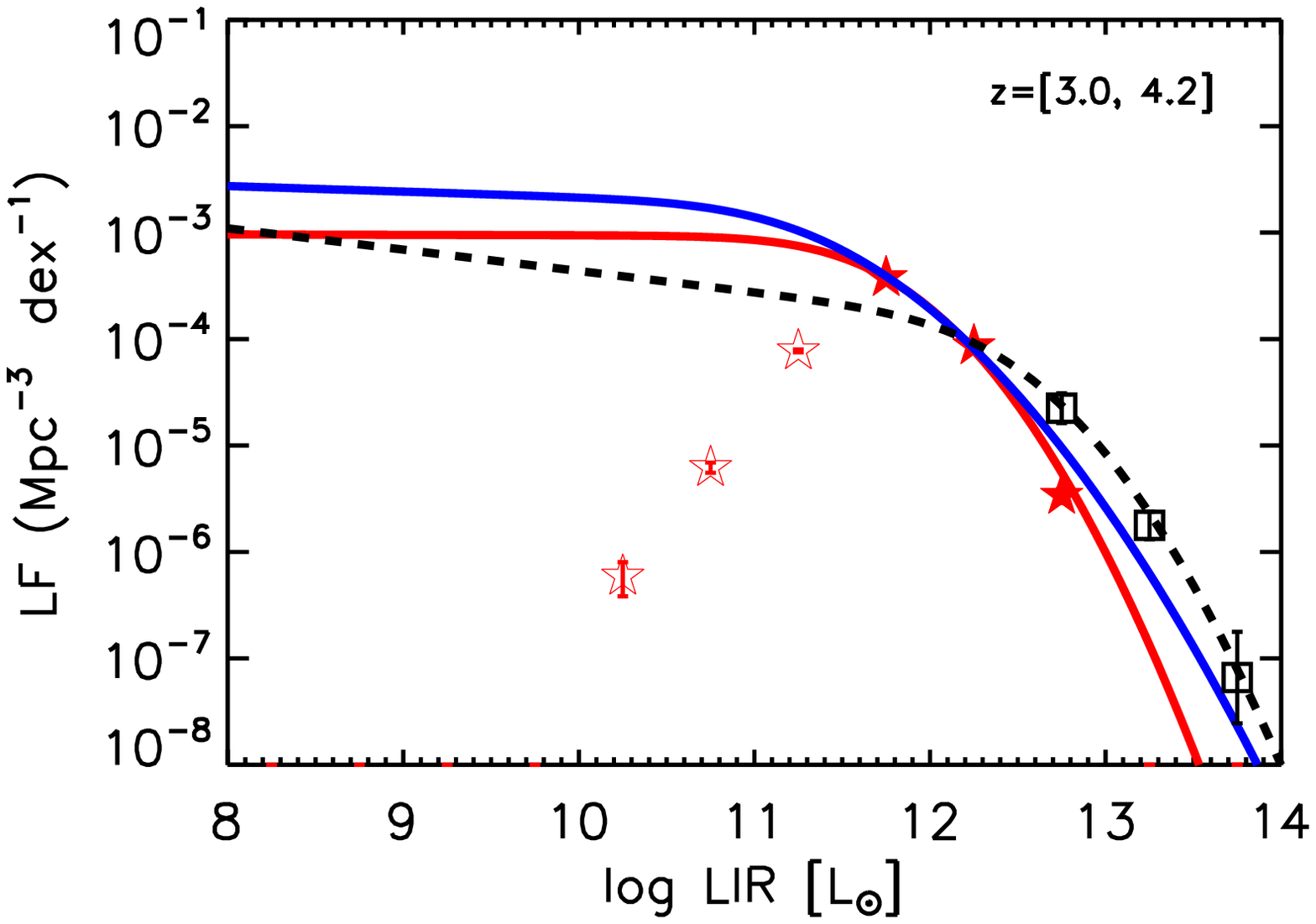}
\includegraphics[height=1.8in,width=2.3in]{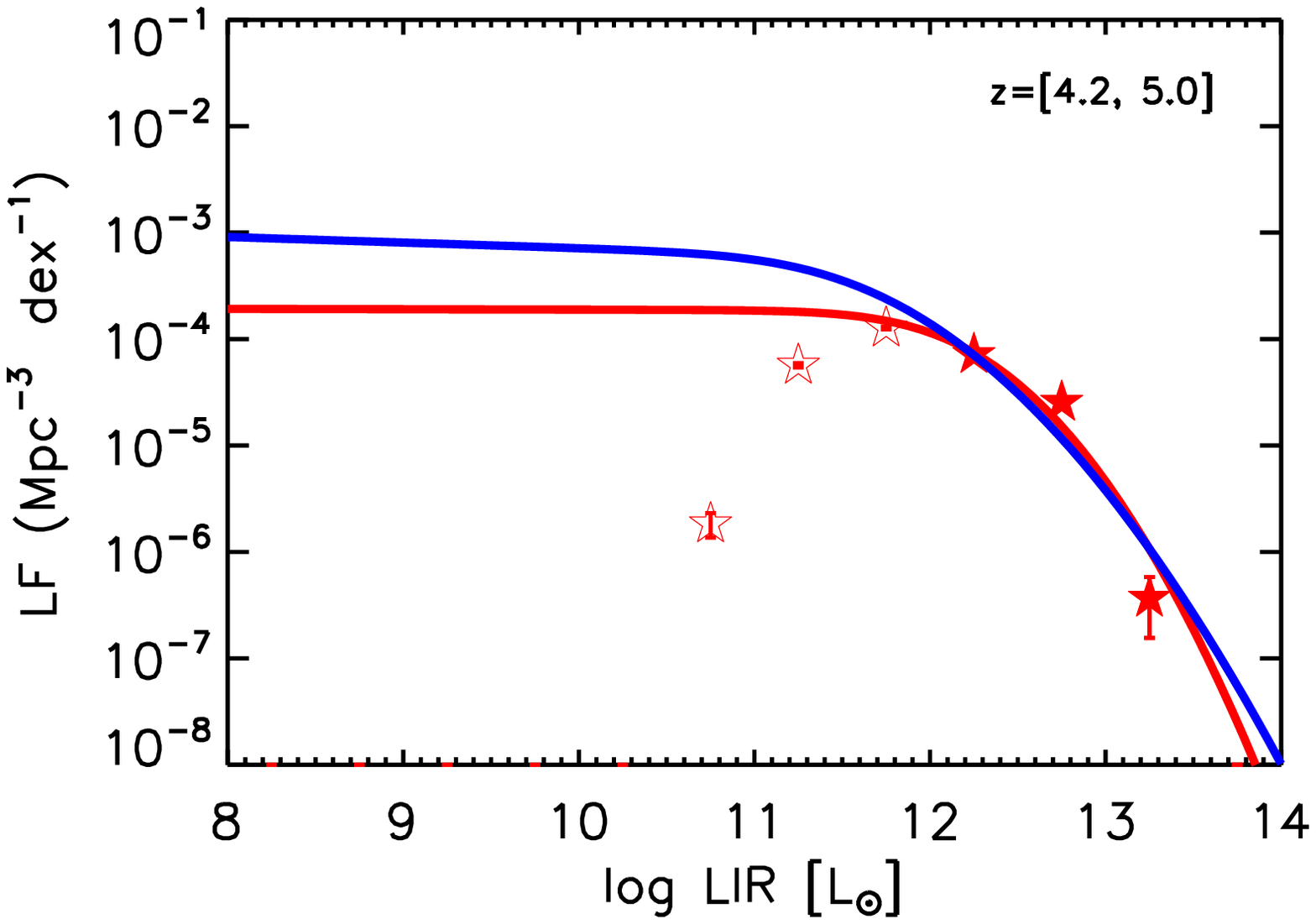}
\includegraphics[height=1.8in,width=2.3in]{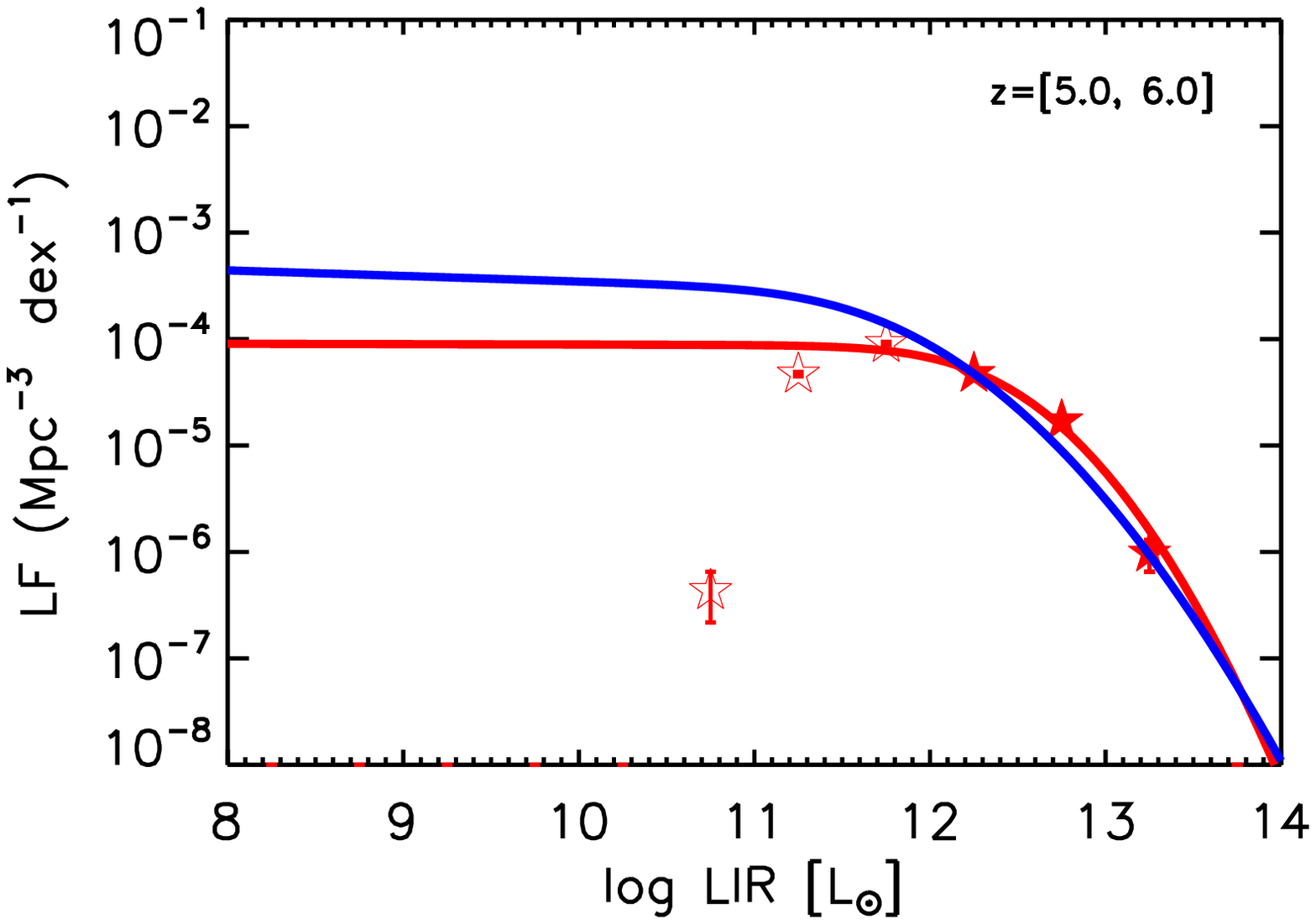}
\caption{The total IR luminosity function out to $z\sim6$. The red stars are derived from our de-blended catalogue in COSMOS (filled red stars: our LF above the completeness limit; empty red stars: our LF below the completeness limit). Error bars on the red stars only represent Poisson errors. The black squares are from Gruppioni et al. (2013). The dashed line is the best-fit modified-Schechter function from Gruppioni et al. (2013). The red line is the best-fit function derived from fitting to our measurements of the IR LF only (i.e., the filled red stars). The blue line is the best-fit function derived from fitting to both our IR LFs and the measurement in Gruppioni et al. (2013) (i.e. the filled red stars and the black squares).}
\label{LF_gru}
\end{figure*}

\begin{figure*}
\centering
\includegraphics[height=2.8in,width=3.6in]{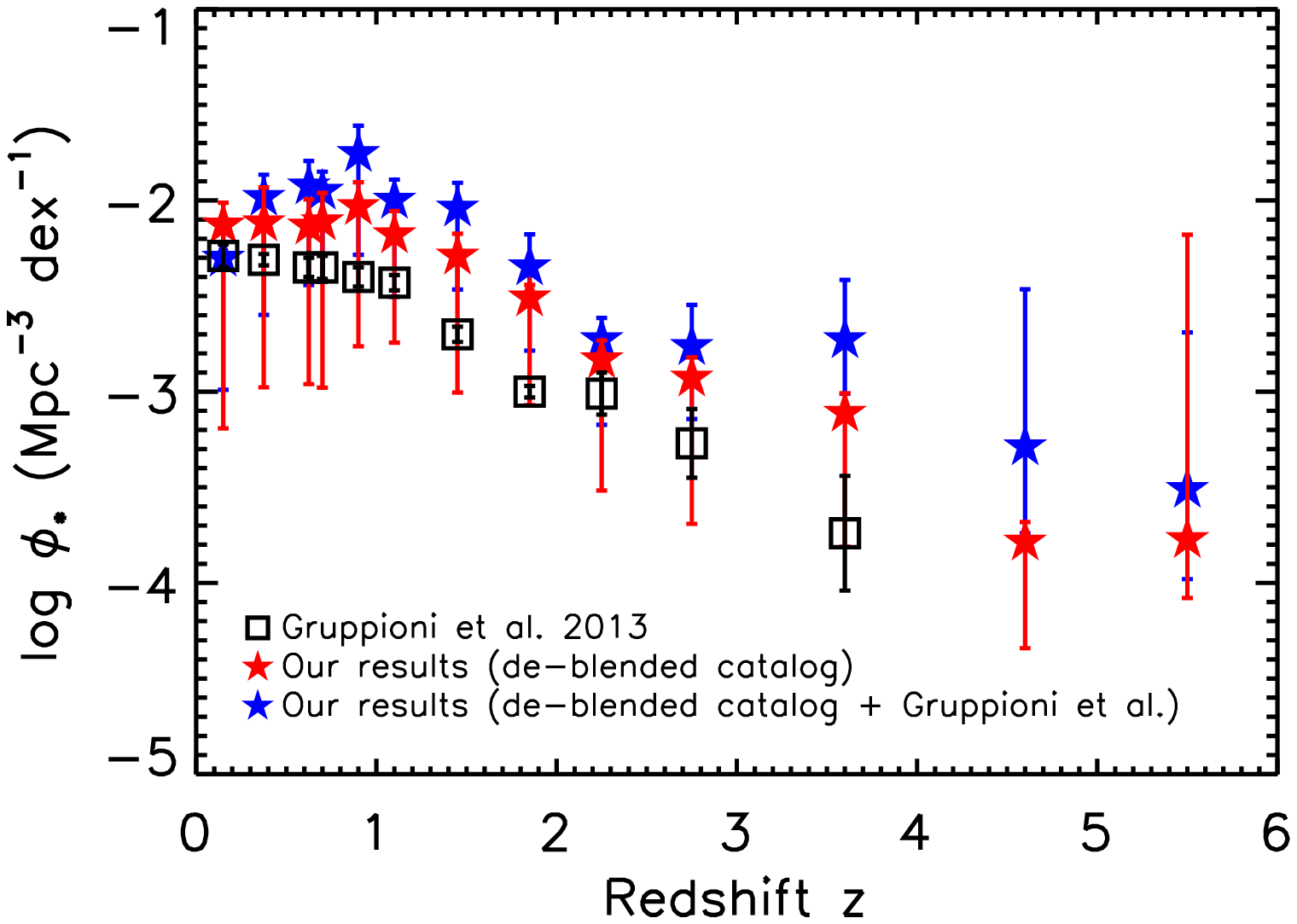}
\includegraphics[height=2.8in,width=3.6in]{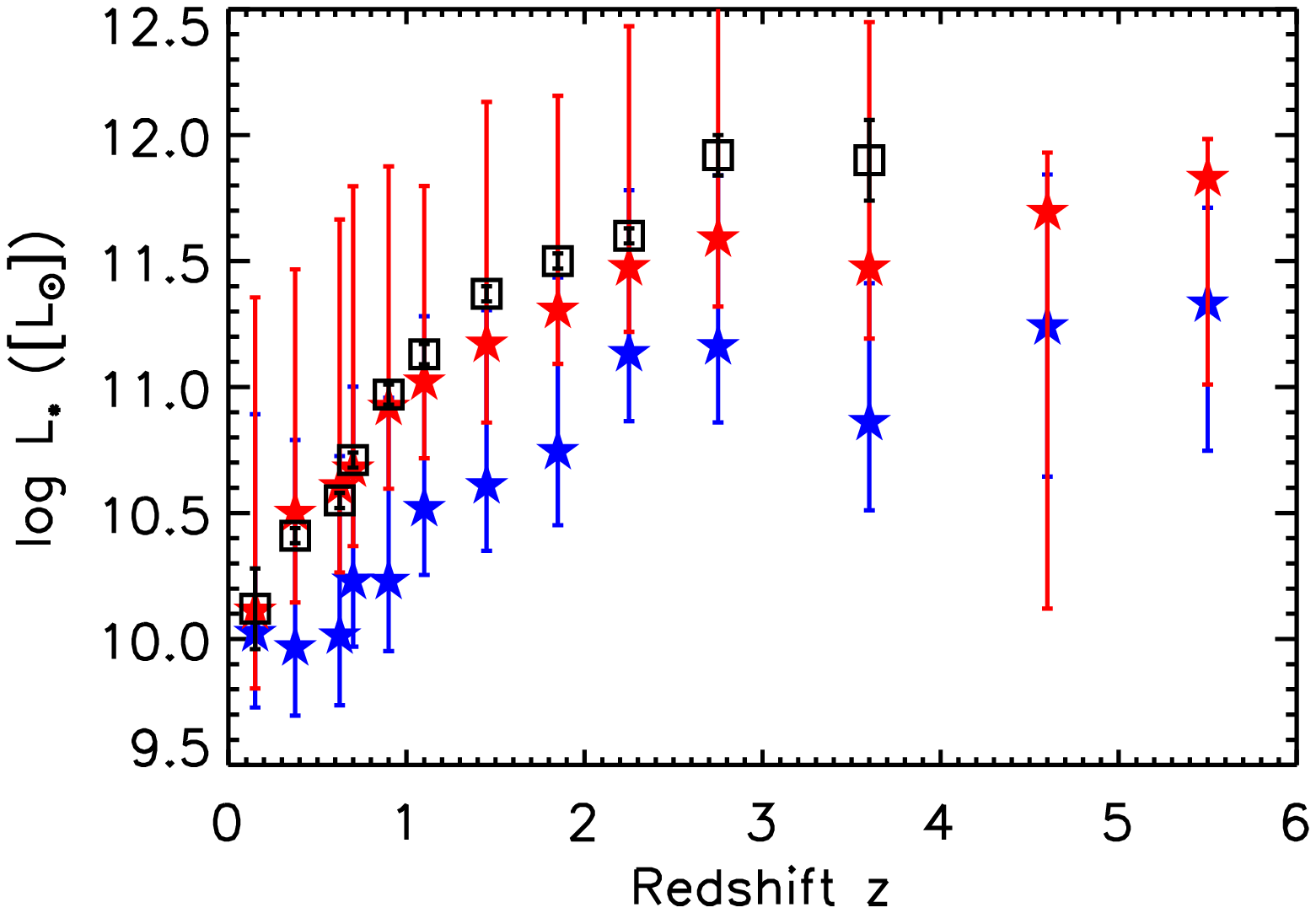}
\includegraphics[height=2.8in,width=3.6in]{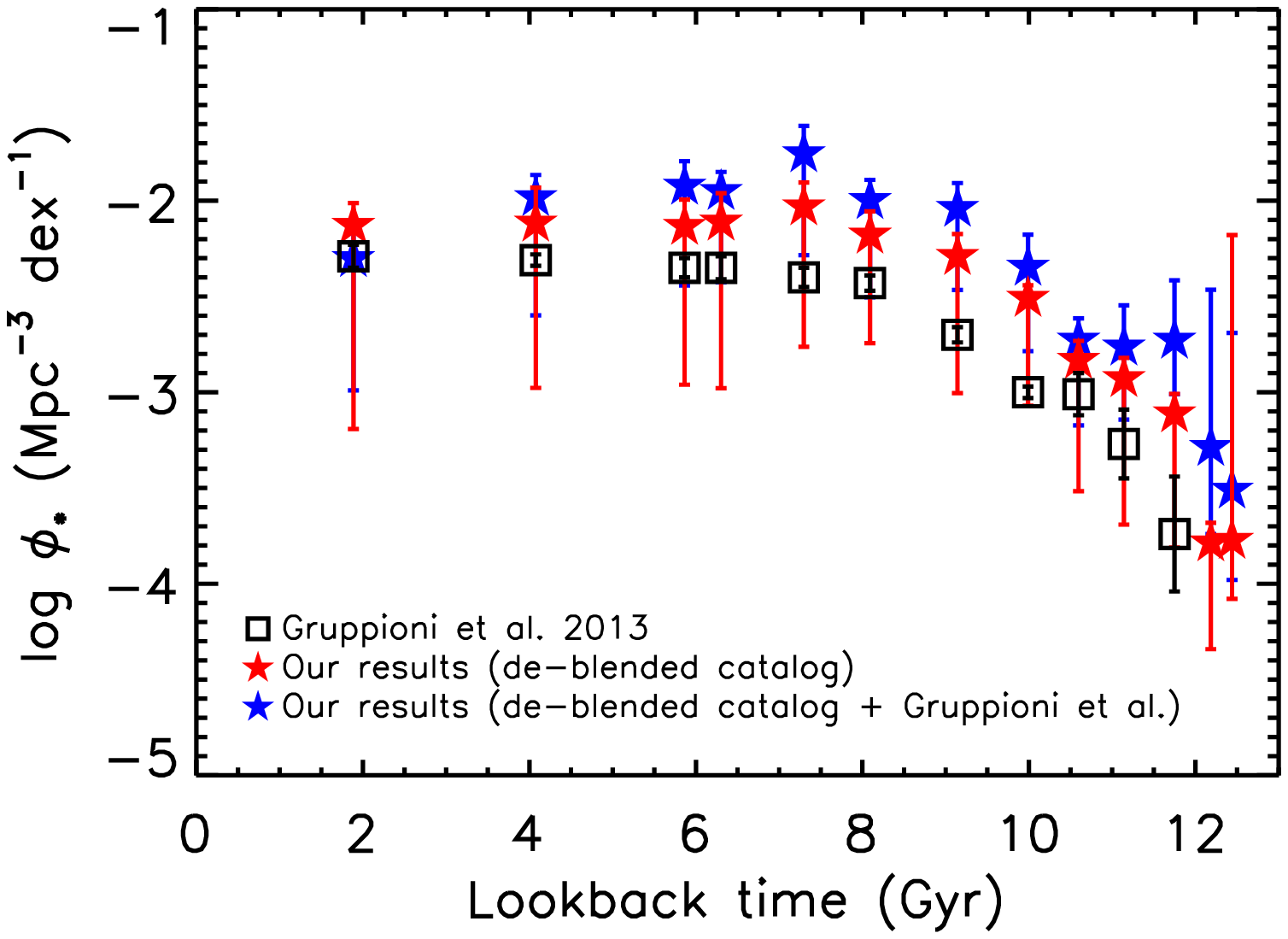}
\includegraphics[height=2.8in,width=3.6in]{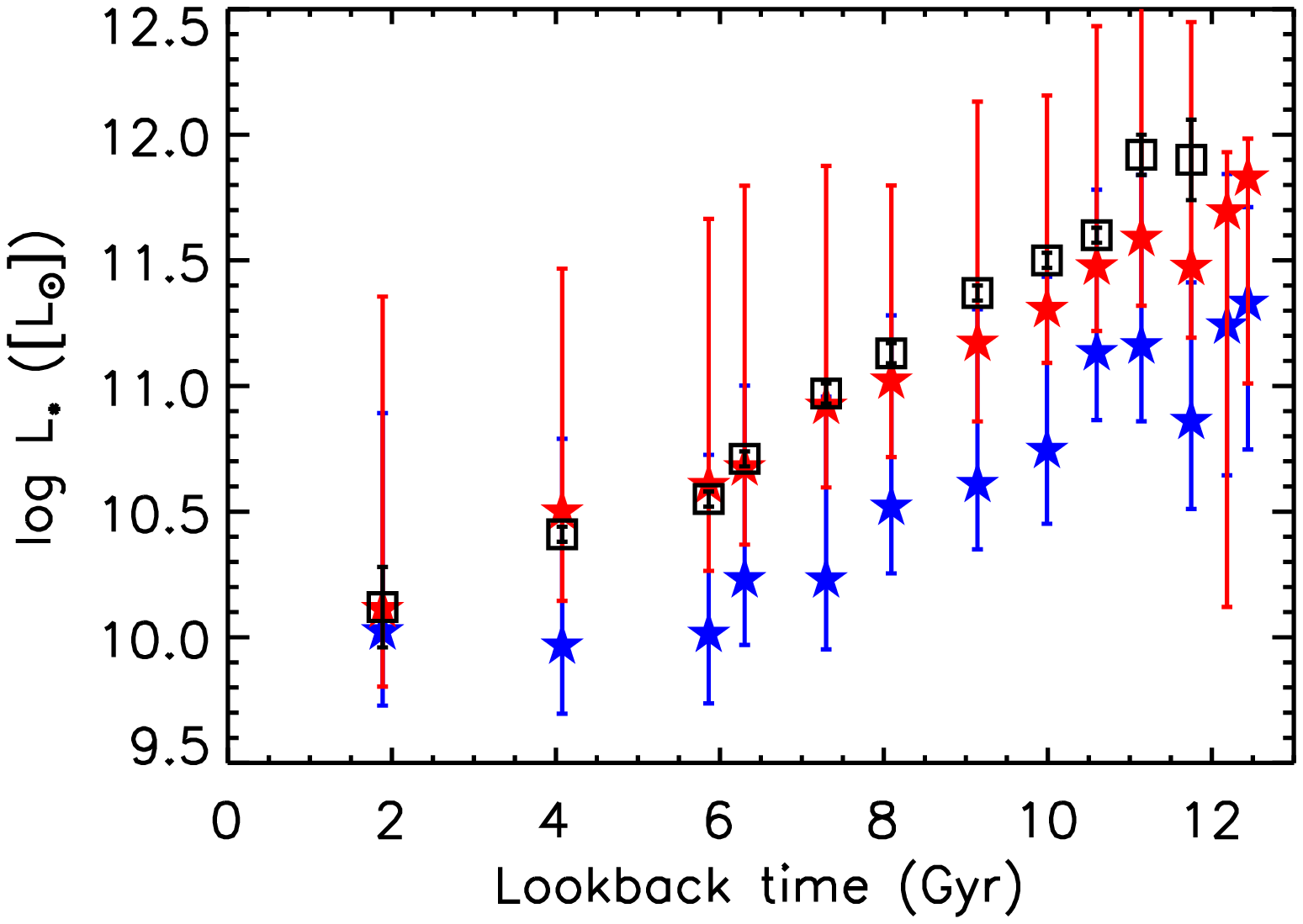}
\caption{Evolution of the characteristic luminosity $L_*$ and normalisation $\phi_*$ in the best-fitting modified Schechter function, i.e. Eq. (4). The red stars are derived from fitting to the measurements based on the de-blended catalogue in COSMOS only. The blue stars are derived from fitting to both the measurements from the de-blended catalogue in COSMOS and the measurements from Gruppioni et al. (2013). The black squares show the measurements taken from Gruppioni et al. (2013). The top panels show the evolution of $L_*$ and  $\phi_*$ as a function of redshift. The bottom panels show the evolution of $L_*$ and  $\phi_*$ as a function of lookback time.}
\label{zevol}
\end{figure*}

In Fig.~\ref{LF_mag}, we compare our total IR LFs with those in Magnelli et al. (2013) derived from the deepest {\it Herschel} PACS surveys at 70, 100 and 160 $\mu$m in the GOODS fields obtained by the PACS Evolutionary Probe (PEP; Lutz et al. 2011) and GOODS-Herschel (GOODS-H; Elbaz et al. 2011). Magnelli et al. (2013) used the positional information of Spitzer/IRAC 3.6 $\mu$m sources to extract sources from the Spitzer/MIPS 24 $\mu$m map, which were in turn used as positional priors for source extraction in the PACS maps. To obtain the required photometric redshift information, they cross-matched the IRAC-MIPS-PACS source catalogues with the shorter wavelength GOODS catalogues (optical + near-IR). LFs are then constructed using the $1/V_{max}$ method but limited to redshifts $z<2.3$, as the PACS data do not extend beyond 160 $\mu$m. Magnelli et al. (2013) fit the IR LFs with a double power-law function which is parameterised as follows,
\begin{equation}
\phi = \phi_{knee} L^{-0.6}, \textrm{when} \log(L/L_{\odot}) < L_{knee},
\end{equation}
and
\begin{equation}
\phi = \phi_{knee} L^{-2.2}, \textrm{when} \log(L/L_{\odot}) > L_{knee}. 
\end{equation}
The free parameters are the normalisation $\phi_{knee}$ and the transition luminosity $L_{knee}$. The fixed power-law slopes are taken from Sanders et al. (2003) which studied the IR LF of IR-bright galaxies selected from the IRAS all-sky survey in the local Universe. Our measurements of the total IR LFs agree quite well with the measurements from Magnelli et al. (2013) and their best-fitting double power-law functions out to redshifts $z<2.3$. In the highest redshift bin $1.8<z<2.3$, our measurements extend to fainter luminosities by almost one dex compared to Magnelli et al. (2013).

In Fig.~\ref{LF_gru}, we compare our LFs with the results from Gruppioni et al. (2013). Gruppioni et al. (2013) used the datasets (at 70, 100 and 160 $\mu$m) from the {\it Herschel} PEP Survey, in combination with the HerMES imaging data at 250, 350 and 500 $\mu$m, to derive the evolution of the rest-frame 35, 60, 90 $\mu$m and the total IR LFs up to $z\sim4$. The inclusion of the SPIRE imaging data allowed Gruppioni et al. (2013) to determine IR luminosities without large uncertainties due to extrapolations. Gruppioni et al. (2013) used a modified-Schechter function (e.g., Saunders et al. 1990; Wang \& Rowan-Robinson 2010; Marchetti et al. 2016; Wang et al. 2016) to fit the total IR LF,
 \begin{equation}
 \phi(L) = \phi_* \left(\frac{L}{L_*}\right)^{1-\alpha} \exp{\left[ -\frac{1}{2\sigma^2}\log^2_{10}\left(1+\frac{L}{L_*}\right)\right]} 
 \end{equation}
which behaves as a power law for $L<L_*$ and as a Gaussian in $\log L$ for $L>L_*$. In principle, there are four free parameters in the modified-Schechter function, i.e., $\alpha$ which describes the faint-end slope, $\sigma$ which controls the shape of the cut-off at the bright end, $L_*$ which is the characteristic luminosity, and $\phi_*$ which is the characteristic density. However, due to a lack of dynamic range, Gruppioni et al. (2013) adopted the faint-end slope value $\alpha=1.2$ and the Gaussian width parameter $\sigma=0.5$ derived in the first redshift bin ($0.0<z<0.3$) for all higher redshift bins. In other words, only $L_*$ and $\phi_*$ are allowed to vary freely in the higher redshift bins.

In general, there is a good agreement between our measurements based on the de-blended catalogue in the COSMOS field and measurements from Gruppioni et al. (2013) in the overlapping luminosity range. Additionally,  we are able to extend the LF measurements down to much fainter luminosities and out to higher redshifts. Our measurements of the LF also seem to suggest that there are fewer sources at the bright end which could be partially caused by the limited size of the COSMOS field. In Fig.~\ref{LF_gru}, we also show the best-fit modified-Schechter function to our measurements only (the red lines) and the best-fit function to both our total IR LFs and the measurements in Gruppioni et al. (2013) (the blue lines). During the fitting process using the MCMC sampler emcee (Foreman-Mackey et al. 2013), we assume both the characteristic luminosity $L_*$ and characteristic density $\phi_*$ evolve with redshift, but the bright-end Gaussian width $\sigma$ and  the faint-end slope $\alpha$ do not depend on redshift. As there are 13 redshift bins in Fig.~\ref{LF_gru}, in total we have 28 free parameters. Table 1 lists the best-fit values and marginalised errors of the parameters in the modified Schechter functions derived from fitting to our measurements only (based on the de-blended catalogue in COSMOS). Table 2 lists the best-fit values and marginalised errors of the parameters in the modified Schechter functions derived from fitting to both our measurements based on the de-blended catalogue and the measurements from Gruppioni et al. (2013) presented in Fig.~\ref{LF_gru}.

In Fig.~\ref{zevol}, we plot the evolution of the characteristic luminosity $L_*$ and normalisation $\phi_*$ in the best-fitting modified Schechter function as a function of redshift or lookback time. The red stars are derived from fitting to our measurements based on the de-blended catalogue in COSMOS only. The blue stars are derived from fitting to both the measurements from the de-blended catalogue in COSMOS and the measurements from Gruppioni et al. (2013). Regarding the evolution of the characteristic density $\phi_*$, the two sets of measurements are  consistent with each other within errors although the red stars (derived based on the de-blended catalogue in COSMOS only) are consistently below the blue stars. Similar to the conclusions reached in Gruppioni et al. (2013), we also find that the characteristic density evolves very mildly for the 8 Gyr or so ($z\sim1$) and then decreases rapidly (by about two orders of magnitude) from $z\sim1$ to $\sim6$. Regarding the evolution of the characteristic luminosity $L_*$, again the two sets of measurements are consistent with each other within errors. The red stars are consistently above the blue stars as a result of anti-correlation between $L_*$ and $\phi_*$. As a function of redshift, $L_*$ increases quickly with redshift out to $z\sim2$ and then seems to more or less flatten out to $z\sim6$. As a function of lookback time, $L_*$ seems to evolve in a simple linear fashion with time. The evolution of $L_*$ is qualitatively similar  to the evolution of the normalisation of the galaxy star-forming main sequence (e.g., Koprowski et al. 2016; Pearson et al. 2018).
                                 
\begin{table}
\caption{Best-fit values and marginalised errors of the parameters in the modified Schechter functions derived from fitting to our measurements only (based on the de-blended catalogue in COSMOS). The faint-end slope parameter $\alpha$ and the bright-end Gaussian width parameter $\sigma$ are assumed to be independent of redshift. The last column $N$ shows the number of galaxies above the completeness limit in the corresponding redshift bin.}\label{table:selection}
\begin{tabular}{llll}
\hline
Redshift range & $\phi_*$   & $L_*$    &$N$ \\
\hline
\vspace{0.15cm}
$0.0<z<0.3$ &   $-2.13^{+0.12}_{-1.06}$      &  $10.11^{+1.24}_{-0.31}$  &2932 \\
\vspace{0.15cm}
$0.3<z<0.45$ &   $-2.12^{+0.18}_{-0.86}$       &  $10.50^{+0.97}_{-0.36}$&7990\\
\vspace{0.15cm}
$0.45<z<0.6$ &  $-2.13^{+0.14}_{-0.83}$       &  $10.61^{+1.06}_{-0.34}$&9113\\
\vspace{0.15cm}
$0.6<z<0.8$ &   $-2.11^{+0.15}_{-0.87}$       &  $10.68^{+1.12}_{-0.31}$ &18660\\
\vspace{0.15cm}
$0.8<z<1.0$ &    $-2.03^{+0.13}_{-0.73}$       & $10.92^{+0.95}_{-0.33}$ &26469\\
\vspace{0.15cm}
$1.0<z<1.2$ &   $-2.18^{+0.12}_{-0.56}$        & $11.02^{+0.76}_{-0.30}$ &21127\\
\vspace{0.15cm}
$1.2<z<1.7$ &   $-2.29^{+0.12}_{-0.71}$       &  $11.17^{+0.96}_{-0.31}$&36380\\
\vspace{0.15cm}
$1.7<z<2.0$ &    $-2.51^{+0.10}_{-0.56}$      &  $11.31^{+0.85}_{- 0.21}$&15491\\
\vspace{0.15cm}
$2.0<z<2.5$ &    $-2.83^{+0.10}_{-0.69}$       &  $11.48^{+0.96}_{-0.26}$&13429\\
\vspace{0.15cm}
$2.5<z<3.0$ &    $-2.93^{+0.11}_{-0.76}$       &  $11.59^{+1.02}_{-0.27}$&9323\\
\vspace{0.15cm}
$3.0<z<4.2$ &   $-3.11^{+0.10}_{-0.70}$        &  $11.47^{+0.97}_{-0.28}$&5037\\
\vspace{0.15cm}
$4.2<z<5.0$ &    $-3.79^{+0.10}_{-0.56}$       &  $11.70^{+0.24}_{-1.57}$&1369\\
\vspace{0.15cm}
$5.0<z<6.0$ &    $-3.77^{+1.59}_{-0.31}$       & $11.83^{+0.16}_{-0.82}$ &599\\
\hline
\vspace{0.25cm}
$\alpha =1.26^{+0.36}_{-0.25} $ &&\\
\vspace{0.15cm}
$\sigma =0.44^{+0.06}_{-0.27} $ &&\\
\hline
\end{tabular}
\end{table}

\begin{table}
\caption{Best-fit values and marginalised errors of the parameters in the modified Schechter functions derived from fitting to both our measurements and the measurements from Gruppioni et al. (2013). The faint-end slope parameter $\alpha$ and the bright-end Gaussian width parameter $\sigma$ are assumed to be independent of redshift.}\label{table:selection}
\begin{tabular}{lll}
\hline
Redshift range & $\phi_*$   & $L_*$     \\
\hline
\vspace{0.15cm}
$0.0<z<0.3$ &   $-2.30^{+0.14}_{-0.69}$      &  $10.02^{+0.87}_{-0.29}$\\
\vspace{0.15cm}
$0.3<z<0.45$ &   $-1.98^{+0.11}_{-0.62}$       &  $9.97^{+0.82}_{-0.27}$\\
\vspace{0.15cm}
$0.45<z<0.6$ &  $-1.92^{+0.13}_{-0.52}$       &  $10.01^{+0.71}_{-0.27}$\\
\vspace{0.15cm}
$0.6<z<0.8$ &   $-1.95^{+0.10}_{-0.48}$       &  $10.23^{+0.77}_{-0.26}$\\
\vspace{0.15cm}
$0.8<z<1.0$ &    $-1.75^{+0.14}_{-0.53}$       & $10.23^{+0.73}_{-0.28}$ \\
\vspace{0.15cm}
$1.0<z<1.2$ &   $-2.00^{+0.11}_{-0.50}$        & $10.52^{+0.76}_{-0.27}$ \\
\vspace{0.15cm}
$1.2<z<1.7$ &   $-2.04 ^{+0.13}_{-0.43}$       &  $10.61^{+0.70}_{-0.26}$\\
\vspace{0.15cm}
$1.7<z<2.0$ &    $-2.35^{+0.17}_{-0.44}$      &  $10.75^{+0.69}_{-0.30}$\\
\vspace{0.15cm}
$2.0<z<2.5$ &    $-2.73^{+0.12}_{-0.44}$       &  $11.13^{+0.65}_{-0.27}$\\
\vspace{0.15cm}
$2.5<z<3.0$ &    $-2.76^{+0.21}_{-0.38}$       &  $11.16^{+0.68}_{-0.30}$\\
\vspace{0.15cm}
$3.0<z<4.2$ &   $-2.73^{+0.31}_{-0.28}$        &  $10.86^{+0.55}_{-0.35}$\\
\vspace{0.15cm}
$4.2<z<5.0$ &    $-3.29^{+0.83}_{-0.45}$       &  $11.24^{+0.60}_{-0.60}$\\
\vspace{0.15cm}
$5.0<z<6.0$ &    $-3.51^{+0.82}_{-0.47}$       & $11.33^{+0.38}_{-0.58}$ \\
\hline
\vspace{0.25cm}
$\alpha =1.28^{+0.39}_{-0.20} $ &&\\
\vspace{0.15cm}
$\sigma =0.65^{+0.04}_{-0.12} $ &&\\
\hline
\end{tabular}
\end{table}

\begin{figure*}
\centering
\includegraphics[height=1.8in,width=2.3in]{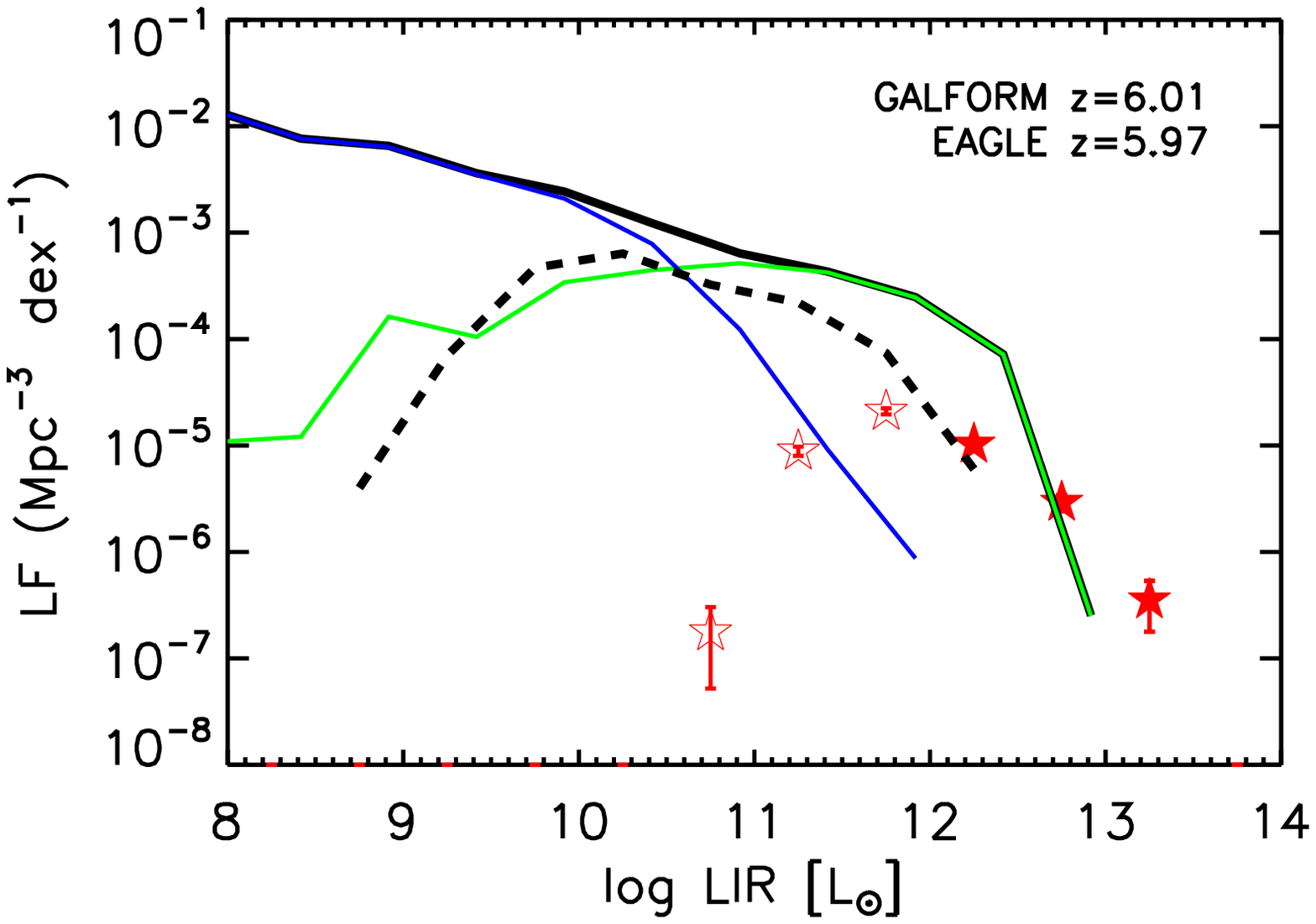}
\includegraphics[height=1.8in,width=2.3in]{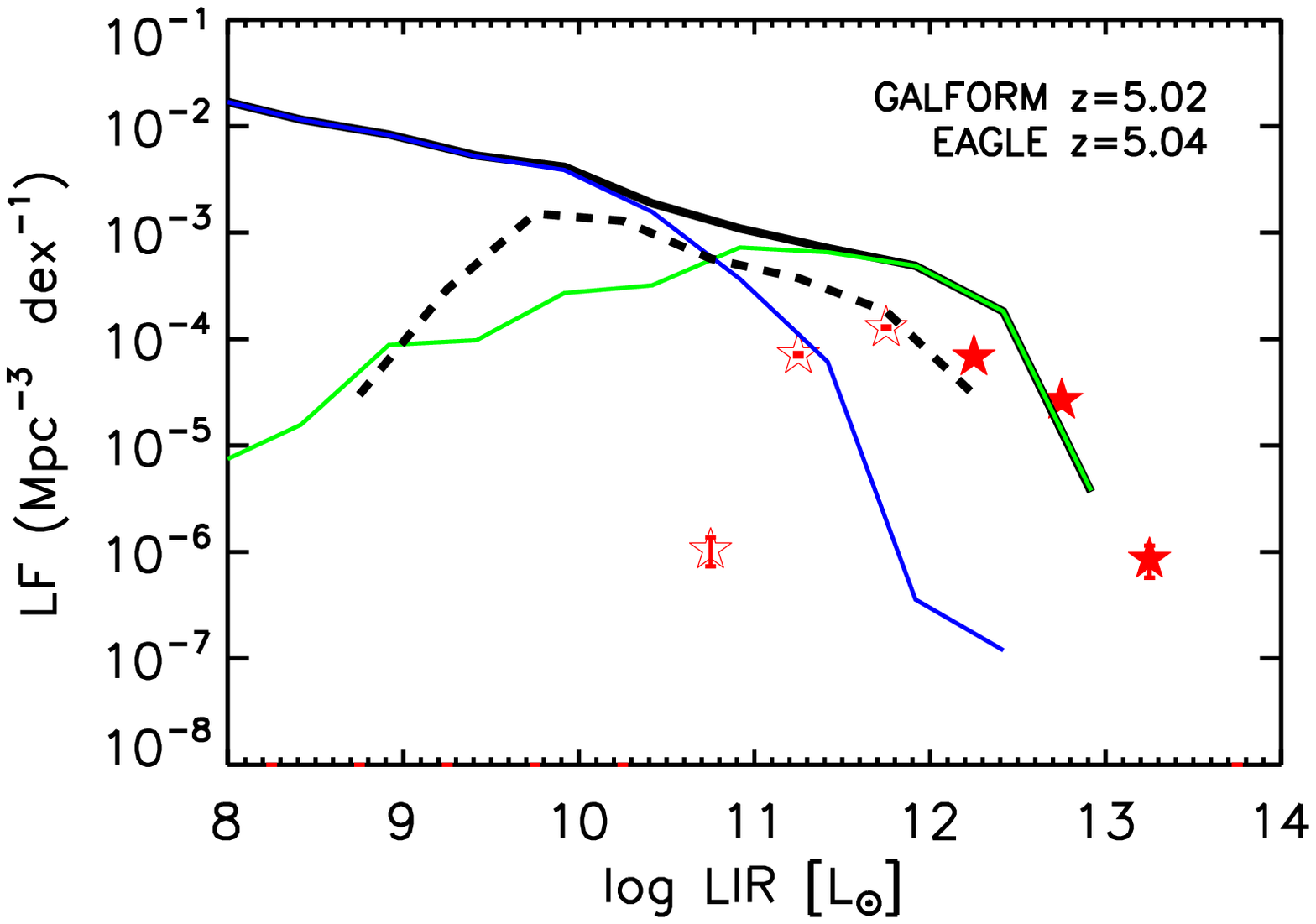}
\includegraphics[height=1.8in,width=2.3in]{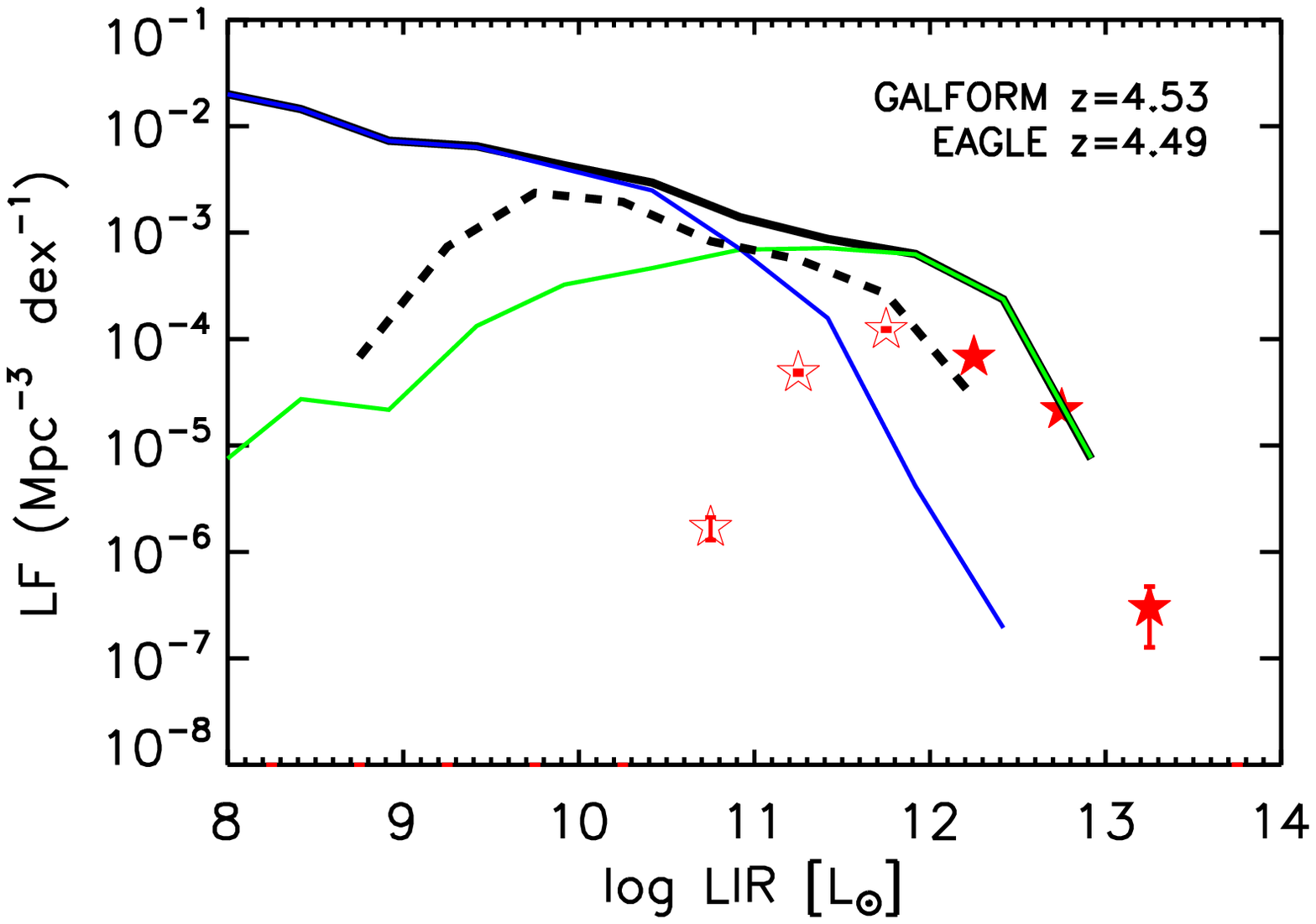}
\includegraphics[height=1.8in,width=2.3in]{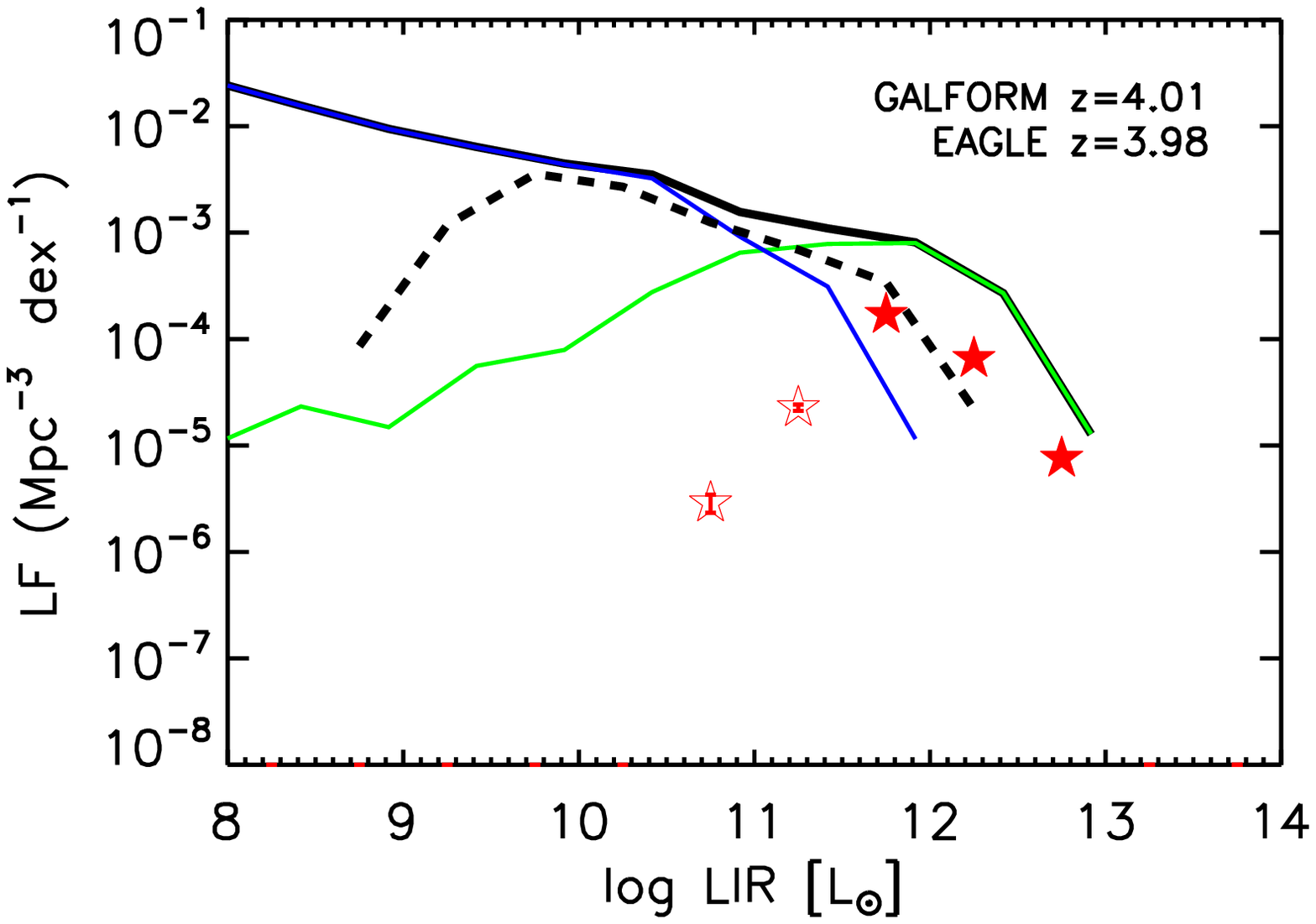}
\includegraphics[height=1.8in,width=2.3in]{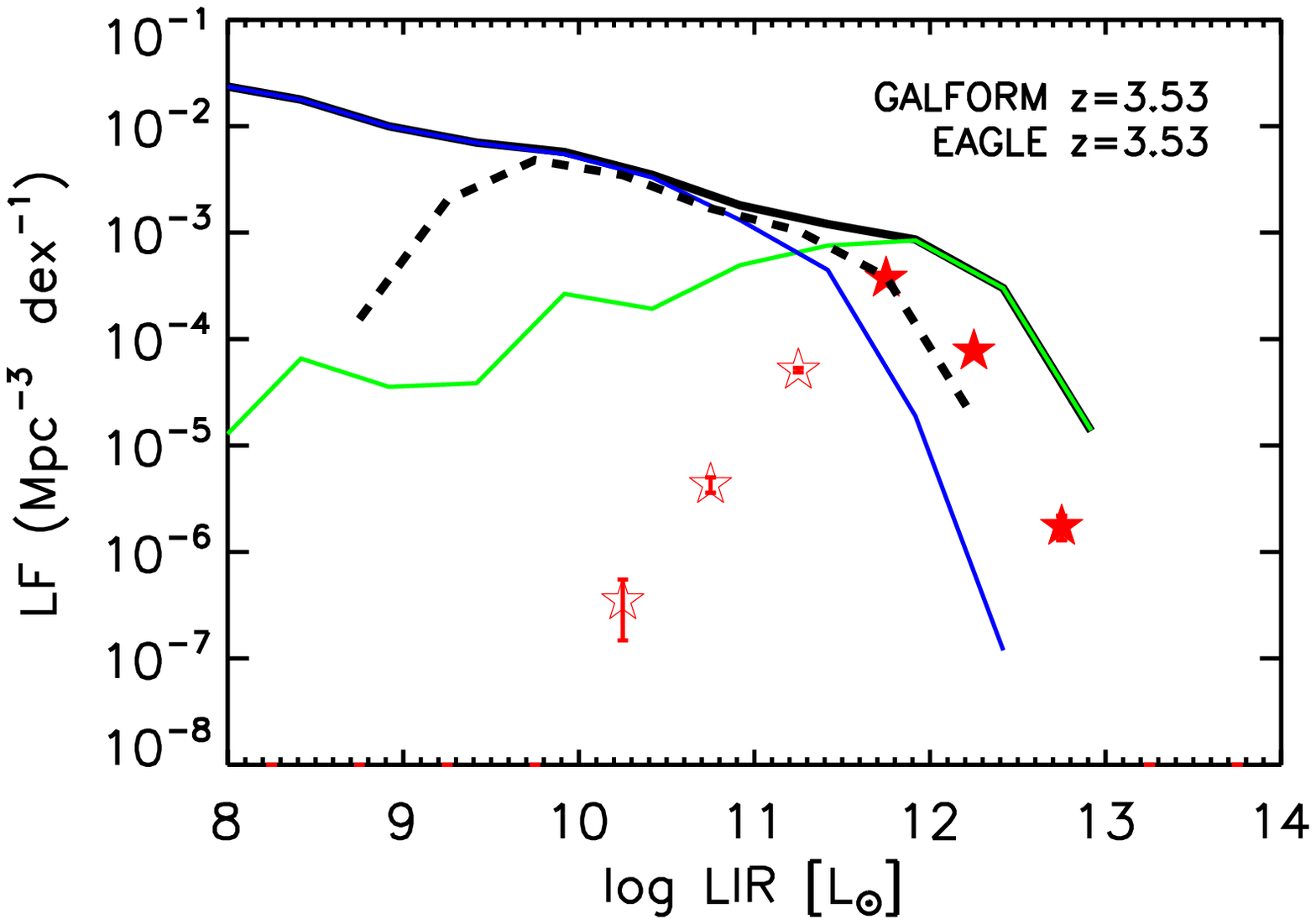}
\includegraphics[height=1.8in,width=2.3in]{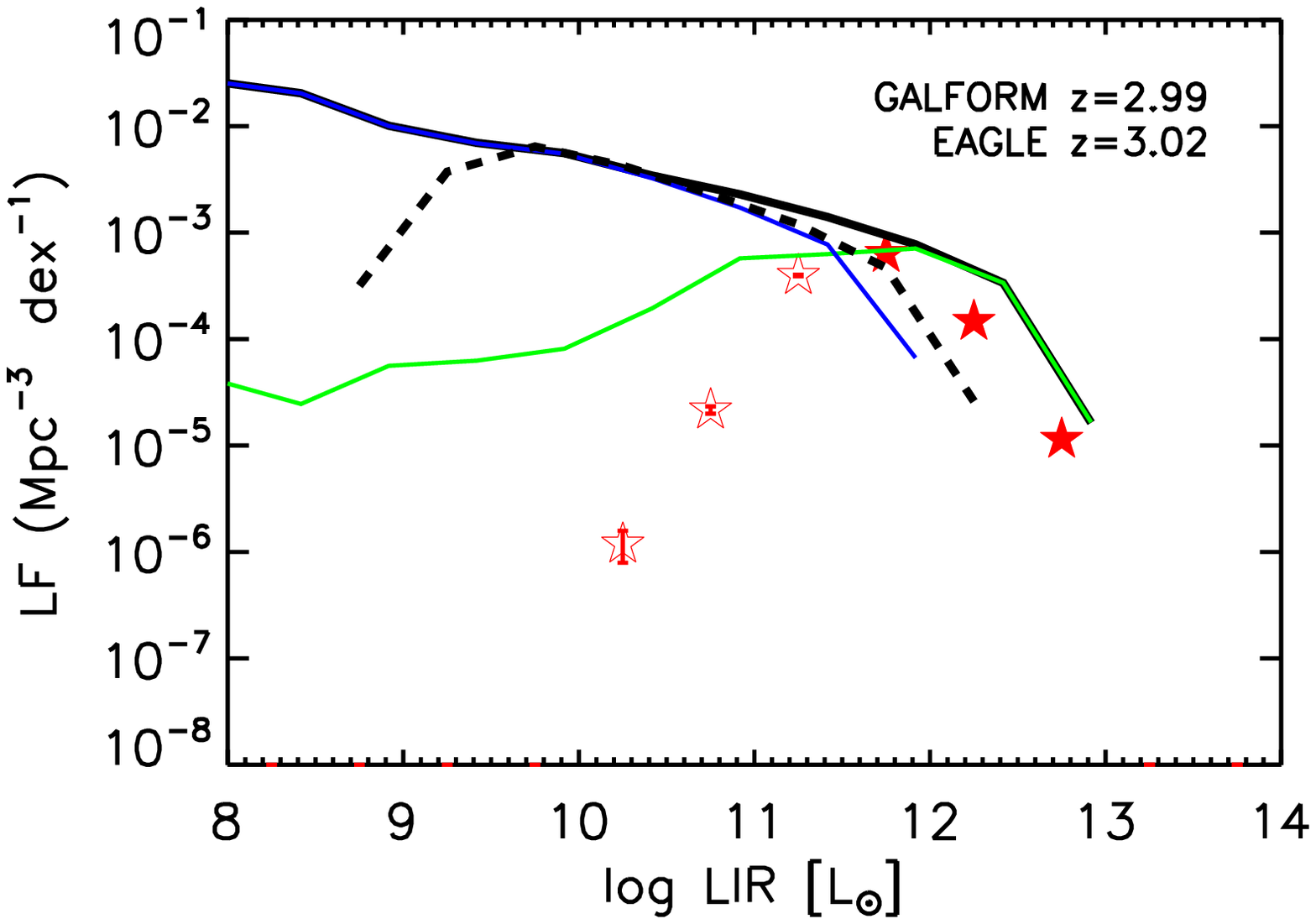}
\includegraphics[height=1.8in,width=2.3in]{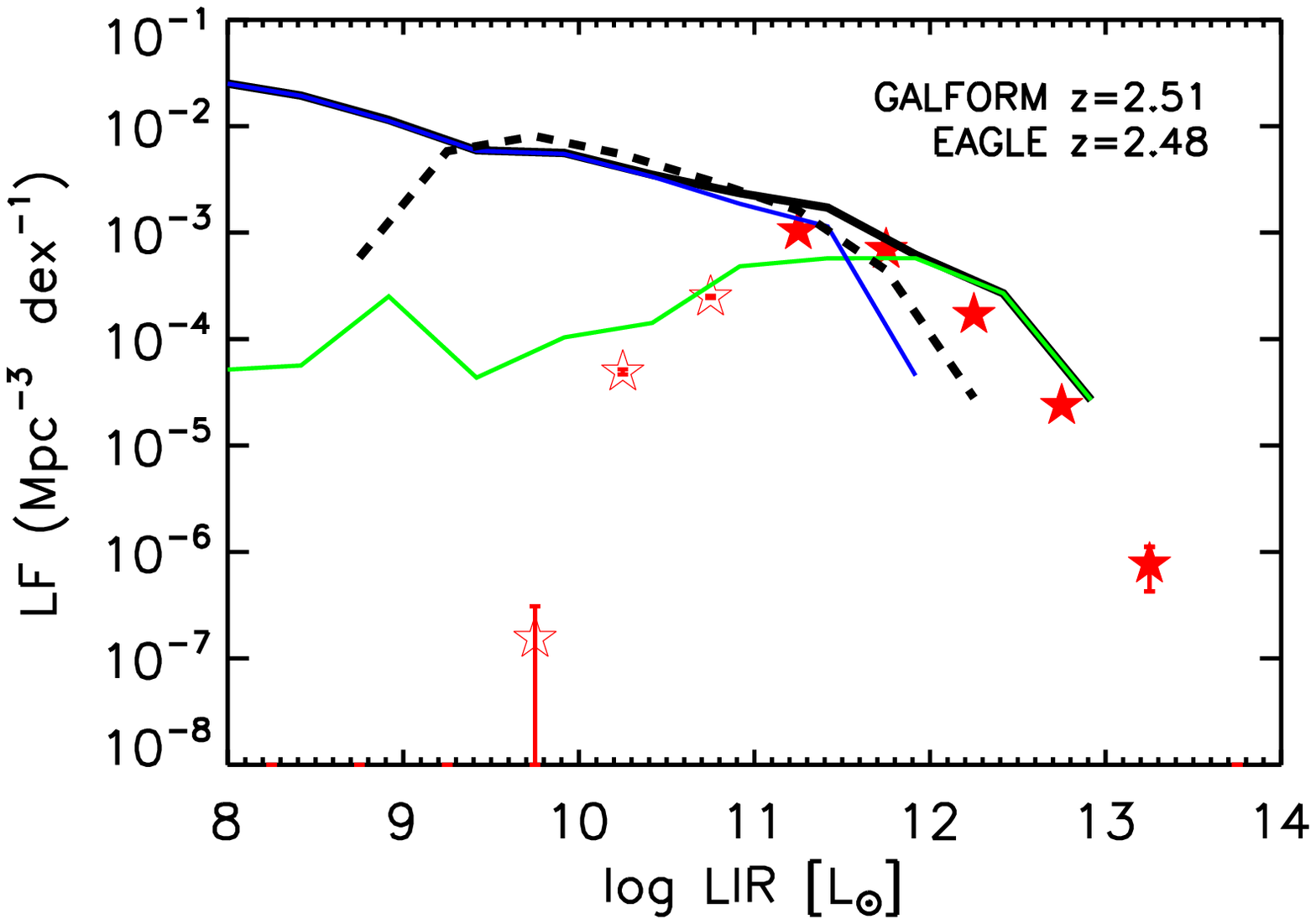}
\includegraphics[height=1.8in,width=2.3in]{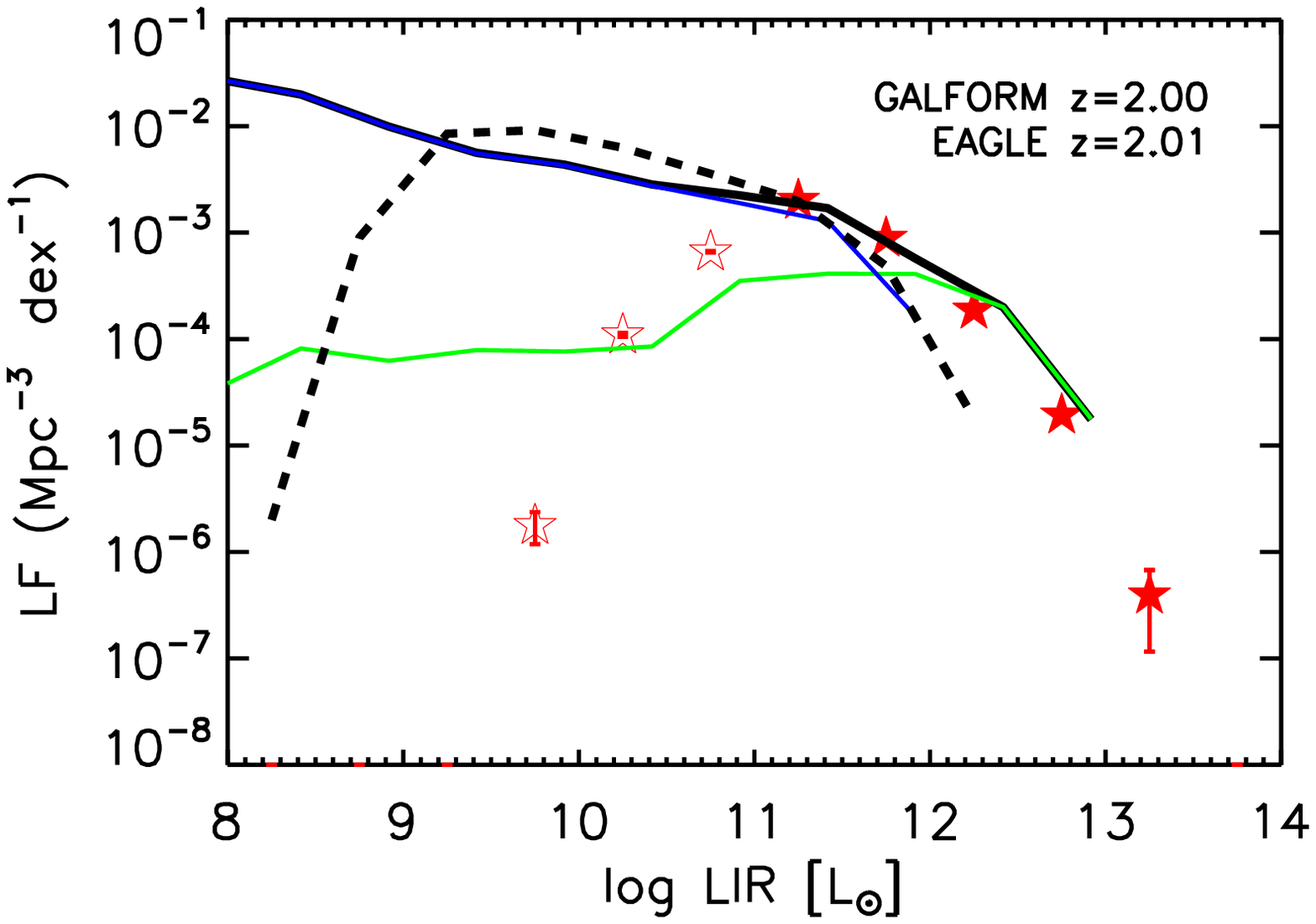}
\includegraphics[height=1.8in,width=2.3in]{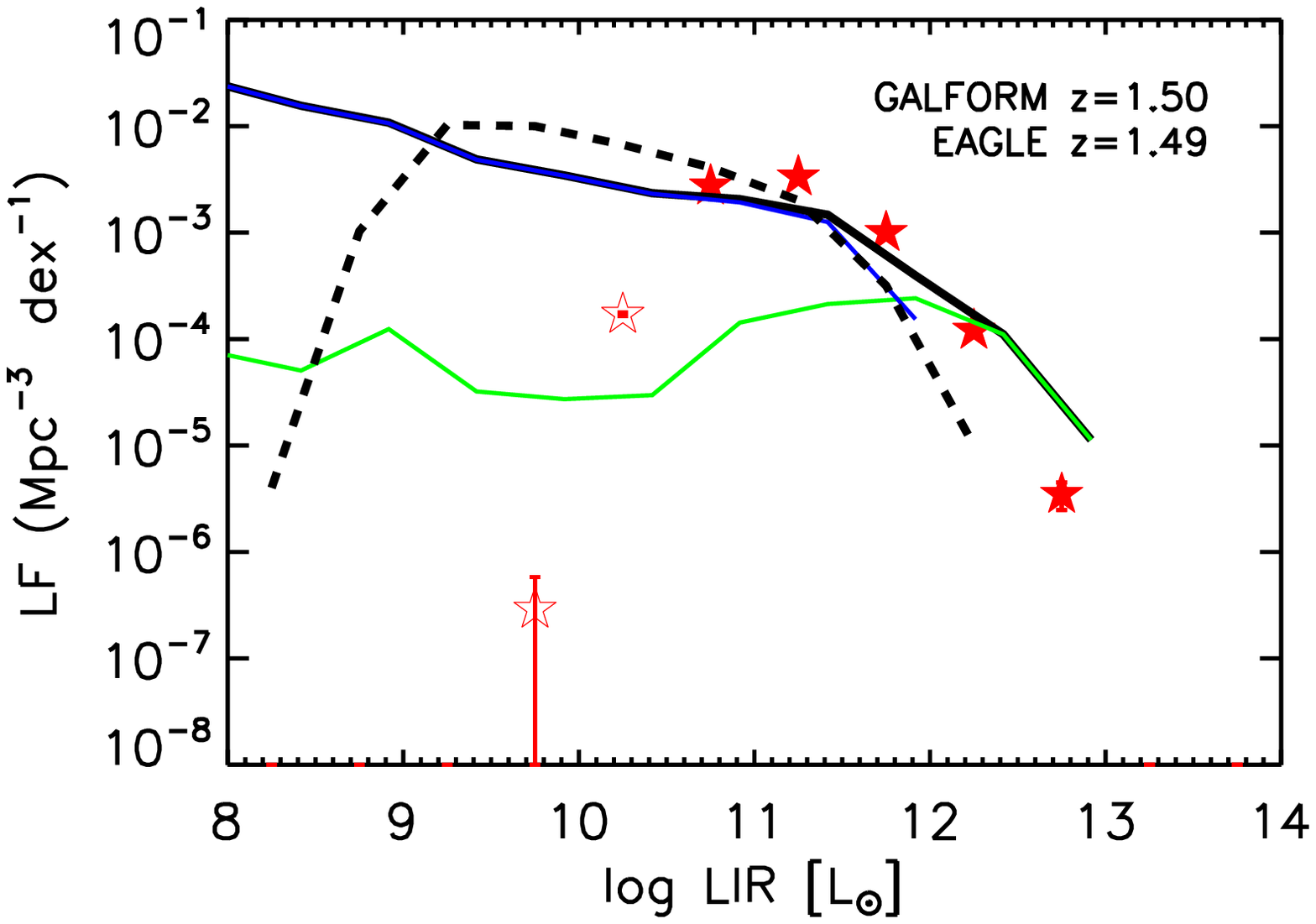}
\includegraphics[height=1.8in,width=2.3in]{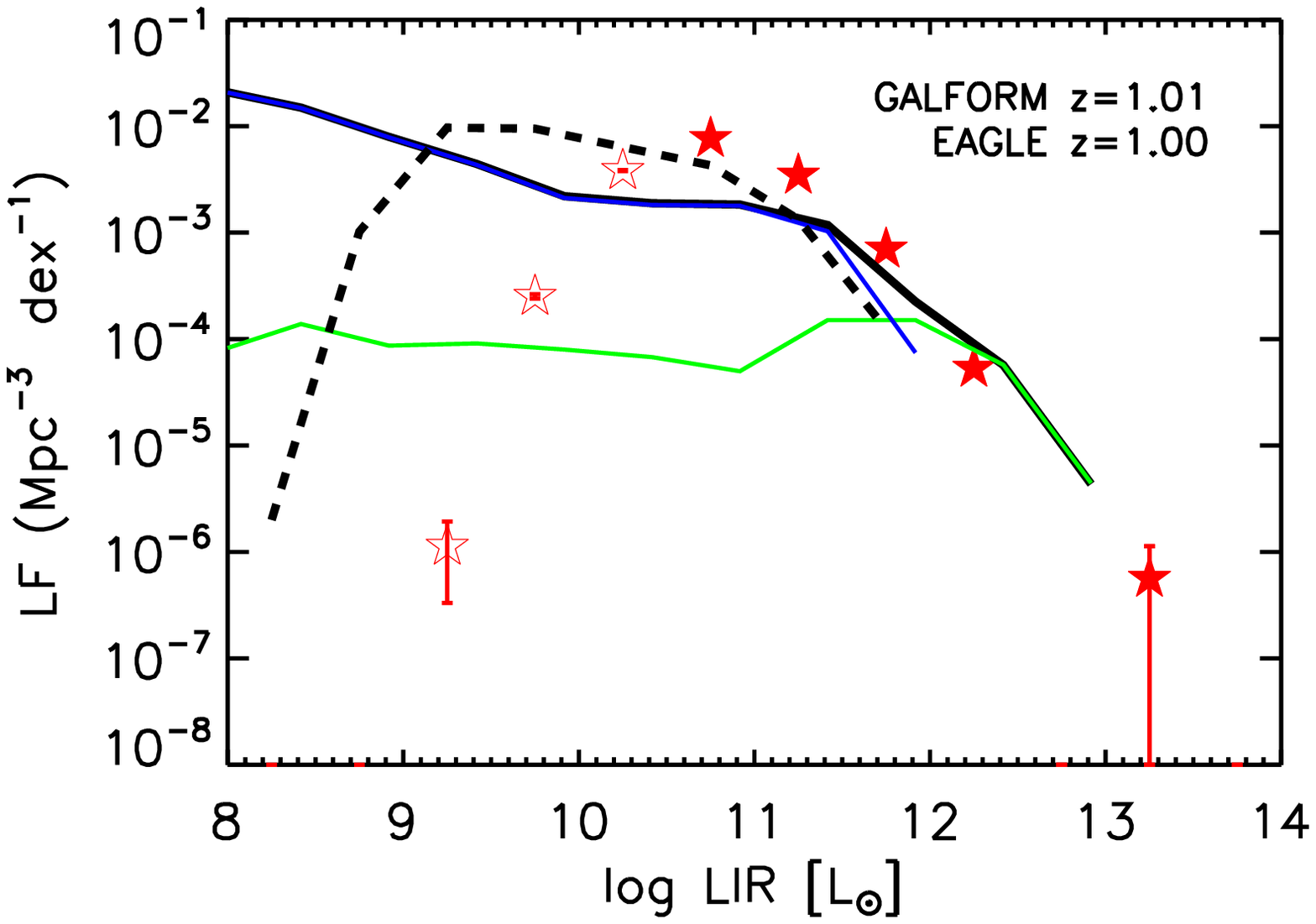}
\includegraphics[height=1.8in,width=2.3in]{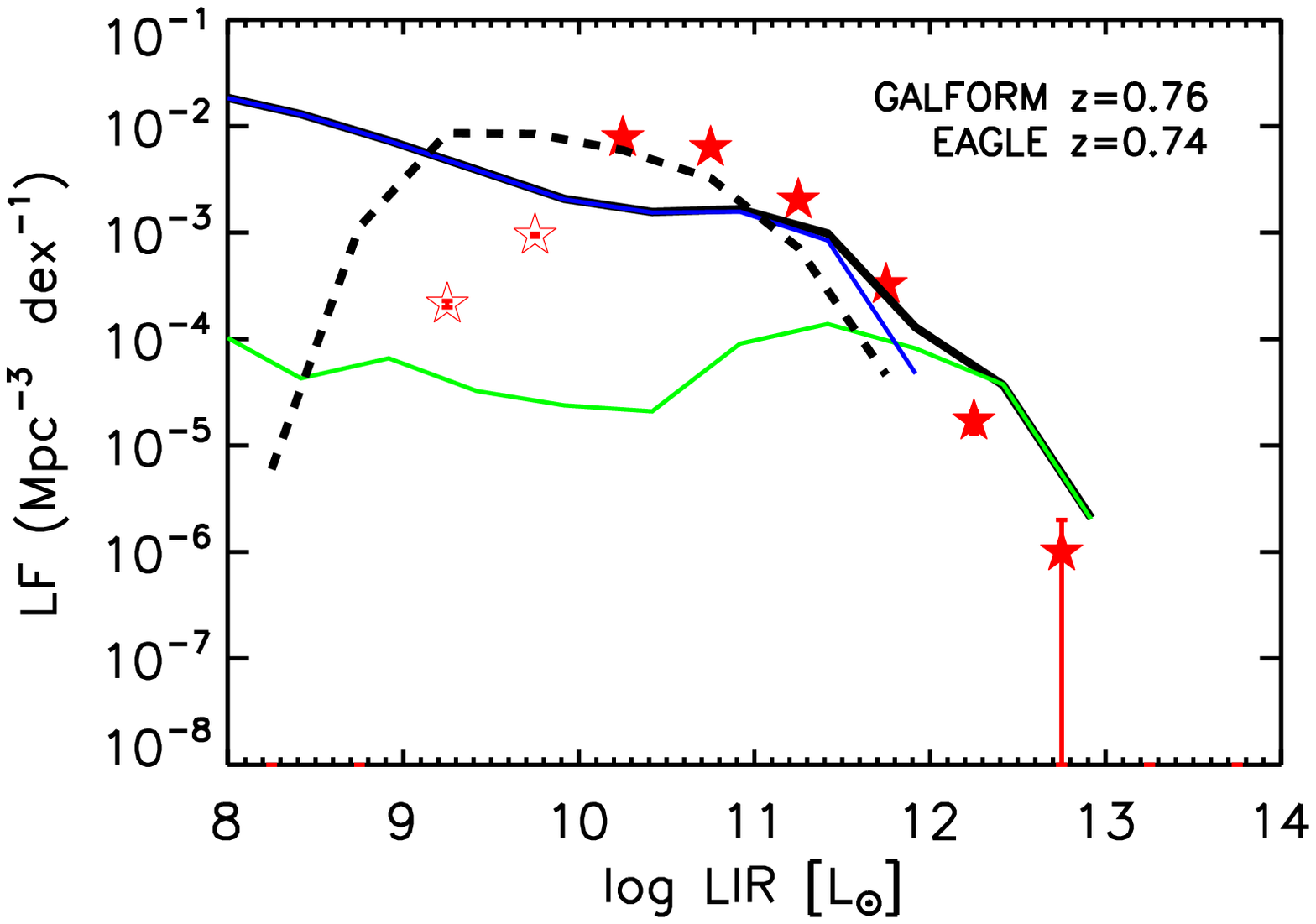}
\includegraphics[height=1.8in,width=2.3in]{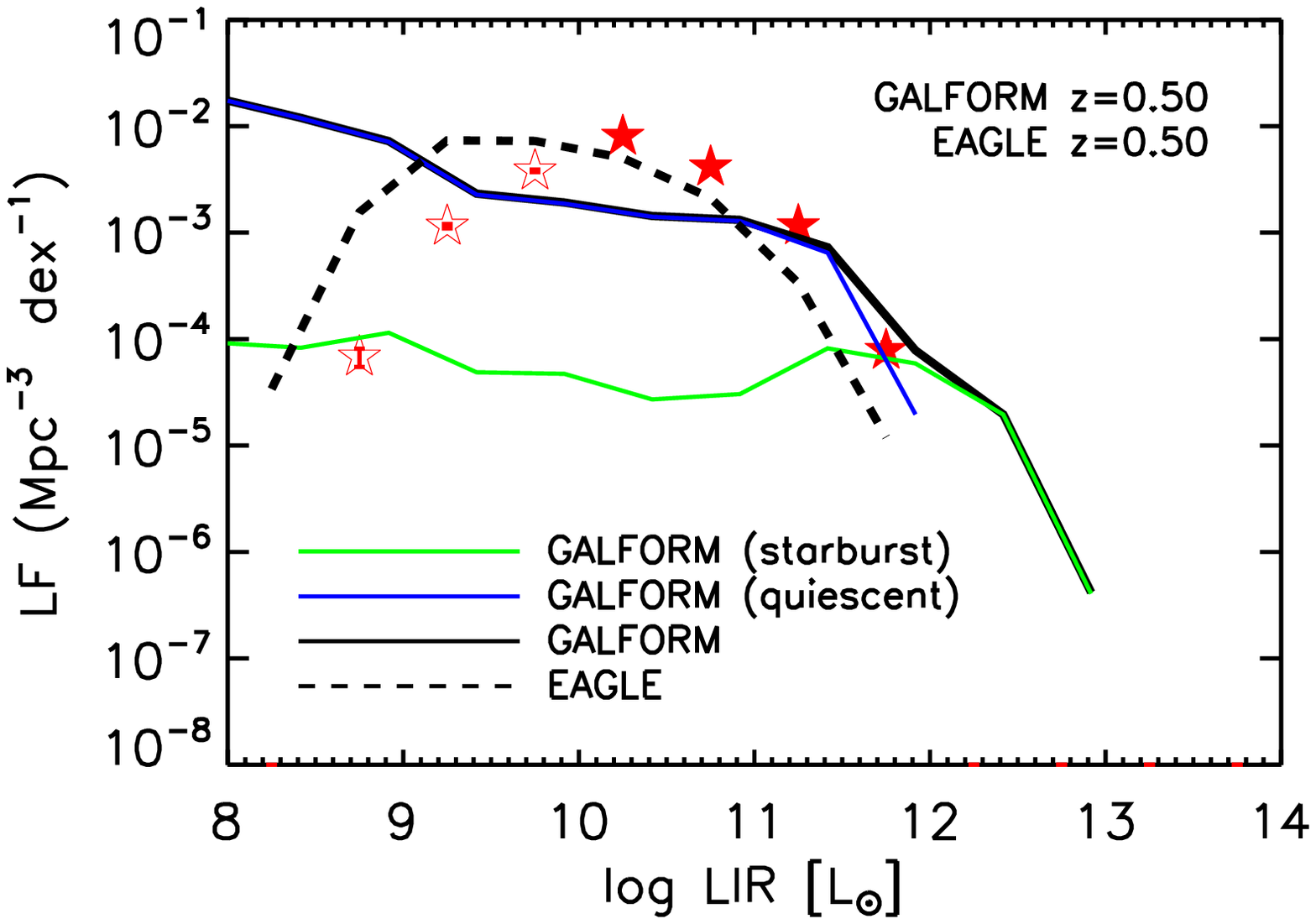}
\caption{The total IR luminosity function. The red stars are derived from our de-blended catalogue in COSMOS (filled red stars: our LF above the completeness limit; empty red stars: our LF below the completeness limit). Error bars on the red stars only represent Poisson errors. The thick black solid lines are from the Durham semi-analytic model. The thin blue solid lines correspond to the LFs of the quiescent galaxies from the Durham SAM. The thin green solid lines correspond to the LFs of the starburst galaxies from the Durham SAM. The black dashed lines are from the EAGLE hydrodynamic simulation.}
\label{LF_dur}
\end{figure*}

In Fig.~\ref{LF_dur}, we compare our total IR LF with predictions from GALFORM out to $z=6$. There is a lack of very bright sources with $L_{IR}>10^{13} L_{\odot}$ which is caused by the limited volume of the simulation (500 Mpc on a side). We further decompose the predicted IR luminosity function from GALFORM into starburst and quiescent populations. The transition between  starburst and quiescent  happens at around $10^{12}L_{\odot}$ at low redshifts and decreases to around $10^{11}L_{\odot}$ towards high redshift. The overall agreement between our measurements and the GALFORM predictions at the bright end ($L_{IR}>10^{11} L_{\odot}$) is reasonably good, especially at $z<2.5$.  It is clear that in order to match the observations at the bright end, the population of starburst galaxies in the simulation is of great importance. On the other hand, the population of quiescent galaxies in the simulation is important in matching the observations at the faint end. However, at $z<1$ GALFORM predictions at the faint end where the quiescent population dominates over the starburst population are much lower compared to our measurements.  At higher redshifts $z>2.5$, GALFORM seems to over-predict the number of bright dusty star-forming galaxies compared to our measurements and this over-prediction generally becomes worse towards higher redshift. Further studies are needed to understand the cause of this prediction, e.g., over production of starburst galaxies or something to do with the top-heavy IMF. Fig.~\ref{LF_dur} demonstrates that it is also informative to compare predictions of the luminosity function as a function of redshift than just the number counts.

In Fig.~\ref{LF_dur}, we also compare our total IR LF with predictions from the EAGLE hydrodynamic simulation out to $z=6$. There is a lack of sources with $L_{IR}>10^{12} L_{\odot}$ which is caused by the limited volume of the EAGLE simulation (100 Mpc on a side). The drop at the faint end is due to the criteria that only galaxies with stellar masses in excess of $10^{8.5} {\rm M_\odot}$ and with dust distributions resolved by $\geq 250$ particles are include in the EAGLE catalogue. In general, the total IR LF predicted from the EAGLE hydrodynamic simulation is lower at the bright end compared to our measurements, which reflects the under-prediction seen in the number counts plots (from Fig. 2 to Fig. 4). This could be caused by a lack of starburst galaxies, the poor sampling of the largest halos in the 100 Mpc EAGLE box, the need of adopting a different IMF (e.g., a top-heavy IMF) for the starburst population, or the need of changing the subgrid physical prescriptions  of the simulation related to star formation, feedback, etc. Interestingly, the total IR LF predicted from EAGLE agrees better with the observations than the predictions from GALFORM in the intermediate luminosity range (between $L_{IR}\sim10^{10} L_{\odot}$ and $L_{IR}\sim10^{11} L_{\odot}$) at $z<1$. At $z>2.5$ where GALFORM over-predicts the IR LF, EAGLE predictions agree reasonably well with the observations in the overlapping dynamic range.

\subsection{The cosmic star-formation history}

Over the past two decades, impressive progress has been made in charting star formation from the local Universe to the epoch of re-ionisation (e.g., Hopkins \& Beacom 2006;  Madau \& Dickinson 2014; Gruppioni et al. 2013; Bouwens et al. 2015), utilising a multitude of SFR tracers. It is a remarkable achievement that there is a reasonable consensus regarding the recent history below $z\sim2$. However, above $z\sim3$, major differences over one order of magnitude still exist. This order of magnitude difference encompasses multiple very different predictions from competing galaxy evolution models (e.g., Gruppioni et al. 2015; Henriques et al. 2015; Lacey et al. 2016), so it is vital to measure the CSFH with much greater precision and accuracy. The cosmic epoch over $3<z<4$ is also very interesting with some studies suggesting that the balance of power may shift from unobscured star formation to dusty star formation (Koprowski et al. 2017; Dunlop et al. 2017; Bourne et al. 2017). The most direct SFR tracer measures UV light which is redshifted to optical and IR for distant galaxies. As a result, very sensitive instruments (such as the Hubble Space Telescope) can be used to probe SFR density (SFRD) in these early cosmic epochs (McLure et al. 2013; Bouwens et al. 2015, 2016; Finkelstein et al. 2015; Parsa et al. 2016). However, large and uncertain dust extinction correction needs to be applied to these UV-only observations. To directly probe the dust obscured star formation, we need far-IR and sub-mm SFR tracers.

\begin{figure*}
\centering
\includegraphics[height=4.in,width=5.6in]{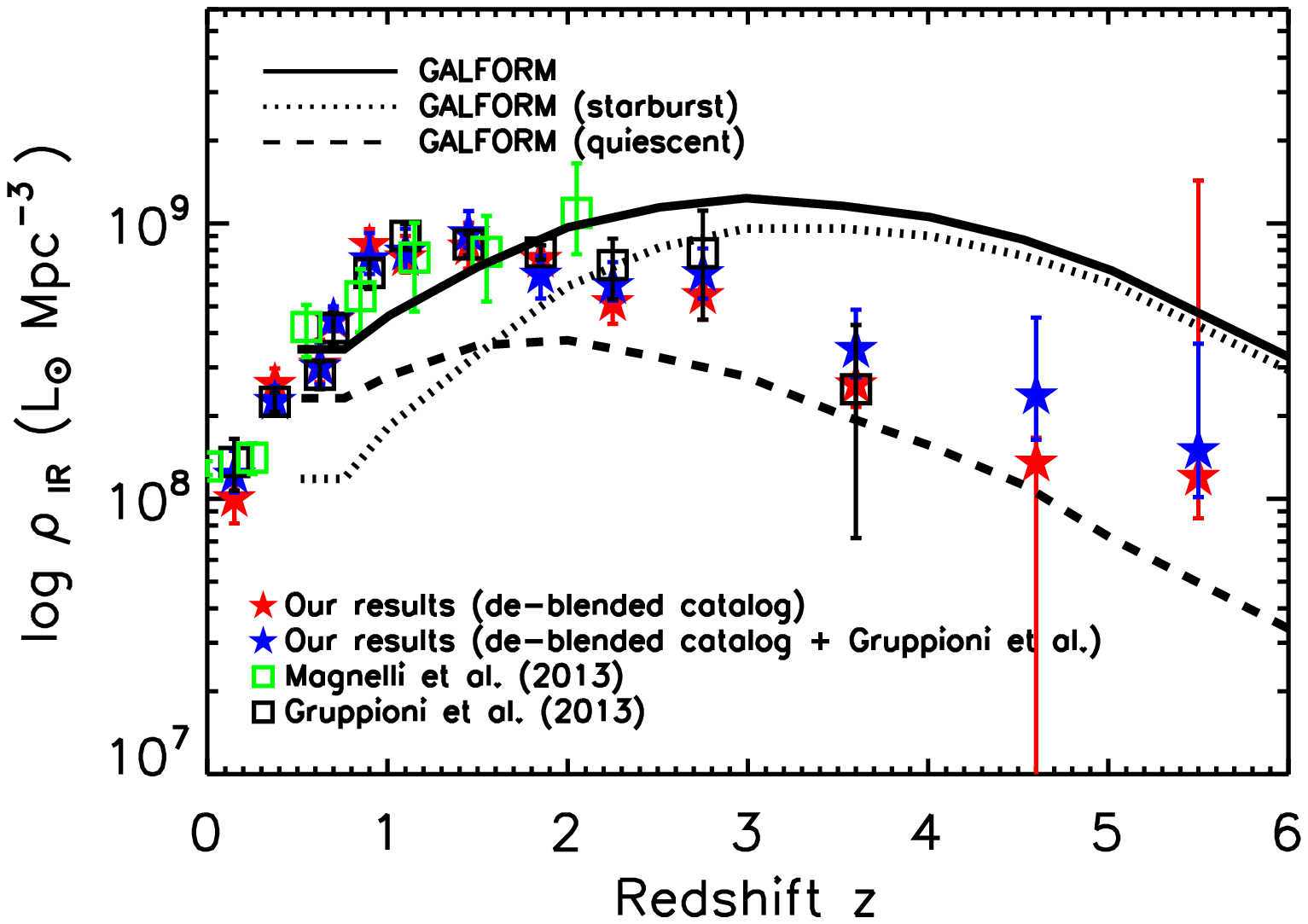}
\includegraphics[height=4.in,width=5.6in]{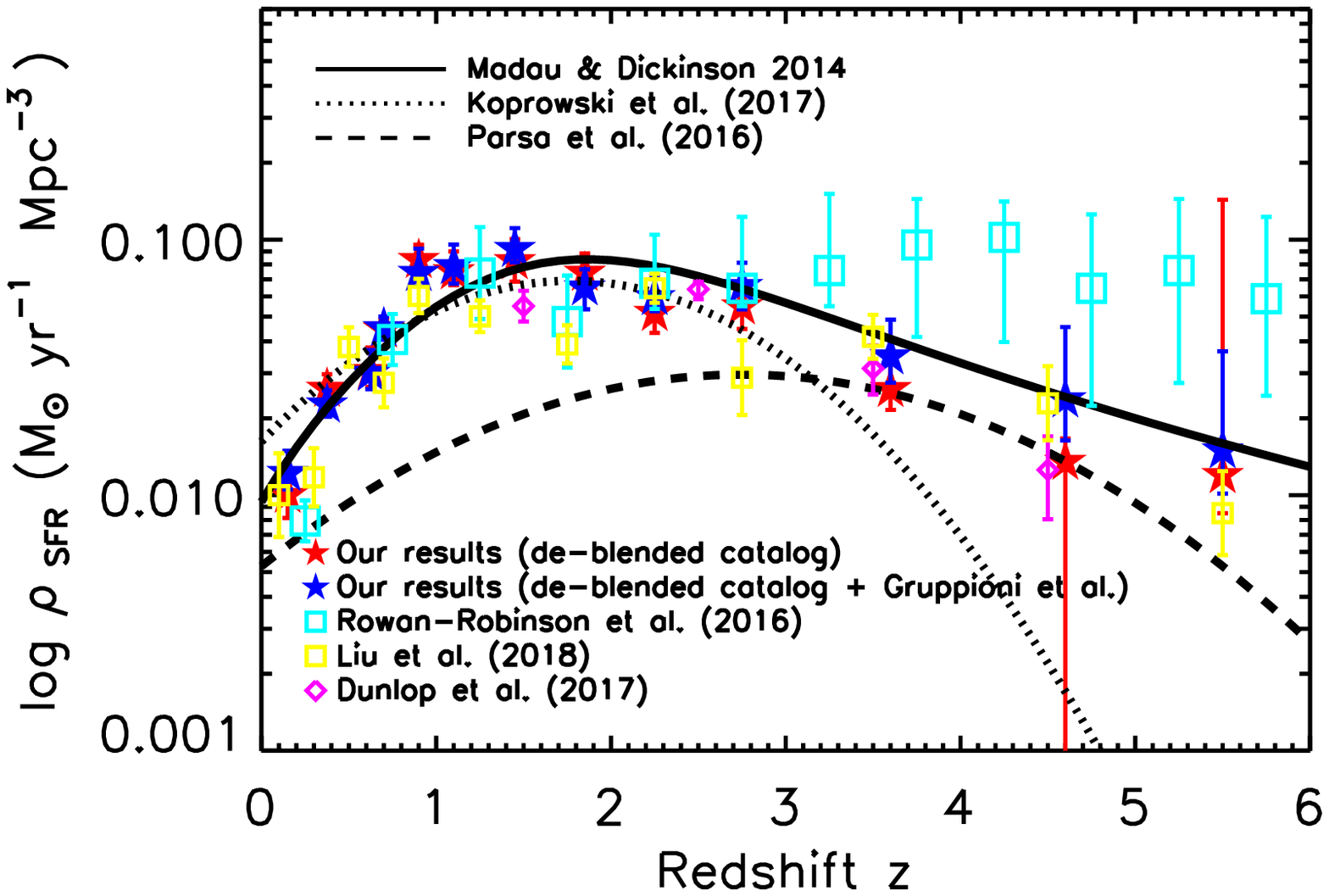}
\caption{Top: Redshift evolution of the total IR luminosity density $\rho_{\rm IR}$ out to $z\sim6$. The results of integrating the best-fit modified Schechter function for our observed total IR LF only (based on the de-blended catalogue in COSMOS) are shown as red stars. The results of integrating the best-fit function for both our observed total IR LF and the Gruppioni et al. (2013) IR LF are shown as blue stars. Error bars on the stars represent the 1$\sigma$ uncertainty. The measurements from Magnelli et al. (2013) are shown as green squares. The measurements from Gruppioni et al. (2013) are shown as black squares. The black solid line shows the predicted $\rho_{\rm IR}$ as a function of redshift from GALFORM. The black dotted line shows the predicted $\rho_{\rm IR}$ of the starburst galaxy population from GALFORM. The black dashed line shows the predicted $\rho_{\rm IR}$ of the quiescent galaxy population from GALFORM. Bottom: The co-moving SFRD $\rho_{\rm SFR}$ as a function of redshift out to $z\sim6$. Estimates of the dust-obscured SFRD $\rho^{\rm obscured}_{\rm SFR}$ based on the best-fit function for our total IR LF only are shown as red stars. Estimates of $\rho^{\rm obscured}_{\rm SFR}$ based on the joint constraints from our total IR LF and the Gruppioni et al. (2013) IR LF are shown as blue stars. We also compare with other {\it Herschel}-based studies such as Rowan-Robinson et al. (2016) and Liu et al. (2018) and the ALMA-based study Dunlop et al. (2017). The black dotted line shows the best-fit function of the evolution of $\rho^{\rm obscured}_{\rm SFR}$ from Koprowski et al. (2017). The black dashed line corresponds to the UV-based unobscured SFRD estimates $\rho^{\rm unobscured}_{\rm SFR}$ from Parsa et al. (2016). The black solid line shows the parametric description of the evolution of the total SFRD $\rho_{\rm SFR}$ provided by Madau \& Dickinson (2014).}
\label{csfh}
\end{figure*}

The deep SPIRE maps contain most of the emission in the cosmic IR background which arises from the integrated dust emission over the entire cosmic history (Puget et al. 1996). However, due to the large beam, SPIRE observations suffer from source confusion and blending which limits our ability to detect faint objects and de-blend neighbouring sources. At $z\sim3$, the SPIRE $5\sigma$ confusion limit corresponds to IR luminosity $10^{13}L_{\odot}$ which is many times brighter than the expected turnover in the LF. The advent of ALMA  finally closed the gap with UV/optical observations in sensitivity and angular resolution. Dunlop et al. (2017) presented the first deep ALMA image of the 4.5 arcmin$^2$ Hubble Ultra Deep Field and measurement of the SFRD using a total sample of 16 sources over $1<z<5$. Even with its extraordinary sensitivity, the small field of view means that it is still impractical to use ALMA to carry out large blank field surveys to address the CSFH controversy which requires statistically large samples of galaxies. To help resolve the SFRD controversy at $z>3$, we can exploit our SED prior enhanced XID+ de-blended {\it Herschel} photometry (which significantly extend the dynamic range probed by previous studies) to probe the knee of the IR luminosity function and derive much tighter constraints of the SFRD. 

Based on the parametric descriptions of the total IR LF, we can carry out luminosity-weighted integration over sufficient dynamic range in each redshift bin to study the time evolution of the total IR luminosity density $\rho_{\rm IR}$. In the top panel of Fig.~\ref{csfh}, we show the evolution of $\rho_{\rm IR}$ out to $z\sim6$. The results of integrating the best-fitting modified Schechter function (down to $L_{\rm IR} = 10^8 L_{\odot}$, following Gruppioni et al. 2013) for our observed total IR LF only (based on the de-blended catalogue in COSMOS) are shown as red stars. The results of integrating the best-fitting function for both our observed total IR LF and the Gruppioni et al. (2013) IR LF are shown as blue stars. The error bars on the red or blue stars represent the $1\sigma$ uncertainty calculated directly from the MCMC chains from emcee. The two sets of measurements are consistent with each other within errors. In addition, the measurements from Gruppioni et al. (2013) are shown as black squares and the measurements from Magnelli et al. (2013) are shown as green squares. It is clear that our measurements (the blue stars and the red stars) are consistent with these two previous {\it Herschel}-based studies in the overlapping redshift range. Due to the large fraction of photometric redshifts and the fact that the PEP selection might miss high-redshift sources, the Gruppioni et al. (2013) estimate in the redshift bin $3.0<z<4.2$ is likely to be a lower limit. We also plot the predicted total IR luminosity density as well as the IR luminosity density from the starburst and quiescent galaxy populations separately from GALFORM as a function of redshift. We find that the starburst population dominates at high redshift and the quiescent population dominates at low redshift. The transition happens at around $z\sim1.5$. The GALFORM predicted total IR luminosity densities agree reasonably well with the observations at $z<3$. At $z>3$, the GALFORM predictions are much higher compared to observed values\footnote{Cowley et al. (2018) discusses the considerable difference between the intrinsic cosmic star formation history predicted by GALFORM and the apparent cosmic star formation history derived by converting the IR luminosity density into SFR volume density.}.

To derive the dust-obscured SFR volume density $\rho^{\rm obscured}_{\rm SFR}$, we multiply our estimates of $\rho_{\rm IR}$ by a constant factor of $10^{-10} M_{\odot}~yr^{-1} / L_{\odot}$ (B{\'e}thermin et al. 2017) which is derived from the Kennicutt (1998) conversion factor after converting to the Chabrier (2003) IMF. In the bottom panel of Fig.~\ref{csfh}, we show our measurements of $\rho^{\rm obscured}_{\rm SFR}$ based on the de-blended catalogue in COSMOS only (the red stars) and based on the joint constraints from our de-blended catalogue and the Gruppioni et al. (2013) measurements (the blue stars). To avoid overcrowding, estimates of $\rho^{\rm obscured}_{\rm SFR}$ based on measurements from Magnelli et al. (2013) and Gruppioni et al. (2013) are not shown as they are consistent with our results based on the good agreement seen in the top panel of Fig.~\ref{csfh} and the fact that the same conversion factor is applied to convert $\rho_{\rm IR}$ into $\rho_{\rm SFR}$. We also compare with two other {\it Herschel}-based studies, i.e., Rowan-Robinson et al. (2016) with small updates from Rowan-Robinson et al. (2018) in some redshift bins and Liu et al. (2018), and the ALMA-based study by Dunlop et al. (2017). Rowan-Robinson et al. (2016) uses a novel approach of selecting 500 $\mu$m sources from a combination of several large {\it HerMES} fields totalling $\sim20$ deg$^2$, in order to extend the measurements of Gruppioni et al. (2013) out to $z\sim6$. We find that the measurements of Rowan-Robinson et al. (2016) are systematically higher than our results at $z>3$, although they are still marginally consistent. The estimates of Liu et al. (2018) are derived by using super-deblended dust emission in galaxies in the GOODS-North field, based on prior catalogues constructed from deep Spitzer 24 $\mu$m and VLA 20 cm detections and progressive de-blending from less to more confused bands. Our results agree well with the $\rho^{\rm obscured}_{\rm SFR}$ estimates derived by Liu et al. (2018). The ALMA-derived measurements of the dust obscured $\rho_{\rm SFR}$ based on a sample of 16 sources from Dunlop et al. (2017) also agree reasonably well with our estimates.

We also compare with the unobscured SFRD estimates $\rho^{\rm unobscured}_{\rm SFR}$ from Parsa et al (2016) which is based on converting the from the rest-frame UV (1500 {\AA}) luminosity to UV-visible SFR. We do not find evidence for a shift of balance between $z\sim3$ and $z\sim4$ in the CSFH from being dominated by unobscured star formation at high redshift to obscured star formation at low redshift\footnote{Our conclusion is of course subject to uncertainties associated with estimates of the unobscured SFRD.}, as found by previous studies of Koprowski et al. (2017), Dunlop et al. (2017) and Bourne et al. (2017). For example, the black dotted line in the right panel of Fig.~\ref{csfh} shows the best-fitting function of the evolution of $\rho^{\rm obscured}_{\rm SFR}$ from Koprowski et al. (2017) which crosses the evolution of $\rho^{\rm unobscured}_{\rm SFR}$ from Parsa et al (2016) (i.e. the dashed line) at $z\sim3$.  However, we do find the redshift range $3<z<4$ to be an interesting transition period as the fraction of the total SFRD that is obscured by dust is significantly lower at higher redshift compared to at lower redshifts. The fraction of dust obscured SF activity is at its highest ($>80\%$) around $z\sim1$ which then decreases towards both low redshift and high redshift.

\section{Discussions and conclusions}

We make use of our state-of-the-art multi-wavelength de-blended {\it Herschel}-SPIRE catalogue in the COSMOS field to study the number counts, the monochromatic and total infrared (integrated from 8 to 1000 $\mu$m) luminosity functions, and the dust-obscured cosmic star-formation history. We compare our results with previous determinations from single dish and interferometric observations and predictions from both empirical models and physically-motivated models (including semi-analytic simulations and hydrodynamic simulations). Our main conclusions are the following:

\begin{itemize}

\item Our number counts at the SPIRE wavelength 250 $\mu$m derived from the multi-wavelength de-blended catalogue  in the COSMOS field show good agreement with previous {\it Herschel} measurements. However, the agreement is increasingly worse towards the longer wavelengths 350 and 500 $\mu$m. At 500 $\mu$m, our number counts can be as much as 0.5 dex lower compared to previous {\it Herschel} studies. This is due to previous {\it Herschel} studies suffering from confusion and source blending issues which are increasingly more severe towards longer wavelengths.

\item Our number counts at 450 $\mu$m from the de-blended catalogue in COSMOS agree very well with the JCMT SCUBA-2 450 $\mu$m measurement (with a factor of $\sim5$ improvement in angular resolution), especially with the most recent SCUBA-2 measurements from Zavala et al. (2017) and Wang et al. (2017). 

\item Excellent agreement are found between our predicted number counts at 870 $\mu$m based on the de-blended catalogue in COSMOS and the SCUBA-2 850 $\mu$m, SMA 860 $\mu$m and ALMA 870 $\mu$m measurements which are derived either from single-dish observations or interferometric observations achieving arcsec or even sub-arcsec angular resolution. 

\item Our monochromatic rest-frame 250 $\mu$m luminosity functions agree well with SCUBA-2 measurements (Koprowski et al. 2017) in the overlapping luminosity and redshift range. We extend the Koprowski et al. (2017) measurements by around 1 dex at the faint end, except in the redshift bin $1.5<z<2.5$ where measurements in the two faintest luminosity bins in Koprowski et al. (2017) are derived from ALMA 1.3 mm observations.

\item Our total infrared luminosity function agree well with previous Herschel PACS and SPIRE measurements (Magnelli et al. 2013; Gruppioni et al. 2013) in the overlapping luminosity and redshift range. Thanks to our de-blending technique and the wealth of multi-wavelength photometric information in the COSMOS field, we can also probe much fainter luminosities and out to higher redshifts. We derive the best-fitting modified Schechter function in a number of redshift bins out to $z\sim6$. We find that the characteristic density evolves very mildly for the 8 Gyr and then decreases rapidly (by about two orders of magnitude) from $z\sim1$ to $\sim6$. The characteristic luminosity $L_*$ increases quickly with redshift out to $z\sim2$ and then seems to more or less flatten out to $z\sim6$. As a function of lookback time, $L_*$ evolves simply in a linear fashion.

\item We find a reasonable agreement between our total IR LF and the predictions from GALFORM. The population of starburst galaxies with top heavy IMF is important in matching the observed LF at the bright end. On the other hand, the population of quiescent galaxies in the simulation is important in matching the observations at the faint end. However, at the faint end, GALFORM predictions are considerably below our measured total IR LF at $z<1$.

\item The predicted total IR LF from the EAGLE hydrodynamic simulation are generally lower at the bright end compared to our measurement, which could be caused by the limited volume of EAGLE and lack of high mass halos, a lack of starburst galaxies or the need of adopting a different IMF (e.g., a top heavy IMF) for the starburst population, or the need of changing the subgrid physical prescriptions of the simulation. Interestingly, towards the faint end the predicted total IR LF from EAGLE agree fairly well with our measurements.

\item Our measurement of the co-moving IR luminosity density and the dust-obscured star-formation rate volume density as a function of redshift agree well with previous {\it Herschel} studies but extends to higher redshifts. By comparing with the SFRD estimates derived from UV-based studies, we find that the fraction of dust obscured SF activity is at its highest ($>80\%$) around $z\sim1$ which then decreases towards both low redshift and high redshift. We do not find evidence for a shift of balance between $z\sim3$ and $z\sim4$ in the CSFH from being dominated by unobscured star formation at high redshift to obscured star formation at low redshift. However, we do find the redshift range $3<z<4$ to be an interesting transition period as the fraction of the total SFRD that is obscured by dust is significantly lower at higher redshift compared to at lower redshifts. 

\end{itemize}

\begin{acknowledgements}
M.J.M.~acknowledges the support of the National Science Centre, Poland, through the POLONEZ grant 2015/19/P/ST9/04010; this project has received funding from the European Union's Horizon 2020 research and innovation programme under the Marie Sk{\l}odowska-Curie grant agreement No. 665778. We thank an anonymous referee for helpful comments which have improved the paper.
\end{acknowledgements}

\newpage

\begin{appendix}

\section{Number counts and luminosity function measurements}

We provide our measurements of number counts in COSMOS at the {\it Herschel}-SPIRE wavelengths (250, 350, 500 $\mu$m) in Table~\ref{table:countsspire250}, Table~\ref{table:countsspire350} and Table~\ref{table:countsspire500}. The uncertainties on the number counts only account for Poisson errors. At the bright end, we expect field-to-field variations to be a larger source of uncertainty. In Table~\ref{table:counts450} and Table~\ref{table:counts870}, we provide number counts in COSMOS at 450 and 870 $\mu$m. Note that the 450 $\mu$m counts are derived from the de-blended 500 $\mu$m flux densities after applying a scaling factor of the 450 to 500 $\mu$m flux ratio. The 870 $\mu$m counts are derived from the predicted 870 $\mu$m flux densities based on the best-fit SED. For more details, please refer to Section 3.1.2 and 3.1.3. In Table~\ref{table:IRLF}, we list our measurements of the total infrared luminosity function in 13 redshift bins.
       
\begin{table}
\caption{Our measurement of the number counts at 250 $\mu$m in COSMOS. The flux density $S$ is the centre of the bin. Uncertainties on the counts only represent Poisson errors.}\label{table:countsspire250}
\begin{tabular}{lll}
\hline
$S$ (mJy) & $S^{2.5}dN/dS$ (Jy$^{1.5}$ sr$^{-1}$)   & Poisson error  (Jy$^{1.5}$ sr$^{-1}$)   \\
\hline
     0.95&      5247.40&      15.8654\\
      1.56&      8670.32&      29.6727\\
      2.58&      11872.1&      50.5201\\
      4.25&      15274.6&      83.3767\\
      7.01&      19670.4&      137.666\\
      11.56&      21567.6&      209.740\\
      19.06&      20242.8&      295.650\\
      31.42&      13876.4&      356.156\\
      51.80&      7198.93&      373.247\\
      85.41&      3277.45&      366.430\\
      140.81&      693.836&      245.308\\
\hline
\end{tabular}
\end{table}

 \begin{table}
\caption{Our measurement of the number counts at 350 $\mu$m in COSMOS. The flux density $S$ is the centre of the bin. Uncertainties on the counts only represent Poisson errors.}\label{table:countsspire350}
\begin{tabular}{lll}
\hline
$S$ (mJy) & $S^{2.5}dN/dS$ (Jy$^{1.5}$ sr$^{-1}$)   & Poisson error (Jy$^{1.5}$ sr$^{-1}$)    \\
\hline
      0.95&      5547.49&      16.3128\\
      1.56&      8550.90&      29.4676\\
      2.58&      10501.0&      47.5132\\
      4.25&      11935.8&      73.7033\\
      7.01&      13523.4&      114.147\\
      11.56&      13996.3&      168.962\\
      19.06&      10061.0&      208.431\\
      31.42&      5576.15&      225.772\\
      51.80&      1199.82&      152.378\\
      85.41&      163.872&      81.9362\\
      140.81&      173.459&      122.654\\
 \hline
\end{tabular}
\end{table}

 \begin{table}
\caption{Our measurement of the number counts at 500 $\mu$m in COSMOS. The flux density $S$ is the centre of the bin. Uncertainties on the counts only represent Poisson errors.}\label{table:countsspire500}
\begin{tabular}{lll}
\hline
$S$ (mJy) & $S^{2.5}dN/dS$ (Jy$^{1.5}$ sr$^{-1}$)   & Poisson error  (Jy$^{1.5}$ sr$^{-1}$)   \\
\hline
     0.95&      5464.69&      16.1906\\
      1.56&      6592.61&      25.8743\\
      2.58&      6054.29&      36.0771\\
      4.25&      5303.91&      49.1313\\
      7.01&      4682.50&      67.1676\\
      11.56&      3606.16&      85.7637\\
      19.06&      2012.19&      93.2130\\
      31.42&      530.191&      69.6175\\
      51.80&      154.816&      54.7356\\
 \hline
\end{tabular}
\end{table}

 \begin{table}
\caption{Our measurement of the number counts at 450 $\mu$m in COSMOS. The flux density $S$ is the centre of the bin. Uncertainties on the counts only represent Poisson errors.}\label{table:counts450}
\begin{tabular}{lll}
\hline
$S$ (mJy) & $S^{2.5}dN/dS$ (Jy$^{1.5}$ sr$^{-1}$)  & Poisson error (Jy$^{1.5}$ sr$^{-1}$)    \\
\hline
     0.95&      55612.8&      160.304\\
      1.56&      78844.2&      277.717\\
      2.58&      75856.4&      396.346\\
      4.25&      66137.9&      538.472\\
      7.01&      59380.2&      742.368\\
      11.5&      48491.2&      976.092\\
      19.06&      31445.7&      1143.67\\
      31.42&      10390.6&      956.535\\
      51.80&      1864.15&      589.496\\
      85.41&      789.281&      558.106\\
 \hline
\end{tabular}
\end{table}

 \begin{table}
\caption{Our measurement of the number counts at 870 $\mu$m in COSMOS. The flux density $S$ is the centre of the bin. Uncertainties on the counts only represent Poisson errors.}\label{table:counts870}
\begin{tabular}{lll}
\hline
$S$ (mJy) & $S^{2.5}dN/dS$ (Jy$^{1.5}$ sr$^{-1}$)   & Poisson error  (Jy$^{1.5}$ sr$^{-1}$)   \\
\hline
      0.58&       6161.7472&       36.673168\\
      0.95&       7420.0241&       58.554406\\
       1.56&       7352.2708&       84.806301\\
       2.58&       6266.4922&       113.91740\\
       4.25&       5523.9150&       155.61868\\
       7.01&       4640.5271&       207.53068\\
       11.56&       2121.9831&       204.18791\\
       19.06&       332.75839&       117.64786\\
 \hline
\end{tabular}
\end{table} 

 \begin{table*}
\caption{Our measurement of the total infrared luminosity function in COSMOS. The total IR luminosity $\log L_{\rm IR}$ is the centre of the bin. Uncertainties on the counts only represent Poisson errors.}\label{table:IRLF}
\begin{tabular}{llll}
\hline
Redshift bin & $\log L_{\rm IR} (L_{\odot})$  & LF (Mpc$^{-3}$ dex$^{-1}$)  & Poisson error  (Mpc$^{-3}$ dex$^{-1}$)  \\
\hline
$      0.0<z<0.3$ &        9.75&   0.0084&  0.0002\\
$      0.0<z<0.3$ &       10.25&   0.0045&  0.0002\\
$      0.0<z<0.3$ &       10.75&   0.0012&  8.4947e-05\\
$      0.0<z<0.3$ &       11.25&  0.0002&  3.2696e-05\\
$      0.0<z<0.3$ &       11.75&  5.7799e-06&  5.7799e-06\\
$     0.3<z<0.45$ &       9.75&    0.0109&  0.0002\\
$     0.3<z<0.45$ &       10.25&   0.0094&  0.0002\\
$     0.3<z<0.45$ &       10.75&   0.0043&  0.0001\\
$     0.3<z<0.45$ &       11.25&  0.0007&  4.5427e-05\\
$     0.3<z<0.45$ &       11.75&  2.3124e-05&  8.1754e-06\\
$     0.45<z<0.6$ &       10.25&   0.0086&  0.0001\\
$     0.45<z<0.6$ &       10.75&   0.0044&  8.7627e-05\\
$     0.45<z<0.6$ &       11.25&   0.0012&  4.5986e-05\\
$     0.45<z<0.6$ &       11.75&  8.5720e-05&  1.2246e-05\\
$     0.45<z<0.6$ &       12.25&  3.4988e-06&  2.4740e-06\\
$     0.6<z<0.8$ &       10.25&   0.0095&  9.2757e-05\\
$     0.6<z<0.8$ &       10.75&   0.0064&  7.5702e-05\\
$     0.6<z<0.8$ &       11.25&   0.0016&  3.7752e-05\\
$     0.6<z<0.8$ &       11.75&  0.0002&  1.4314e-05\\
$     0.6<z<0.8$ &       12.25&  1.1722e-05&  3.2510e-06\\
$     0.6<z<0.8$ &       12.75&  9.0166e-07&  9.0166e-07\\
$     0.8<z< 1.0$ &       10.75&   0.0091&  7.9075e-05\\
$     0.8<z< 1.0$ &       11.25&   0.0038&  5.1005e-05\\
$     0.8<z< 1.0$ &       11.75&  0.0007&  2.1651e-05\\
$     0.8<z< 1.0$ &       12.25&  4.1722e-05&  5.3420e-06\\
$      1.0<z<1.2$ &       10.75&   0.0070&  6.3435e-05\\
$      1.0<z<1.2$ &       11.25&   0.0038&  4.6495e-05\\
$      1.0<z<1.2$ &       11.75&  0.0007&  2.0429e-05\\
$      1.0<z<1.2$ &       12.25&  5.9408e-05&  5.8254e-06\\
$      1.0<z<1.2$ &       12.75&  5.7123e-07&  5.7123e-07\\
$      1.0<z<1.2$ &       13.25&  5.7123e-07&  5.7123e-07\\
$      1.2<z<1.7$ &       11.25&   0.0036&  2.6300e-05\\
$      1.2<z<1.7$ &       11.75&   0.0010&  1.3871e-05\\
$      1.2<z<1.7$ &       12.25&  0.0001&  4.6736e-06\\
$      1.2<z<1.7$ &       12.75&  3.2517e-06&  7.8865e-07\\
$      1.7<z<2.0$ &       11.25&   0.0026&  2.7386e-05\\
$      1.7<z<2.0$ &       11.75&   0.0011&  1.7924e-05\\
$      1.7<z<2.0$ &       12.25&  0.0002&  7.2243e-06\\
$      1.7<z<2.0$ &       12.75&  1.4403e-05&  2.0368e-06\\
$      1.7<z<2.0$ &       13.25&  2.8805e-07&  2.8805e-07\\
$      2.0<z<2.5$ &       11.25&   0.0015&  1.5965e-05\\
$      2.0<z<2.5$ &       11.75&  0.0007&  1.1118e-05\\
$      2.0<z<2.5$ &       12.25&  0.0002&  5.4133e-06\\
$      2.0<z<2.5$ &       12.75&  2.1059e-05&  1.8761e-06\\
$      2.0<z<2.5$ &       13.25&  5.0141e-07&  2.8949e-07\\
$      2.5<z<3.0$ &       11.75&  0.0008&  1.1329e-05\\
$      2.5<z<3.0$ &       12.25&  0.0002&  5.7064e-06\\
$      2.5<z<3.0$ &       12.75&  2.4608e-05&  2.0296e-06\\
$      2.5<z<3.0$ &       13.25&  5.0220e-07&  2.8995e-07\\
$      3.0<z<4.2$ &       11.75&  0.0004&  5.2906e-06\\
$      3.0<z<4.2$ &       12.25&  8.6658e-05&  2.5324e-06\\
$      3.0<z<4.2$ &       12.75&  3.4042e-06&  5.0192e-07\\
$      4.2<z<5.0$ &       12.25&  7.1666e-05&  2.9656e-06\\
$      4.2<z<5.0$ &       12.75&  2.5525e-05&  1.7698e-06\\
$      4.2<z<5.0$ &       13.25&  3.6815e-07&  2.1255e-07\\
$      5.0<z<6.0$ &       12.25&  4.7607e-05&  2.2721e-06\\
$      5.0<z<6.0$ &       12.75&  1.7026e-05&  1.3588e-06\\
$      5.0<z<6.0$ &       13.25&  9.7599e-07&  3.2533e-07\\
 \hline
\end{tabular}
\end{table*} 

\end{appendix}
\end{document}